\documentclass[a4paper,11pt]{article}
\pdfoutput=1 

\usepackage{jinstpub} 

\usepackage{lineno}

\RequirePackage[T1]{fontenc}

\RequirePackage{graphicx}

\usepackage[config=altsf]{subfig}

\RequirePackage{mathptmx}      

\DeclareSymbolFont{cmlargesymbols}{OMX}{cmex}{m}{n}

\let\sumop\relax

\DeclareMathSymbol{\sumop}{\mathop}{cmlargesymbols}{"50}

\RequirePackage{flushend}

\usepackage{amsmath}

\usepackage{multirow}

\usepackage{color}

\usepackage[utf8]{inputenc}

\usepackage[english]{babel}

\usepackage[nottoc]{tocbibind}

\RequirePackage[numbers,sort&compress]{natbib}

\title{Bibliography management: \texttt{natbib} package}

\usepackage{csquotes}

\title{\boldmath Jet performance at the Circular electron-positron Collider}

\author[a]{Pei-Zhu Lai,\note{Corresponding author.}}
\author[b,1]{Manqi Ruan,}
\author[a]{and Chia-Ming Kuo}

\affiliation[a]{National Central University,\\No. 300, Zhongda Rd., Zhongli District, Taoyuan City 32001, Taiwan}
\affiliation[b]{Institute of High Energy Physics,\\19B Yuquan Road, Shijingshan District, Beijing, China}

\emailAdd{Manqi.ruan@ihep.ac.cn}

\abstract{Jet reconstruction is critical for the precision measurement of Higgs boson properties and the electroweak observables at the CEPC. We analyze the jet energy and angular responses of benchmark 2- and 4-jet processes with fully simulated samples with the CEPC baseline detector geometry. We observe a relative resolution of 3.5$\%$ and 1$\%$ on the jet energy and angular measurement for jets in the detector barrel region ($|cos{\theta}| < 0.6$) with energy greater than 60 GeV. Meanwhile, the jet energy/angular scale can be controlled within 0.5/0.01$\%$. The differential dependences of the jet response on the jet direction and energy are extracted. We also analyze the impact on the jet responses induced by different jet clustering algorithms and matching criteria, which yields a relative difference of up to 8$\%$.}

\def\Plus{\texttt{+}}

\def\Minus{\texttt{-}}

\def\Slash{\texttt{/}}

\begin{document}
\maketitle
\flushbottom

\section{Introduction}
The Higgs boson, discovered in 2012 at the Large Hadron Collider (LHC)~\cite{R1, R2, R3, R4, R5, R6, R7}, renders new possibilities to probe the mysteries of the universe, including the origin of matter, nature of dark matter, Hierarchy problem, naturalness problem, and the stability of the vacuum. Compared to the LHC, the currently existing Higgs factory, electron-positron Higgs factories can provide enormous critical information on the Higgs boson properties. Their collision circumstances are free of QCD background. Furthermore, they can meticulously record Higgs boson events with an efficiency of nearly 100$\%$, and can clearly identify those events from the SM backgrounds. Via the recoil mass method, it is possible to determine the absolute value of the generation cross section of the Higgsstrahlung process (a.k.a. ZH process), one of the major production modes of the Higgs boson for electron-positron collisions with a center-of-mass energy $\left(\sqrt{s}\right)$ of 240 GeV. Anchored by the absolute measurement of $\sigma \left(ZH\right)$, electron-positron colliders can then measure the absolute value of Higgs boson production cross sections, decay branching ratios, total width, and the coupling constant with its decay products.
The Circular Electron Positron Collider (CEPC)~\cite{R8} is one of the proposed electron-positron Higgs factories. Others include the International Linear Collider (ILC)~\cite{Djouadi:1056637}, the Compact LInear Collider (CLIC)~\cite{Robson:2652846}, and the Future Circular Colldier (FCC)~\cite{Benedikt:2651299,Benedikt:2651300}. Because of the comparative advantage of the electron-positron Higgs factory, its complementarity with respect to the competitiveness of the LHC, and its prospection of a consequent upgrade, the particle physics European strategy update has identified that “an electron-positron Higgs factory is regarded as the highest-priority next collider for the field.”~\cite{CERN-ESU-016}

Installed in a tunnel with a main ring circumference of 100 km, the CEPC will operate at $\sqrt{s}$ = 240 GeV as a Higgs factory, at $\sqrt{s}$ = 91.2 GeV as a Z factory, and operate the W threshold scan at $\sqrt{s}$ around 160 GeV. 
The baseline operation scheme and corresponding boson yields are summarized in Table~\ref{Operation_Scheme}~\cite{R8}.
The CEPC is anticipated to provide unprecedented accuracies on the measurement of Higgs couplings and the electroweak (EW) parameters, on top of the ultimate accuracy of the HL-LHC and the current accuracies~\cite{R9, R10}.

\begin{table*}
\caption{The CEPC operation scheme of different modes, including center-of-mass energies ($\sqrt{s}$), corresponding expected instantaneous luminosities ($L$) per interaction point (IP), total integrated luminosities and event yields. (*) The maximum reachable instantaneous luminosity depends on the solenoid magnetic field of the detector, here considering 2 Tesla. ($\dagger$) Another $5 \times 10^{8}$ single $Z$ events would be brought at the Higgs factory through initial-state-radiation return process. ($\ddagger$) Extra $9.4 \times 10^{7}$ $W^{ \Plus }W^{ \Minus }$ pairs would be collected at $\sqrt{s}$ = 240 GeV.} 
\label{Operation_Scheme}
\begin{tabular*}{\textwidth}{@{\extracolsep{\fill}}cccccc@{}}
\hline
Operation mode & \multicolumn{1}{c}{$\sqrt{s}$ (GeV)} & \multicolumn{1}{c}{$L$ per IP ($10^{34} cm^{-2} s^{-1}$)} & \multicolumn{1}{c}{Years} & \multicolumn{1}{c}{Total $L$ ($ab^{-1}$, 2 IPs)} & \multicolumn{1}{c}{Event yields} \\
\hline
$H$ 				& 	240		  	& 3 			& 7 & 	5.6 	& $1 \times 10^{6}	$\\
$Z$ 					& 	91.2 			& 32(*) 	& 2 &	 16 	& $7 \times 10^{11}	\left((\dagger\right)$\\
$W^{ \Plus }W^{ \Minus }$ & 	158-172 	& 10 		& 1 & 	2.6 	& $2 \times 10^{7}	\left(\ddagger\right) $\\
\hline
\end{tabular*}
\end{table*}

The reconstruction of the hadronic system is critical for the precise Higgs and EW measurements at the CEPC for two reasons.
Firstly, the majority of Higgs and EW processes have hadronic (multi-jet) final states.
The main Higgs boson production mode for $e^{ \Plus }e^{ \Minus }$ collisions at $\sqrt{s} = 240$ GeV is the ZH process, with about 30$\%$ of its events decaying into 2-jet final states, 60$\%$ decaying into 4-jet final states, and 8$\%$ decaying into 6-jet final states.
Meanwhile, the major di-boson processes, WW and ZZ, also have 42$\%$ and 49$\%$ of their events decaying into 2-jet and 4-jet final states, respectively.
Secondly, the CEPC has a significant advantage in identifying and measuring hadronic final states with respect to the LHC since it is free of QCD backgrounds and pileups.

The reconstruction of the hadronic system can be divided into two different stages: the first one is the identification of the hadronic system and the measurement of its 4-momenta; the second stage is to identify jets from the hadronic system via a jet clustering algorithm. At the CEPC, the hadronic system is composed of up to 100 final-state particles~\cite{R8}. Ref~\cite{Yu_2017} shows that the leptons at the CEPC detector can be identified with high efficiency and purity. Therefore, at the CEPC, the hadronic system can be, in principle, well identified. Furthermore, the accuracy of 4-momentum measurements of the hadronic system can be characterized by the relative mass resolution. 
The boson mass resolution (BMR), defined as the relative mass resolution of the massive boson decays into hadronic final states, is therefore introduced.

\begin{figure} 
\begin{minipage}{\columnwidth}
\centering
  \includegraphics[width=0.45\columnwidth]{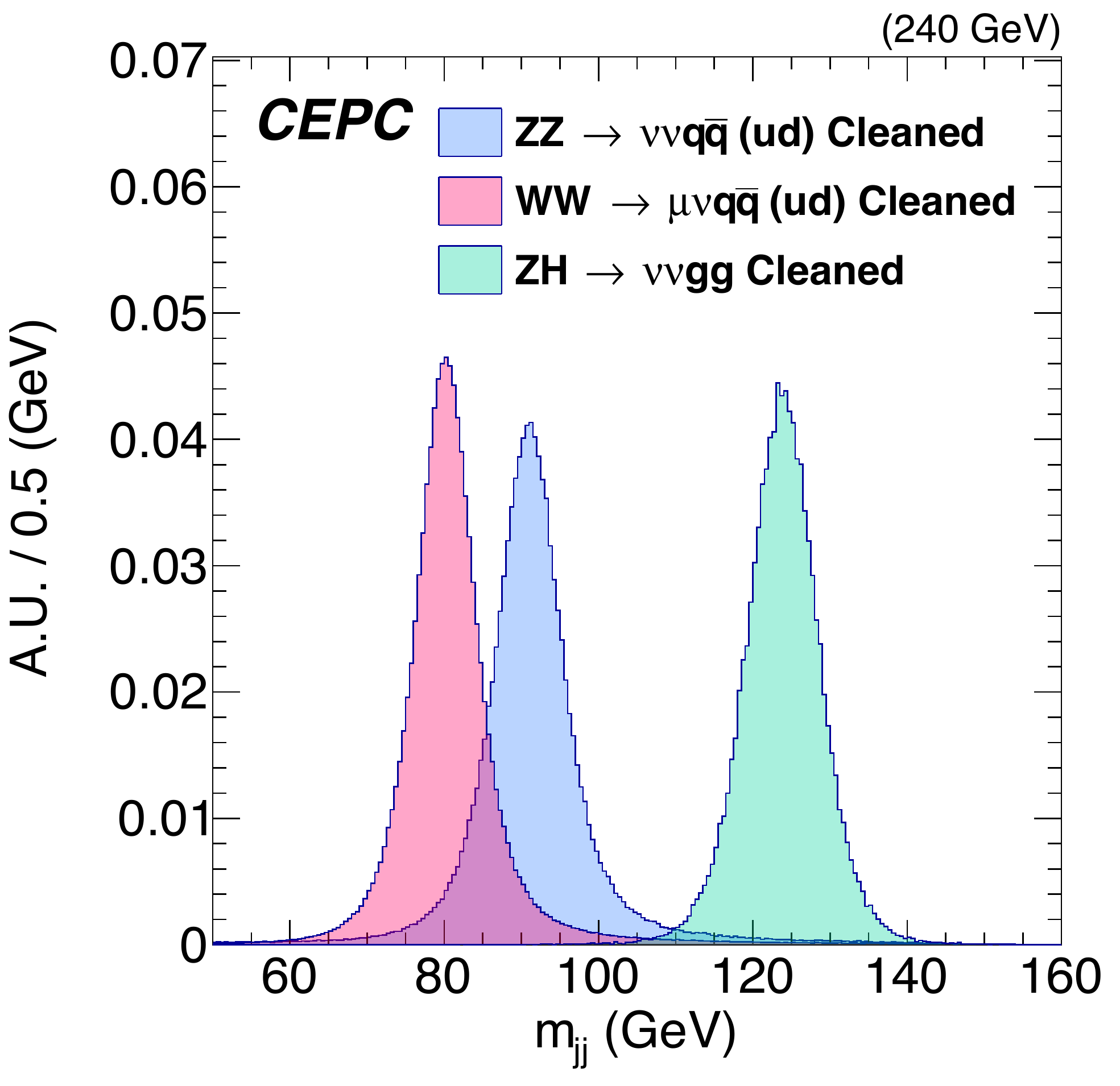}  
\end{minipage}
\caption{Reconstructed di-jet mass distribution. The Higgs boson mass resolution is achieved to be 3.8$\%$ at the CEPC.}
\label{WZH_mass}
\end{figure}

\begin{figure} 
\begin{minipage}{\columnwidth}
\centering
  \includegraphics[width=0.42\columnwidth]{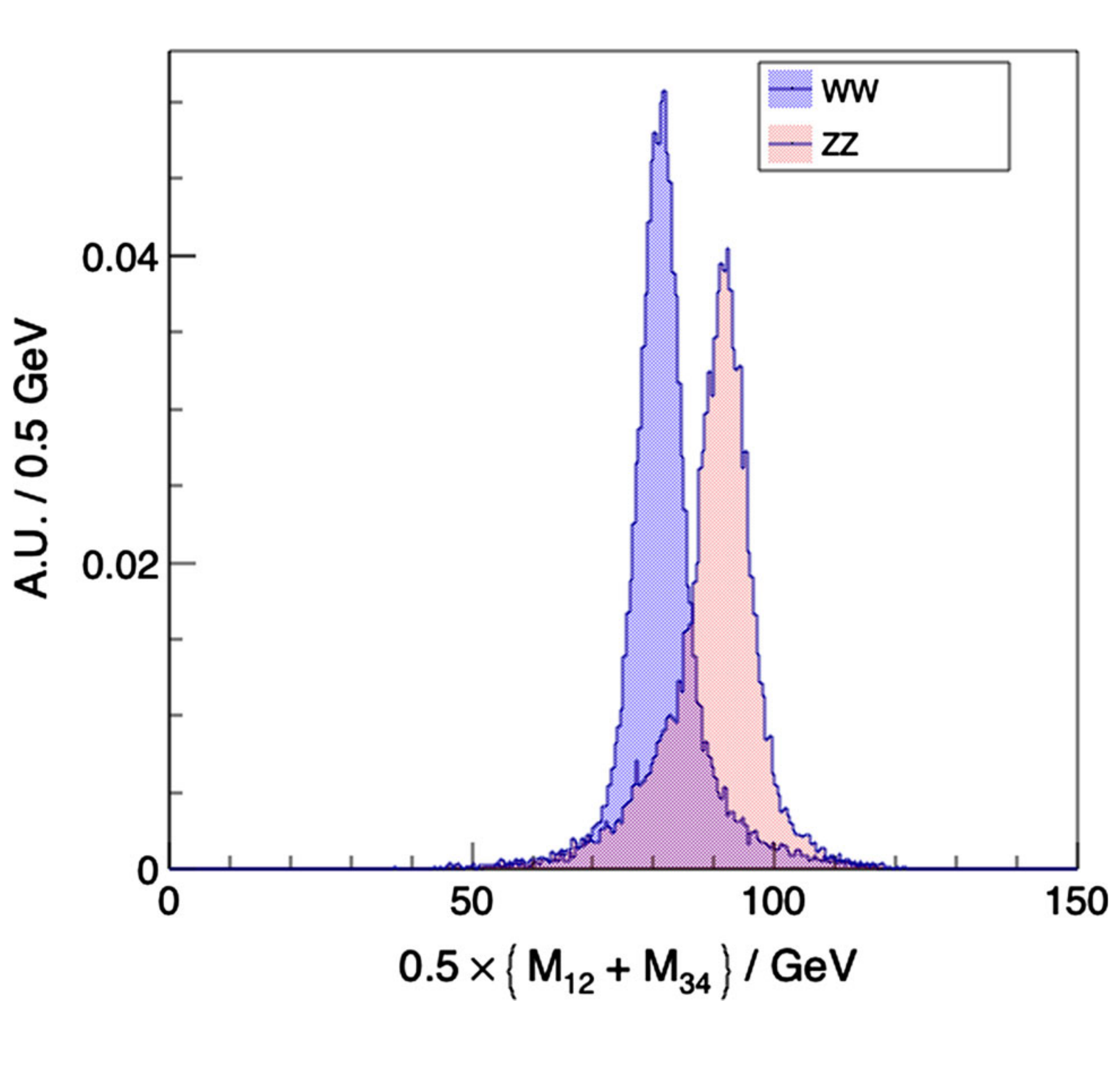}  
\end{minipage}
\caption{The RecoJet invariant mass distribution of the WW and ZZ fully-hadronic final states with equal mass condition. The overlapping fraction (region) between the W- and Z-boson distributions is 39.9$\%$ $\pm$ 0.40$\%$ with $2.5\sigma$ separation of their mass peak. More detailed descriptions are in~\cite{zhu2018performance}.}
\label{WWZZ_mass}
\end{figure}

According to the CEPC Conceptual Design Report (CDR)~\cite{R8}, a boson mass resolution better than 4$\%$ is required at the CEPC to successfully separate the qqH signal from the ZZ (to qqX) background using the recoil mass to the di-jet system. Therefore, a Higgs BMR of 3.8$\%$ leads to a neat separation between the W, the Z, and the Higgs bosons with hadronic final states in their reconstructed invariant mass spectrum 
(see Fig.~\ref{WZH_mass}~\cite{R8}).
Another important benchmark is the WW-ZZ separation, where the overlapping fraction is defined as the overlapping region between the two distributions. An overlapping fraction of 39.9$\%$ $\pm$ 0.40$\%$ is achived for the mass separation of WW and ZZ fully-hardonic final states and it leads to 2.5$\sigma$ separation in their main mass peak, as shown in Fig.~\ref{WWZZ_mass}~\cite{zhu2018performance}.

The hadronic system can be divided into jets via a jet clustering algorithm. Conventionally, those jets represent the color triplet/octet (quarks and gluons) or their combinations at the parton-level, and the identification of jets enable many important physics measurements, such as the strong coupling constant~\cite{Olive_2016}, electroweak parameters~\cite{Ofierzynski_2008, CACCIA1999246}, the triple gauge boson couplings~\cite{gounaris1996triple, Heister_et_al__2001, Diehl_1993br}, and many differential cross sections~\cite{collaboration2018measurements, triple_gluon_LEP}. 
The jet flavor and charge measurements identify the nature of the parton. The identification of jet color/charge is essential for the g(Hbb), g(Hcc), and g(Hgg) measurements as well as the weak mixing angle ($sin^{2} \theta_{W}$) measurement. 
Meanwhile, for the hadronic final states composed of more than three partons such as the fully-hadronic final state of WW, ZZ, and ZH events~\cite{zhu2018performance}, the identification and pairing of jets is critical for the identification of the physics process as the invariant mass of the jet pair provides cardinal information of the intermediate boson state.
In short, the jet clustering, and the further operations upon identified jets, is critical for the physics measurements concerning hadronic final states.  
On the other hand, the representation of the reconstructed jet as a parton is certainly not ideal: any combination of the final-state particles is a color singlet, and it does not carry the fractional charge as the quark. 

\begin{figure}[!ht]
\centering
\subfigure[]{
\begin{minipage}[t]{0.45\linewidth}
\includegraphics[width=1.0\columnwidth]{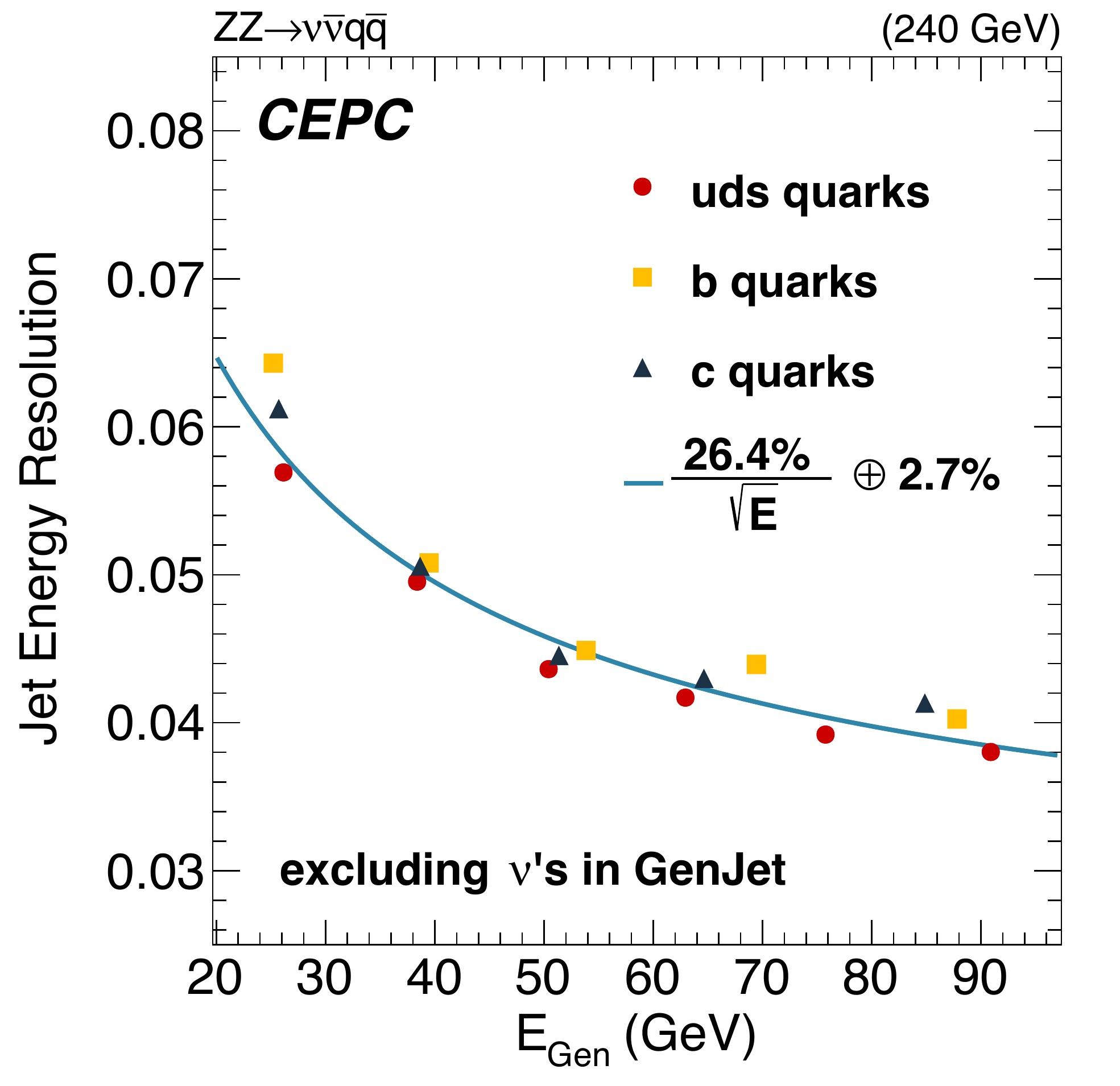}    
\label{JER_Flavor_Wether_nu_1}
\end{minipage}%
}%
\subfigure[]{
\begin{minipage}[t]{0.45\linewidth}
\includegraphics[width=1.0\columnwidth]{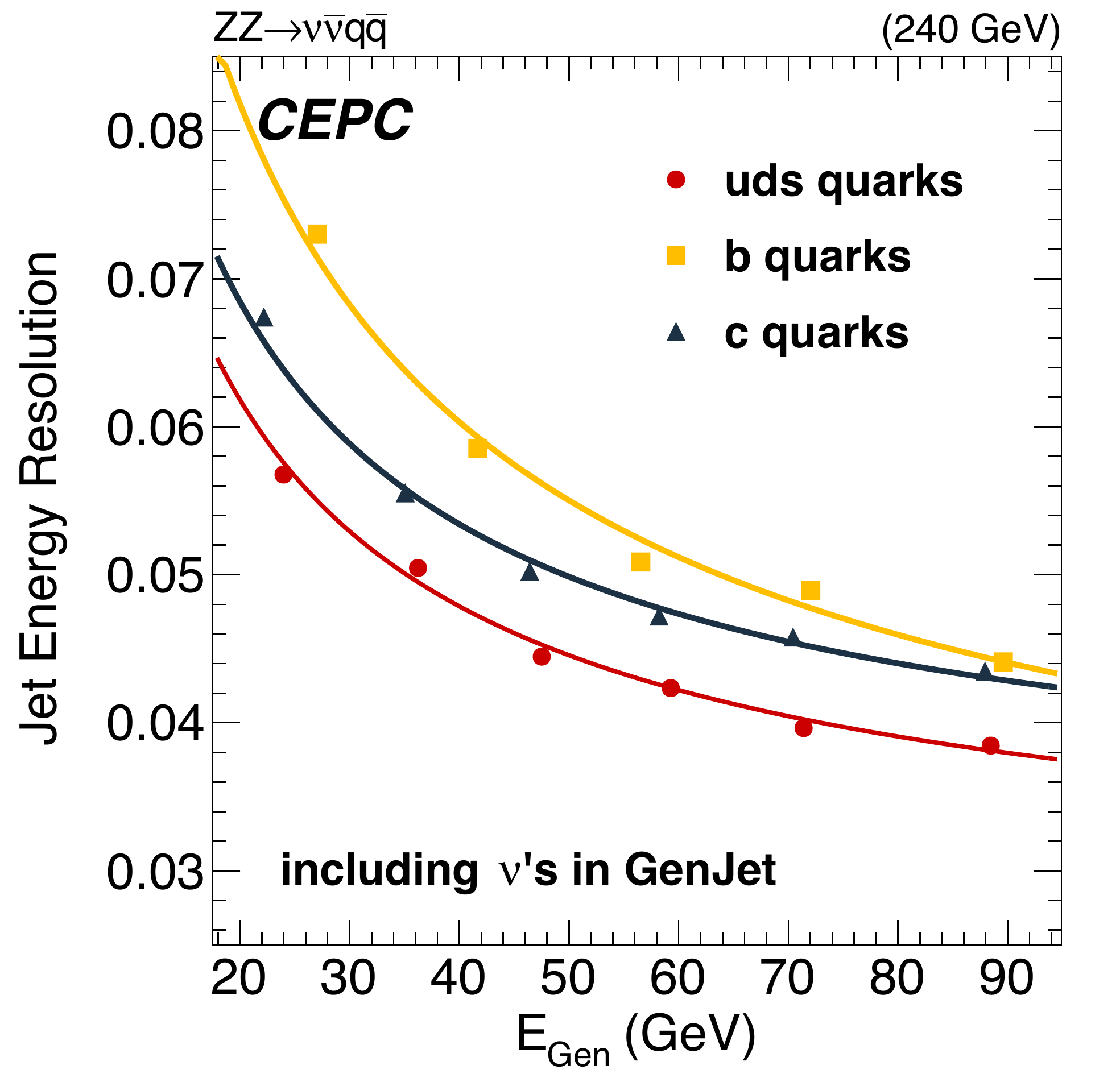}    
\label{JER_Flavor_Wether_nu_2}
\end{minipage}%
}%
\caption{The jet energy resolution is shown using GenJet with (a) excluded neutrinos and (b) included neutrinos to illustrate the fluctuation induced from b- and c-jets. The errors shown are only statistical. The intrinsic detector responses to c- and b-jets are similar to the light-flavored jet when excluding the neutrinos from the GenJet, as seen in (a). When neutrinos are included in the GenJet, the neutrinos produced from the semi-leptonic decay of the heavy-flavored quarks cause the energy loss in RecoJet and the jet energy resolutions degrade as shown in (b).
}
\label{JER_Flavor_Wether_nu}
\end{figure}

The $e^{ \Plus }e^{ \Minus }k_{t}$ algorithm (a.k.a. Durham algorithm~\cite{CATANI1991432}) is used as the baseline jet clustering algorithm at the CEPC using the FastJet package~\cite{Fast_Jet}. The $e^{ \Plus }e^{ \Minus }k_{t}$ algorithm has a single distance criterion: 
\begin{center}
\begin{equation} \label{eq:3}
d_{ij} =  2 min \left( E^{2}_{i}, E^{2}_{j} \right) \left( 1 - cos \theta_{ij} \right)
\end{equation}
\end{center}

for each particle pair $i$ and $j$ with energy $E$, and separated by an angle $\theta_{ij}$.
To differentiate the impact induced by the jet clustering algorithm, and the detector performance, we introduce three objects:  
\begin{itemize}
  \item "Partons" represent the initial quarks or gluons provided by the generator software. In our study, the parton-level information is calculated with the Whizard software~\cite{R13_Whizard, R14_Whizard}.
  \item "GenJets" are the clustered true-level Monte Carlo (MC) particles that were produced from the hadronization of the partons, simulated by Pythia~\cite{Pythia_ref}, including the subsequent decay products such as photons, leptons, or other lighter hadrons. Neutrinos are excluded in the clustering, while only the decay products of hadrons with $c\tau > 1$ cm are included.
   \item "RecoJets" are clustered from the reconstructed final-state particles by the $e^{ \Plus }e^{ \Minus }k_{t}$ algorithm in the same way as the "GenJet".
\end{itemize}
The relative difference between the Parton and the GenJet represents the theoretical uncertainties of a given jet clustering algorithm, 
while the relative difference between the GenJet and RecoJet characterizes the detector performance. 
The GenJet and RecoJet are mapped to each other with the combination that minimizes the sum of angles between the GenJet-RecoJet pairs.
For a given pair, the relative difference is then expressed in terms of the jet energy resolution (JER), the jet energy scale (JES), the jet angular resolution (JAR), and the jet angular scale (JAS). A more detailed elucidation of the methodology is elaborated in Sect. \ref{2.1}.
The GenJets are clustered from visible final-state MC particles. When the neutrinos are included in GenJets, the JER to the light-flavored jets and heavy-flavored jets are highly different, see the Fig. ~\ref{JER_Flavor_Wether_nu}.

\begin{figure} 
\begin{minipage}{\columnwidth}
\centering
  \includegraphics[page=1, width=0.46\columnwidth]{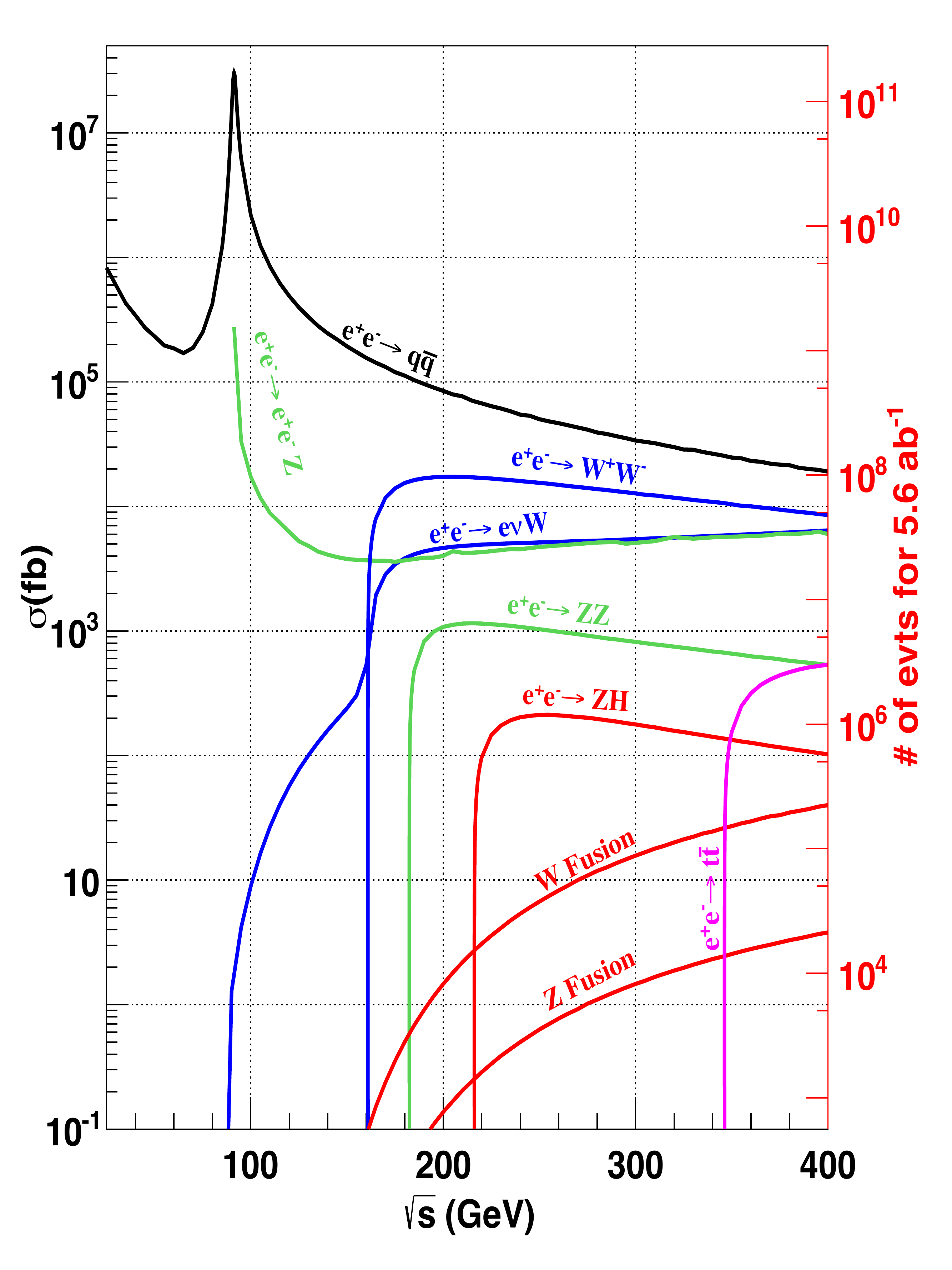}  
\end{minipage}
\vspace{0.2cm}
\caption{The cross section of physical process as a function of the center-of-mass energy for electron-positron colliders~\cite{R8}. The performance of the jets which decay from several benchmark standard model processes are studied in this paper.}
\label{XS_as_Energy}
\end{figure}

This paper analyzes the jet reconstruction performance of several major standard model hadronic processes at the CEPC, 
including $Z \rightarrow q\bar{q}$ at Z-pole, $ZZ \rightarrow \nu \bar{\nu} q \bar{q}$ or $q \bar{q} q \bar{q} $, $WW \rightarrow \mu \nu q \bar{q}$ or $q \bar{q} q \bar{q} $, and $ZH \rightarrow \nu \bar{\nu} \left( q \bar{q} \; or \; gg \right)$ or $q \bar{q} \left( q \bar{q} \; or \; gg \right)$ at $\sqrt{s}$ = 240 GeV, as shown in Fig.~\ref{XS_as_Energy}. 
The simulated statistic for this paper and expected number of events in the final 5$ab^{-1}$ integrated luminosity through the official simulation chain (elaborated in Sect. \ref{2}) are summarized in Table~\ref{Statistic_Info}. To characterize the detector performance, focus will be given on the comparison between GenJet and RecoJet.

\begin{table*}
\caption{Expected number of events in the final 5$ab^{-1}$ integrated luminosity and the simulated statistic for this paper with the official simulation chain (elaborated in Sect. \ref{2}). ($\dagger$) One of the Z bosons is forced to decay to neutrinos in the final state.} 
\label{Statistic_Info}
\begin{tabular*}{\textwidth}{@{\extracolsep{\fill}}clcc@{}}
\hline
Operation mode & \multicolumn{1}{l}{ Process } & \multicolumn{1}{c}{Simulated} & \multicolumn{1}{c}{Expected}\\
\hline
$Z$ 		& $e^{ \Plus }e^{ \Minus } \rightarrow Z \rightarrow q\bar{q}$		  																& 10M 		& 	0.7T \\
\hline
               & $e^{ \Plus }e^{ \Minus } \rightarrow WW \rightarrow \mu \nu q\bar{q}$                  								&  5.8M   &     6.8M  \\
               & $e^{ \Plus }e^{ \Minus } \rightarrow WW \rightarrow q\bar{q}q\bar{q}$                  								&  5.2M   & 		44M \\                           
$H$     	& $e^{ \Plus }e^{ \Minus } \rightarrow ZZ \rightarrow \nu \bar{\nu} q\bar{q}$            								&  1.1M    & 		0.9M\\
               & $e^{ \Plus }e^{ \Minus } \rightarrow ZZ \rightarrow q\bar{q}q\bar{q}$                     								&  2.5M   &		2.9M\\
               & $e^{ \Plus }e^{ \Minus } \rightarrow ZH \rightarrow \nu \bar{\nu}\left(q\bar{q} \; or \; gg\right)\left(\dagger\right)$ 	&  233K   &		145K\\
               & $e^{ \Plus }e^{ \Minus } \rightarrow ZH \rightarrow q\bar{q}\left(q\bar{q} \; or \; gg\right)$                    			&  680K   & 	479K\\
                           
\hline
\end{tabular*}
\end{table*}

The CEPC detector has generally 3.5\% and 1\% JER and JES, respectively, for jets within the detector barrel region ($|cos_{\theta}| < 0.6$) and energy larger than 60 GeV when the jets are clustered by the $e^{ \Plus }e^{ \Minus }k_{t}$ algorithm. To map each GenJet to a RecoJet, each possible pairing was checked and the sum of the $\Delta R$ for each combination was computed. The mapping chosen will be the combination with the least sum of $\Delta R$ (labeled as "Sum $\Delta R$ Minimum" in Appendix~\ref{app}).
Although JER and JAR are found to degrade in the detector forward region and with decreasing energy, the jet energy/angular scale can be controlled within 0.5/0.01$\%$.

\begin{figure}[!ht]
\centering
\subfigure[]{
\begin{minipage}[t]{0.42\linewidth}
\includegraphics[width=1.0\columnwidth]{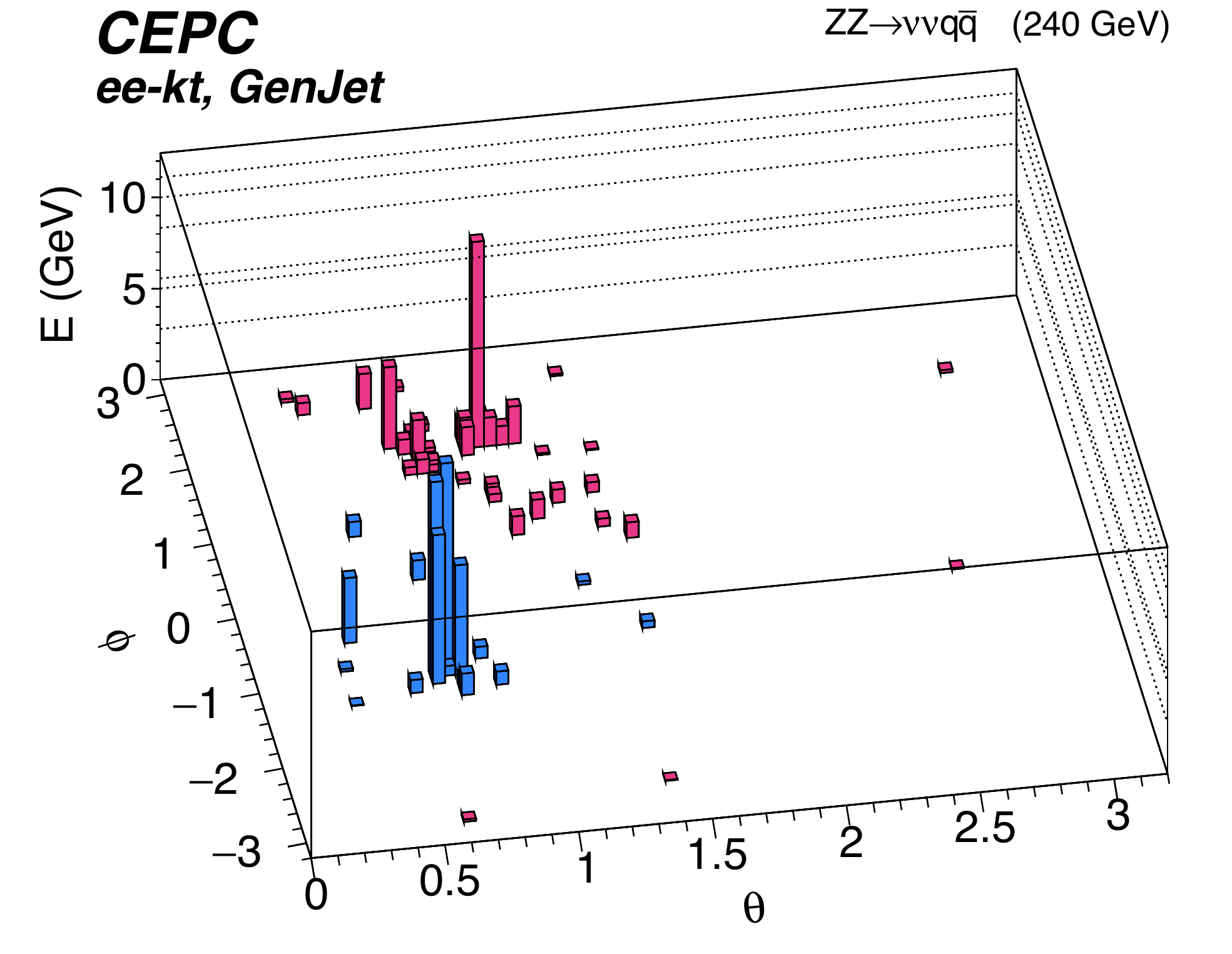}    
\end{minipage}%
}%
\subfigure[]{
\begin{minipage}[t]{0.42\linewidth}
\includegraphics[width=1.0\columnwidth]{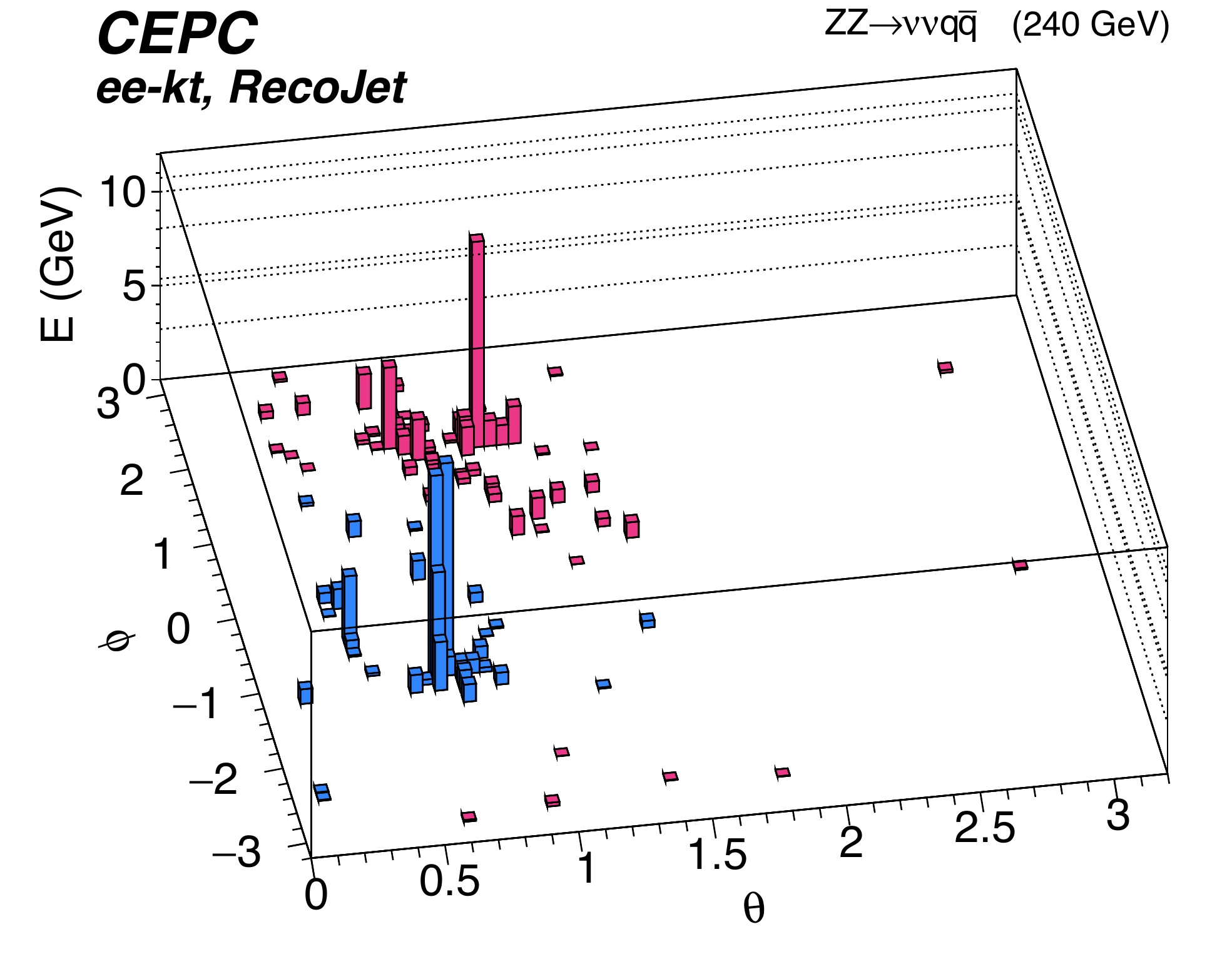}    
\end{minipage}%
}%
\vspace{-0.6cm}
\subfigure[]{
\begin{minipage}[t]{0.42\linewidth}
\includegraphics[width=1.0\columnwidth]{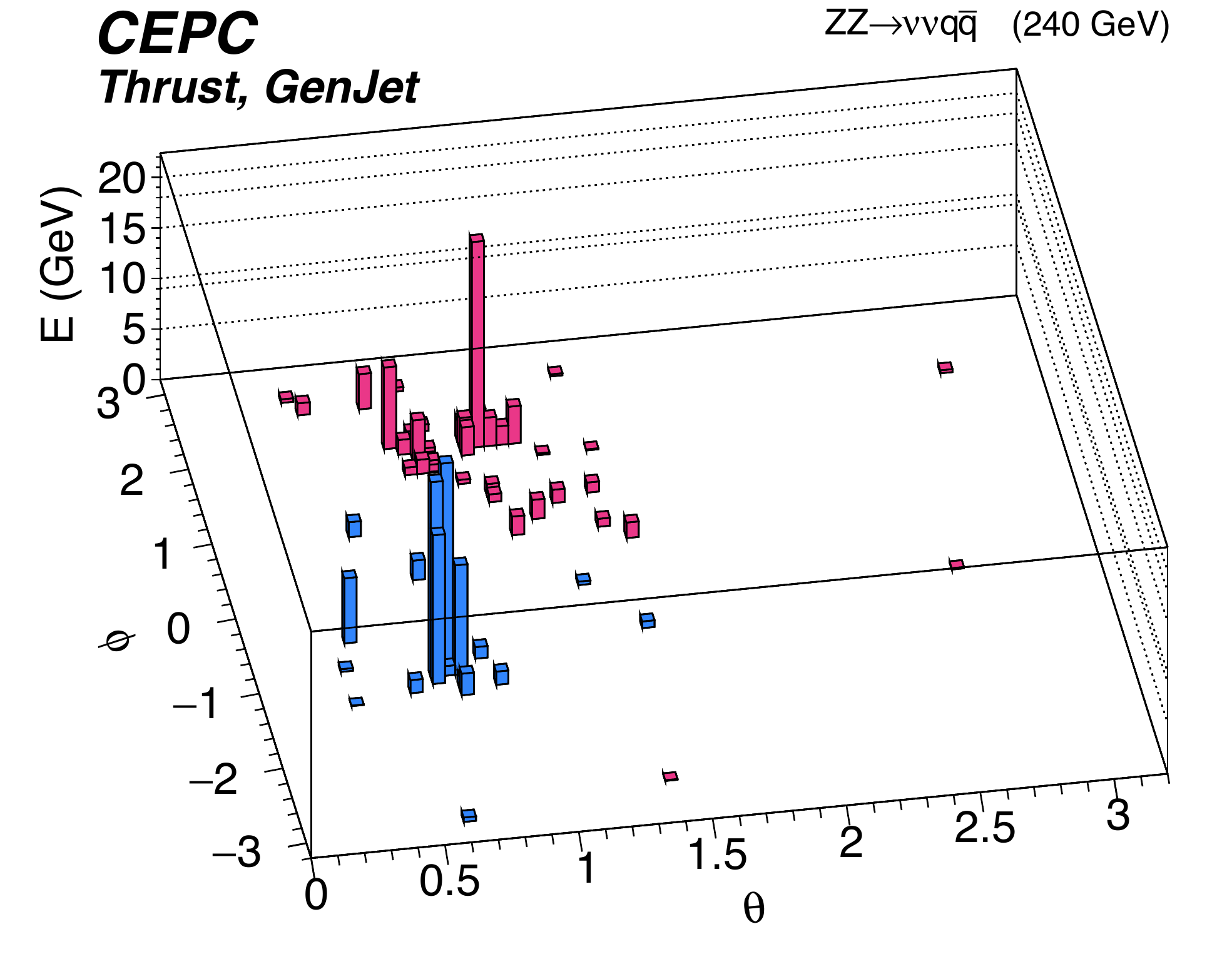}    
\end{minipage}%
}%
\subfigure[]{
\begin{minipage}[t]{0.42\linewidth}
\includegraphics[width=1.0\columnwidth]{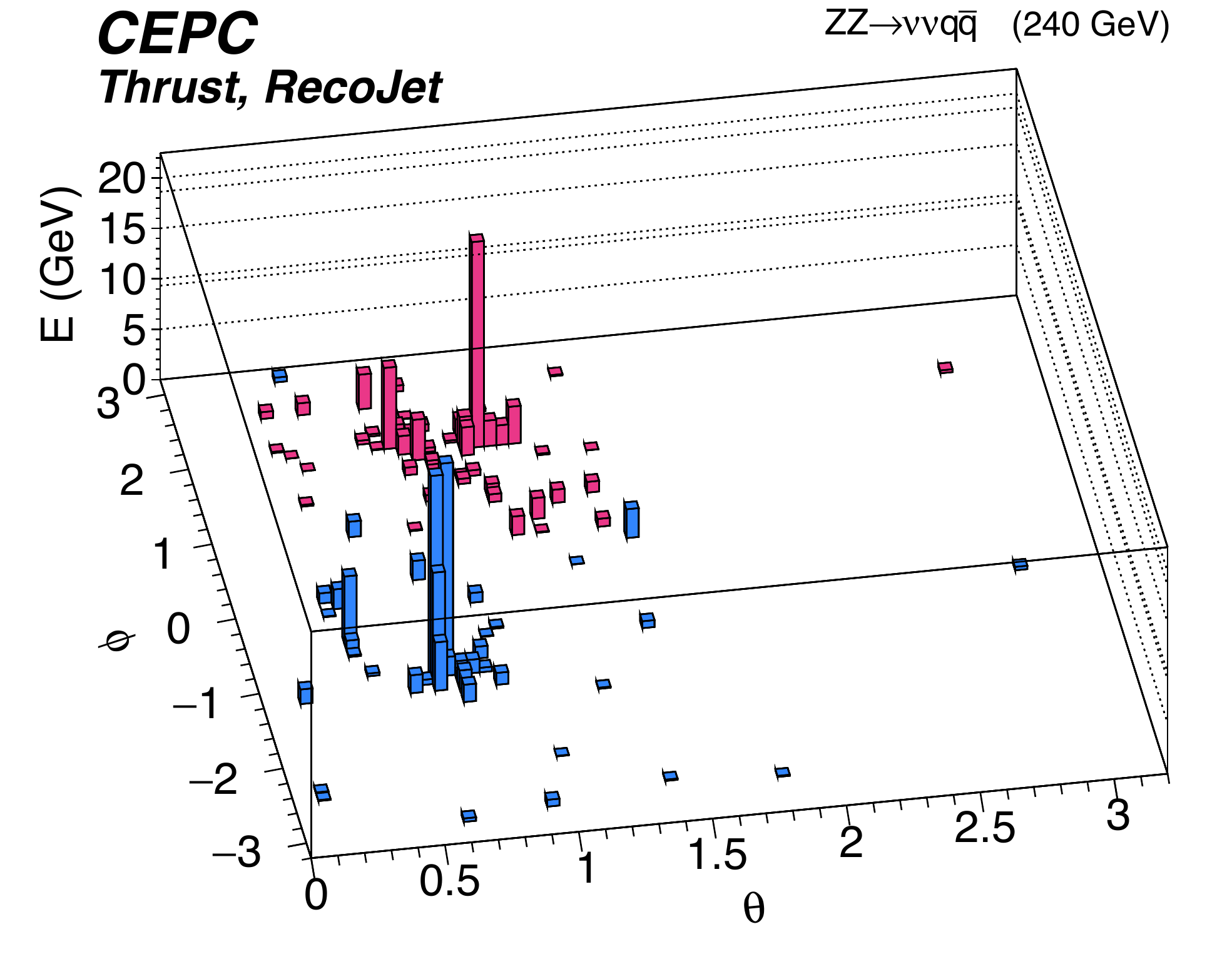}    
\end{minipage}%
}%
\vspace{-0.6cm}
\subfigure[]{
\begin{minipage}[t]{0.42\linewidth}
\includegraphics[width=1.0\columnwidth]{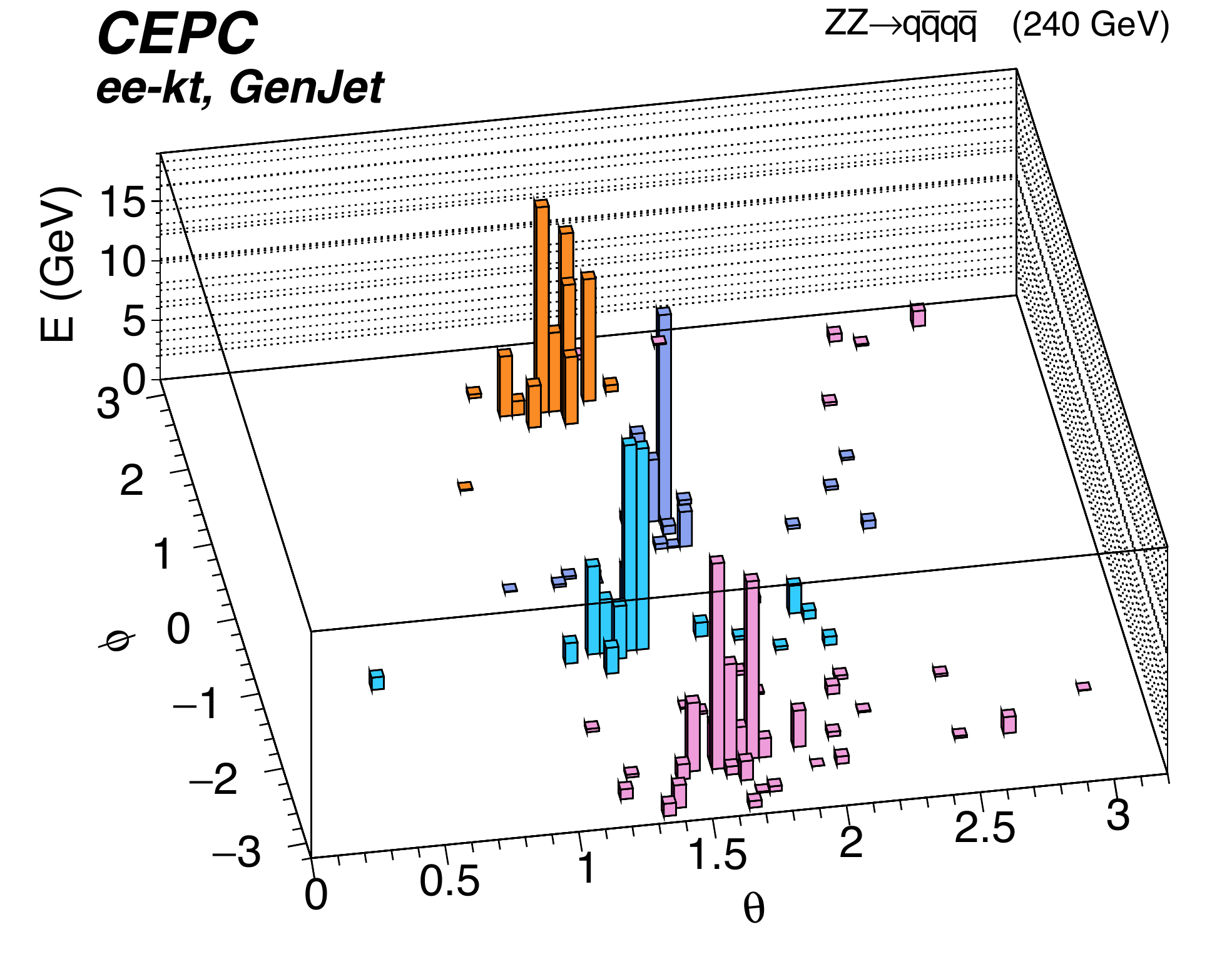}    
\end{minipage}%
}%
\subfigure[]{
\begin{minipage}[t]{0.42\linewidth}
\includegraphics[width=1.0\columnwidth]{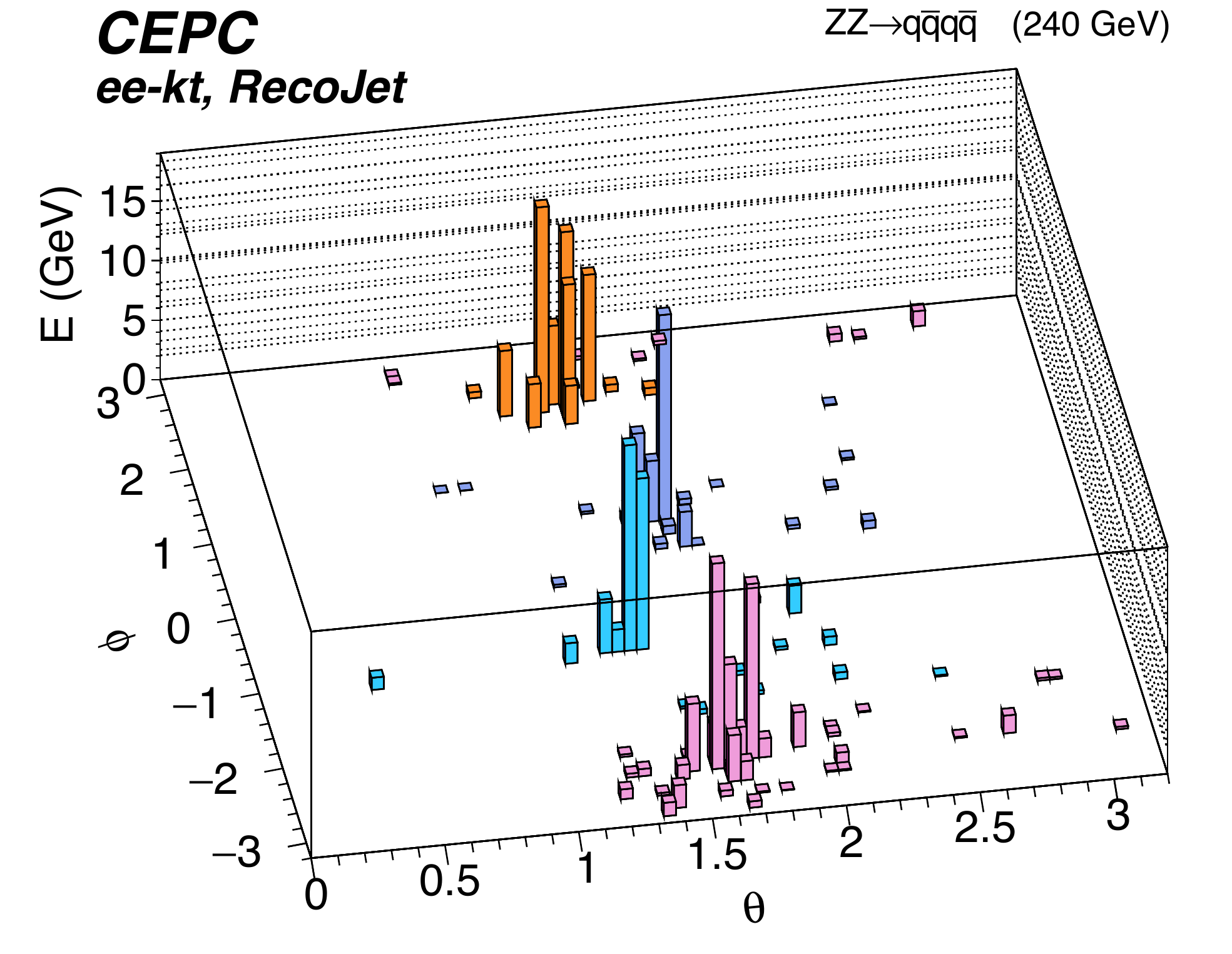}    
\end{minipage}%
}%
\vspace{-0.4cm}
\caption{The event displays of the jet clustering results and jet confusions in 2- and 4-jet final states. The event chosen contains at least one jet with $E_{Gen} > 60$  GeV and within $|cos\theta| < 0.6$. The resulting jets clustered for 2-jet final states with the $e^{\Plus}e^{\Minus}k_{t}$ algorithm are shown in the upper plots (a) and (b). Those clustered with the thrust-based algorithm are presented in the middle plots (c) and (d). The jets clustered for the 4-jet final states can be seen in the bottom plots (e) and (f). The clustering is performed on GenJets for plots on the left, while RecoJets are used for plots on the right.}
\label{Event_Display}
\end{figure}

We observed the jet clustering algorithm has a significant impact on the corresponding jet performance.
To evaluate this impact, another jet clustering algorithm based on the event-shape variable, thrust~\cite{Becher_2008,Ali_2011,thrust_ref_1,thrust_ref_2,thrust_ref_3}, was also studied and compared to the baseline algorithm. The thrust allows to obtain one thrust axis with the maximum momentum efflux projection in an event. The event can be divided into two hemispheres by a plane perpendicular to the thrust axis, indicating this method of the jet clustering is dedicated for 2-jet final states. 
Examples of the event displays showing the jet clustering results and jet confusions in 2- and 4-jet final states are demonstrated in Fig.~\ref{Event_Display}. Despite the similarities between the two clustering algorithms for GenJets for this particular event, their performance are expected to be different as seen in the case of RecoJets, where some particles would be grouped into different jets as the clustering algorithm is changed.
The jet response of the thrust-based jet clustering algorithm has a significantly different performance compared to the $e^{ \Plus }e^{ \Minus }k_{t}$ algorithm for the high-energy jets in the barrel region, and it can improve the jet energy resolution by up to 8$\%$ (as presented in Sect. \ref{4.2}). Because two jets have higher chances of being collinear in this phase space, the clustering ability would manifest in this purview. On the other hand, for most events where the jets are back-to-back against each other, their separation guarantees that both clustering algorithms will perform similarly well.

The structure of this paper is organized as follows. The conceptual detector geometry, the samples, the software toolkits, and the methodology to determine the jet performance are elucidated in Sect. \ref{2}. The baseline jet energy and angular resolution/scale and energy correction are shown in Sect. \ref{3}. The baseline performance and thrust-based algorithm are compared in Sect. \ref{4} and the conclusions are summarized in the Sect. \ref{5}. 

\section{Software tools, samples, and methodology}
\label{2}
\begin{figure} 
\begin{minipage}{\columnwidth}
\centering
  \includegraphics[width=0.65\columnwidth]{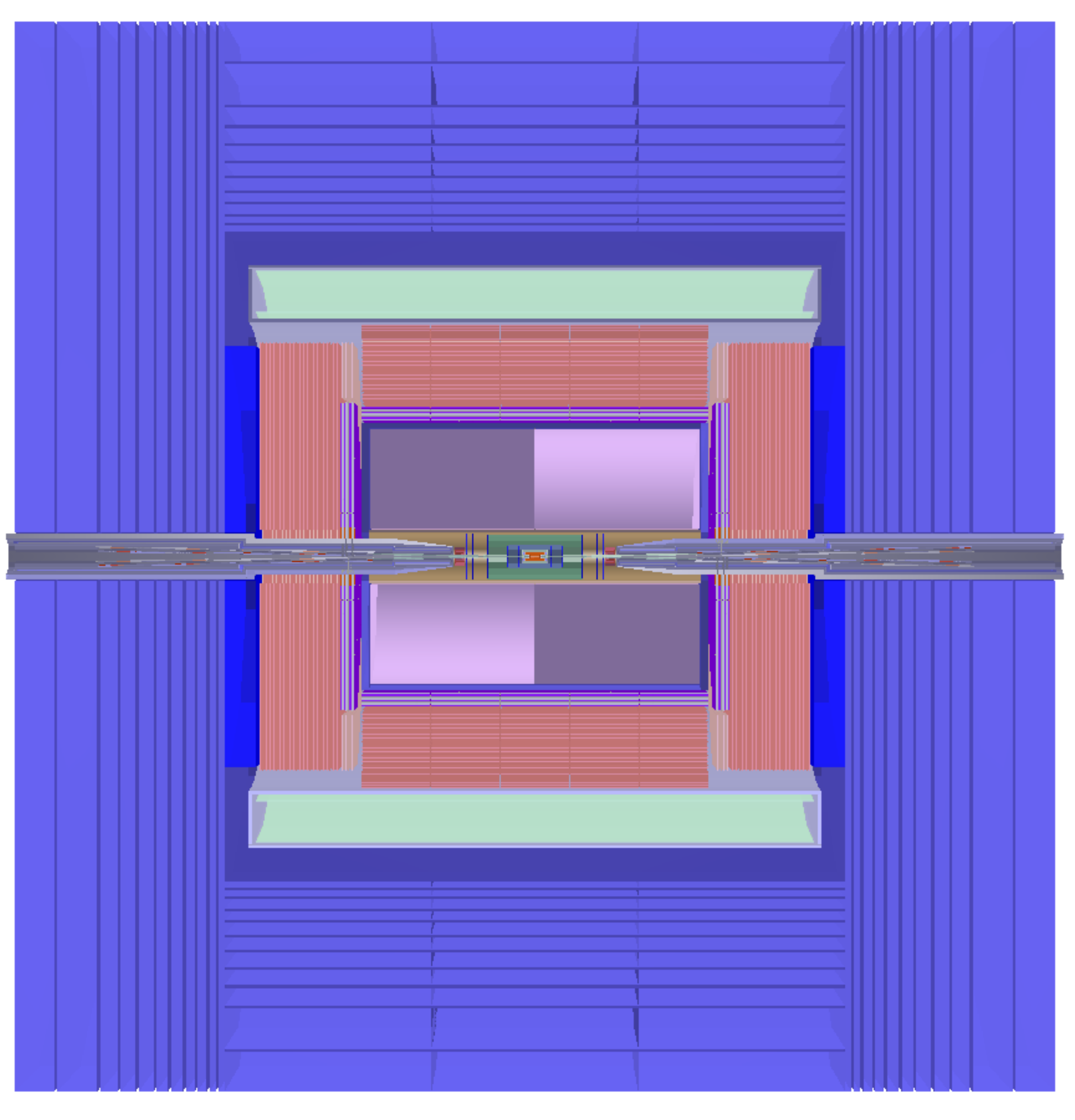}
\end{minipage}
\caption{A cross-sectional view of the CEPC baseline conceptual detector, starting from the vertex detector (Red), silicon tracker (Navy blue), TPC (the region within blue square), ECAL (Purple), HCAL (Red), 3T superconducting solenoid (Green), and muon detector (Blue), from inside to outside~\cite{R8}.}
\label{Overview_CEPC_detector}
\end{figure}

The CEPC conceptual detector apparatus features a 3.0T superconducting solenoid of 6m internal diameter, 
where the B-Field can be decreased down to 2.0T at the Z factory operation, with which the instantaneous luminosity can be doubled~\cite{R8}. 
Within the magnetic field is a vertex detector, a silicon tracker, a time projection chamber (TPC), a calorimetry system, including a silicon-tungsten electromagnetic calorimeter (ECAL), and a steel-glass resistive plate chamber (GRPC) sampling hadron calorimeter (HCAL). 
A muon detector is installed outside the solenoid embedded in the steel flux-return yoke. 
The overview of the CEPC conceptual detector is in Fig.~\ref{Overview_CEPC_detector}. The CEPC conceptual detector uses a right-handed coordinate system, the z-axis pointing along the direction of the counter-clockwise beam, the y-axis pointing upward, and the x-axis is fixed to fit a right-handed coordinate system (pointing toward the center of the CEPC ring). The polar angle, $\theta$ is measured from the positive z-axis, and the azimuthal angle, $\phi$ is measured in the x-y plane in radians.

The vertex detector system consists of three cylindrical and concentric double-layers mounted on ladders with full polar angle coverage within $|cos\theta|$ < 0.97 and high spatial resolution (2.8-6 $\mu m$) pixel sensors on both sides. 
The silicon tracker composed of silicon inner tracker (SIT), forward tracking disks (FTDs), silicon external tracker (SET), and endcap tracking disks (ETDs), with full polar angle coverage within $|cos\theta|$ < 0.985 with a designed spatial resolution of $\sigma\left(1 \Slash p_{T}\right) \sim 10^{-3}\Minus10^{-4} $. 
The main tracker, TPC, has a coverage in polar angles of $|cos\theta|$ < 0.995 with an expected position accuracy of 50 $\mu m$. 
The calorimeter system, including 30 layers of high granularity ECAL and 40 layers of high granularity HCAL, has a coverage in polar angles $|cos\theta|$ < 0.995. 
In the forward region (0.991 < $|cos\theta|$ < 0.999), the luminosity calorimeter is installed not only to measure luminosity but also to monitor the condition of the accelerator and detector. 
The muon system design covers the range $|cos\theta|$ < 0.998 as a baseline.

\begin{figure} 
\begin{minipage}{\columnwidth}
\centering
  \includegraphics[width=0.65\columnwidth]{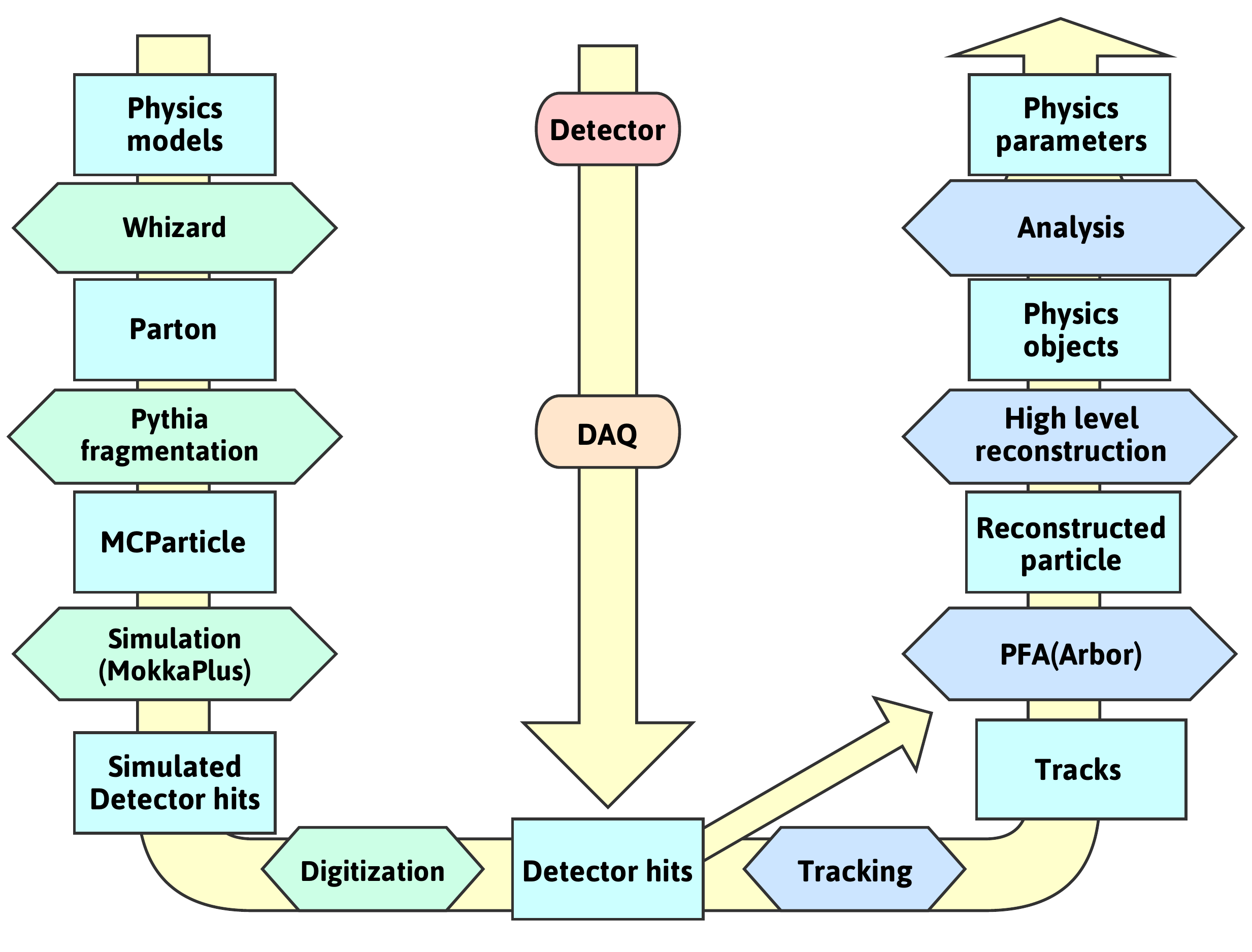}
\end{minipage}
\caption{The flow of the CEPC simulation and reconstruction software. The light blue boxes represent the produced physical objects and the hexagons depict software modules~\cite{R8}.}
\label{Working_flow_Simulation}
\end{figure}

The data flow of the CEPC simulation-reconstruction is expounded in Fig.~\ref{Working_flow_Simulation}. 
The true-level parton information is calculated with Whizard~\cite{R13_Whizard, R14_Whizard}.
If the parton carries a color charge, Pythia~\cite{Pythia_ref} sequentially imitates its fragmentation and hadronization, which is currently the best tuned fragmentation according to the LEP data.
The GEANT4-based~\cite{GENT4_ref} full simulation framework, MokkaC~\cite{Mokka_plus_ref}, calculates the interaction to the CEPC conceptual baseline detector inside the detector effective volume. After that, the corresponding simulated hits of the energy deposit in the calorimeter and trajectory of tracks in the tracker system are obtained. The digitization module converts the simulated hits into the detector hits to mimic realistic raw data considering the sub-detector responses. The residual part is the realistic reconstruction part, where the tracking algorithm, Clupatra~\cite{Gaede_2014}, reconstructs the digital trajectory hits, and then the particle flow algorithm (PFA), Arbor~\cite{R12}, clusters the digital calorimeter hits and matches the trajectory to the calorimeter to reconstruct visible particles into particle flow candidates. 

To obtain the correct energy scale for the jet, two consecutive stages of energy calibration are applied in the CEPC baseline setup, the single-particle stage and data-driven stage, corresponding to the two stages of the hadronic system reconstruction. The single-particle calibration stage is derived through several dedicated single particle samples for each sub-detector. For instance, the ECAL and HCAL are globally calibrated according to the energy difference between the reconstructed particles and the true-level particles, with the ECAL calibrated using the single $\gamma$ sample, and the HCAL calibrated with the $K^{0}_{L}$ sample. In the simulation, single-particle gun events are used to do the calibration, while in real experiments, it would be essential to calibrate the detector using control samples such as Bhabha scattering or the Z-boson decays to leptons, jets, etc.
The data-driven level corrects the Arbor PFA double-counting behavior caused by the fragmentation of charged hadrons mis-reconstructed as several additional neutral hadrons. Some splitting calorimeter hits of a charged hadron may be away from the center of the HCAL deposit, and be clustered as new neutral hadrons in the HCAL. However, the tracker still provides a more accurate momenta measurement for the charged hadrons, while an event is given some extra energy due to the additional neutral hadrons. Without data-driven calibration, the Arbor PFA double counting would bring a percentage level overestimation on the boson mass scale and jet energy scale~\cite{the_paper_ref}. In order to calibrate the Arbor PFA double counting, the ratio of the Z-boson mass affected by double counting to the anticipated Z-boson mass is used as a candle for RecoJet energy calibration. The performance of the calibrated detector response, and the uncertainties involved in its measurement, will mainly depend on the stability of the detector during the course of its operation. As long as the detector is stable enough, and with enough data, the differential behavior of the detector can be well-controlled.

High-level reconstruction algorithms are applied to identify all the physics objects, including leptons, photons, taus, jets, and missing energy/momentum from the PF candidates. The objects undergo two stages of reconstruction: first, the PFA (Arbor) uses the information on the tracks and calorimeter hits to identify each final-state particle; second, the high-level objects are identified from all the reconstructed particles.
Once jets are identified, a Boosted Decision Tree--based~\cite{TMVA_ref} toolkit, LCFIPlus Package~\cite{LCFIPlus_ref}, is used to identify the c- and b-jets.

\subsection{Methodology of measurement of the jet energy/angular resolution and scale}\label{2.1}
Once the GenJet/RecoJet are reconstructed from the true-level particles and reconstructed particles, respectively, a matching procedure is employed to establish the GenJet-RecoJet pairs. This matching procedure is vital in extracting the jet response from these GenJet-RecoJet pairs. We compare different matching approaches, and the one minimizing the accumulated angular difference is chosen (labeled as "Sum $\Delta R$ Minimum" in Appendix~\ref{app}). Once the GenJet-RecoJet pairs are identified, the relative energy difference and angular difference are calculated and modelled. 

The relative energy difference ($\mathsf{R}$) and the angular (polar and azimuth angle) difference ($\mathsf{D}$) of the GenJet-RecoJet pairs, given by:
\begin{equation}
  \begin{aligned}
  & \mathsf{R_{\mathrm{R-G}}}  =\frac{E_{\mathrm{RecoJet}}-E_{\mathrm{GenJet}}}{E_{\mathrm{GenJet}}}\\
  & \mathsf{D_{\mathrm{R-G}}}  =\theta_{\mathrm{RecoJet}}-\theta_{\mathrm{GenJet}} \quad or \quad \phi_{\mathrm{RecoJet}}-\phi_{\mathrm{GenJet}}\\
\end{aligned}
\label{eq:2}
\end{equation}
are modelled with the double-sided crystal ball (DBCB) function, 
\begin{equation}
\begin{aligned}[l]
    &f\left(x| \alpha_{1}, \alpha_{2}, n_{1}, n_{2}, \bar{x},\sigma\right)= \\
    &\left\{
    \begin{aligned}[c]
    & \left(\frac{n_{1}}{|\alpha_{1}|}\right)^{n_{1}}e^{-\frac{|\alpha_{1}|^{2}}{2}} \left( \frac{n_{1}}{|\alpha_{1}|}-|\alpha_{1}|-\frac{x-\bar{x}}{\sigma} \right)^{-n_{1}}\qquad  \frac{x-\bar{x}}{\sigma} < -\alpha_{1} \\
  & \quad e^{-\frac{1}{2}\left(\frac{x-\bar{x}}{\sigma}\right)^{2}}\qquad \qquad \qquad \qquad \qquad \qquad \qquad   -\alpha_{1}<\frac{x-\bar{x}}{\sigma} < \alpha_{2} \\
   & \left(\frac{n_{2}}{|\alpha_{2}|}\right)^{n_{2}}e^{-\frac{|\alpha_{2}|^{2}}{2}} \left( \frac{n_{2}}{|\alpha_{2}|}-|\alpha_{2}|-\frac{x \Plus \bar{x}}{\sigma} \right)^{-n_{2}}\qquad   \alpha_{2}<\frac{x-\bar{x}}{\sigma}
\end{aligned}   \right.
\end{aligned}
\label{eq:3}
\end{equation}
and Gaussian function, respectively.

\begin{figure}[!ht]

\begin{minipage}{\columnwidth}
\centering
  \includegraphics[width=0.45\columnwidth]{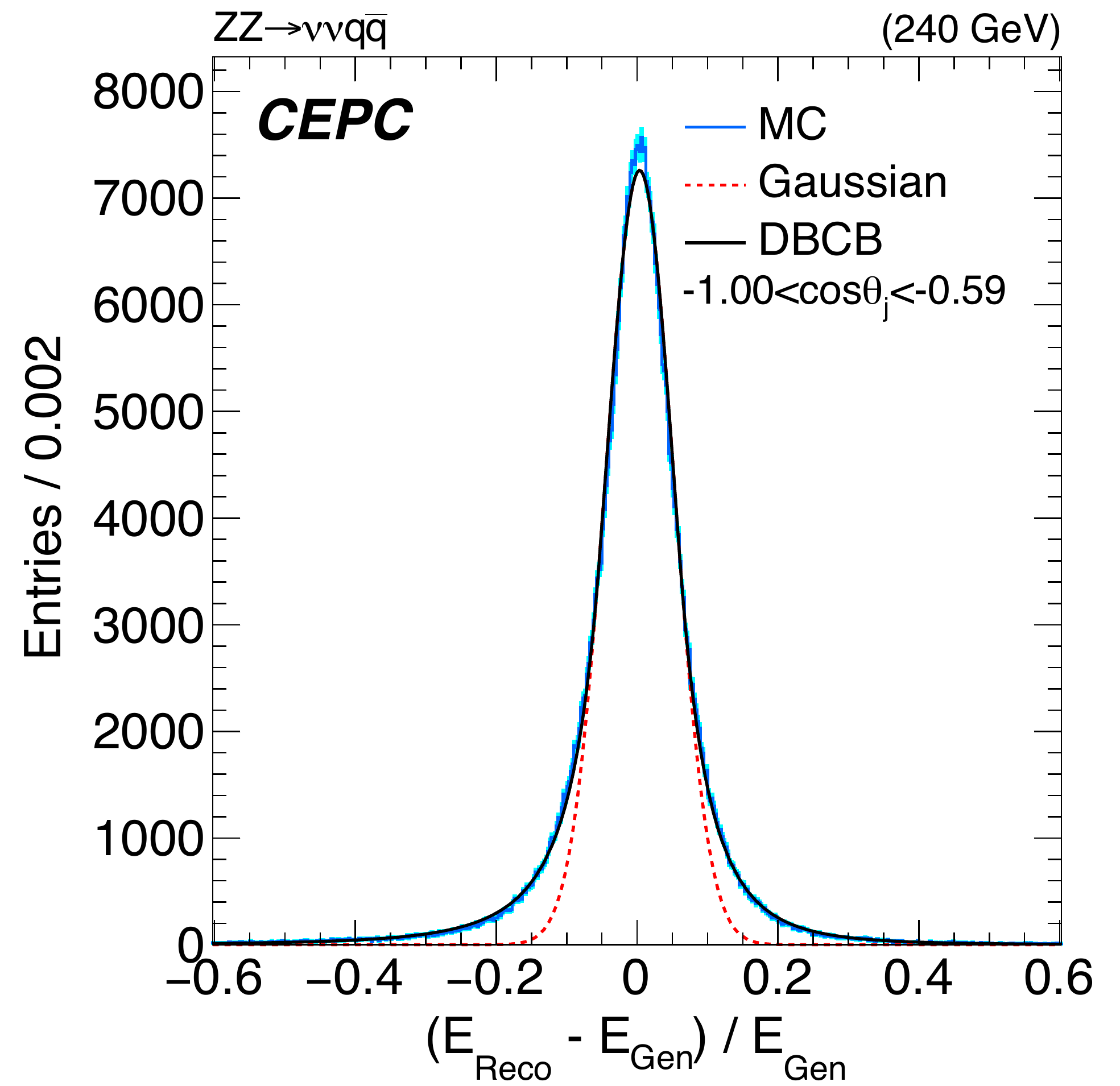}
\end{minipage}

\caption{The energy difference between the RecoJet and GenJet relative to the GenJet energy. The distribution is fitted by DBCB, where the width and mean of its core are extracted to define the jet energy resolution and jet energy scale, respectively. The uncertainties are shown as a shade on the distribution. The $cos\theta_{j}$ in the legend refers to the true-level parton polar angle.}
\label{Fit_demo}
\end{figure}

Based on the good agreement with the models and the distributions, as shown in Fig.~\ref{Fit_demo}, the JER/S are extracted from the standard deviation ($\sigma$) and mean ($\bar{x}$) of the DBCB fitting to the relative energy difference between RecoJet and GenJet while JAR/S are the $\sigma$/$\bar{x}$ of the Gaussian fitting within $\bar{x} \pm 1.5 \sigma$ of angular difference. The Gaussian part is induced by the intrinsic detector resolution, while the lower and higher exponential tail of the DBCB distribution are caused by the imperfect effect on jet clustering/matching, the overestimation from PFA reconstruction, and the initial state radiation (ISR) pollution. When overestimation of PFA reconstruction and mis-clustering of the ISR in the RecoJet happens, it will cause the RecoJet energy to deviate considerably from the original GenJet energy. Next, the core fraction is defined as the ratio of the number of events within the region of the fitted Gaussian $\sigma$ to the total events in the distribution. The core fraction quantifies the percentage of events within the energy or angular resolutions. For instance, as in Fig.~\ref{Fit_demo}, some events are within the Gaussian $\sigma$ region, while others are out of Gaussian estimations. To study the differential JER/S and JAR/S as functions of the angle and energy, every differential bin is ensured to have a statistical uncertainty of less than 1$\%$.

\section{Performance with the CEPC baseline setup}
\label{3}
The performance of the baseline detector under the baseline reconstruction (Arbor and $e^{ \Plus }e^{ \Minus }k_{t}$ algorithm) is presented in this section, using fully simulated semi-leptonic and fully-hadronic di-boson processes (WW, ZZ, ZH) at the CEPC Higgs factory mode ($\sqrt{s}=240$ GeV). 
\subsection{Jet energy resolution and scale}
\label{3.1}
\begin{figure}[!ht]
\centering
\subfigure[]{
\begin{minipage}[t]{0.35\linewidth}
\includegraphics[width=1.0\columnwidth]{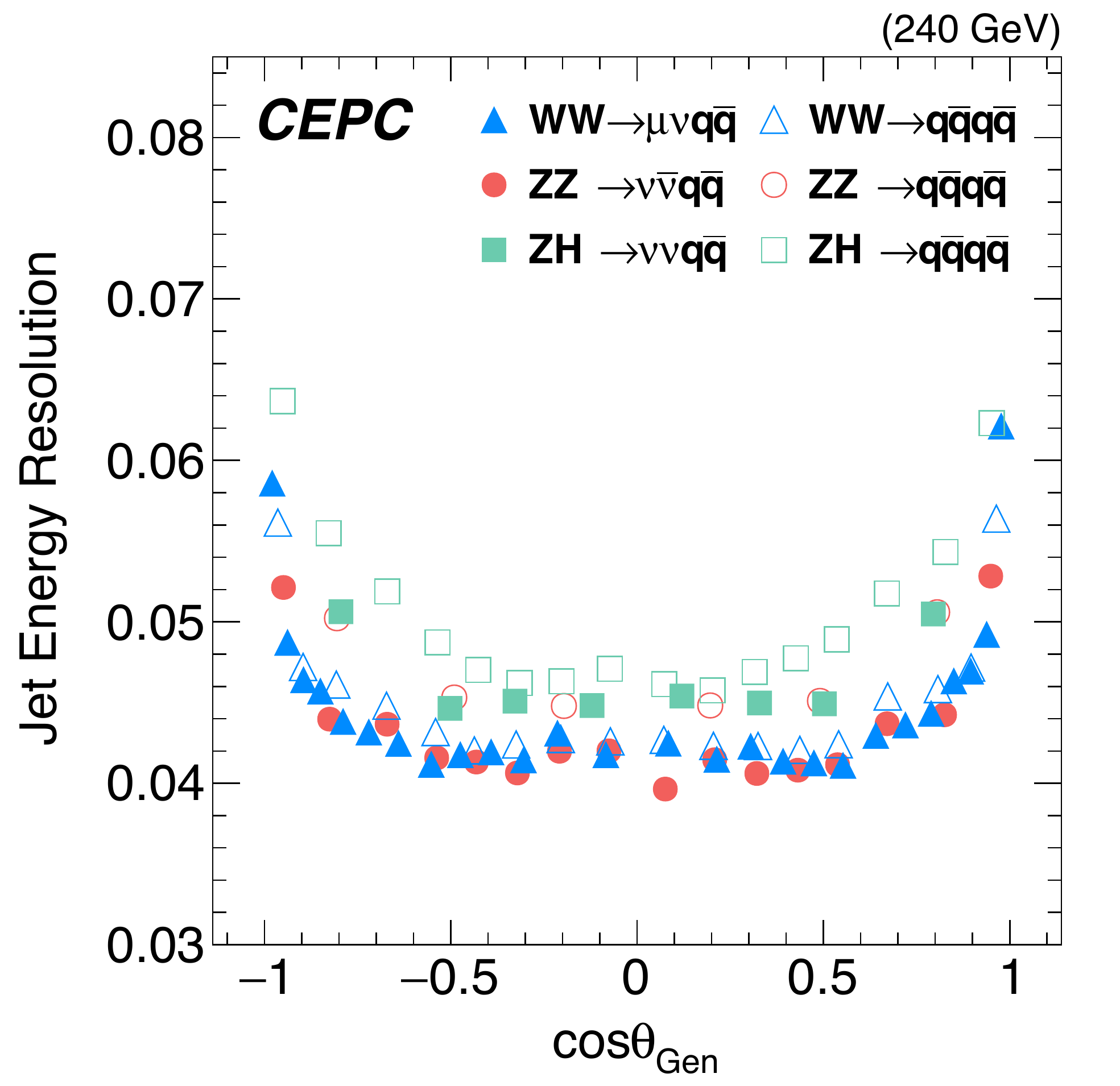}    
\end{minipage}%
}%
\subfigure[]{
\begin{minipage}[t]{0.35\linewidth}
\includegraphics[width=1.0\columnwidth]{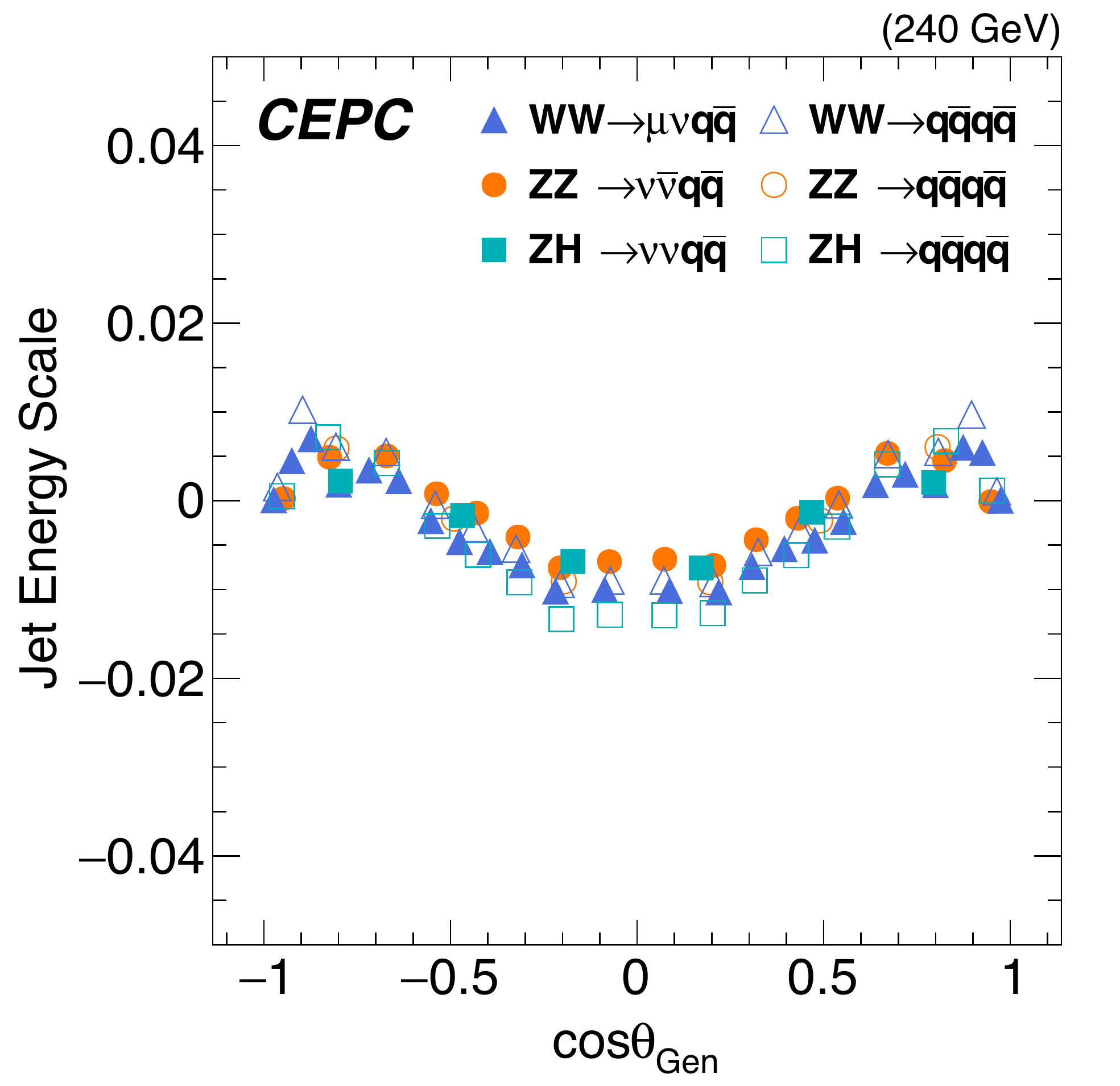}    
\end{minipage}%
}%
\vspace{-0.6cm}
\subfigure[]{
\begin{minipage}[t]{0.35\linewidth}
\includegraphics[width=1.0\columnwidth]{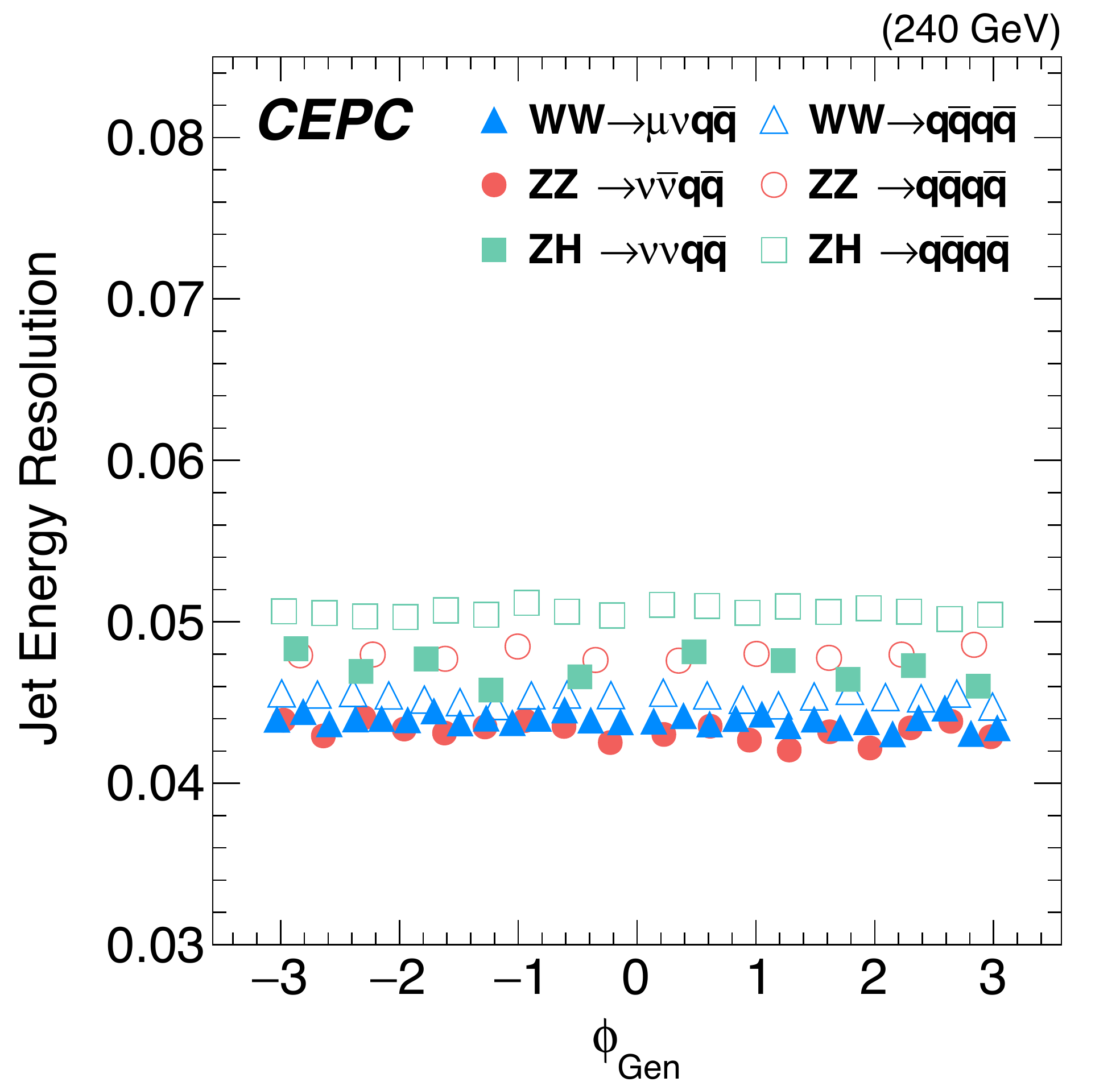}    
\end{minipage}%
}%
\subfigure[]{
\begin{minipage}[t]{0.35\linewidth}
\includegraphics[width=1.0\columnwidth]{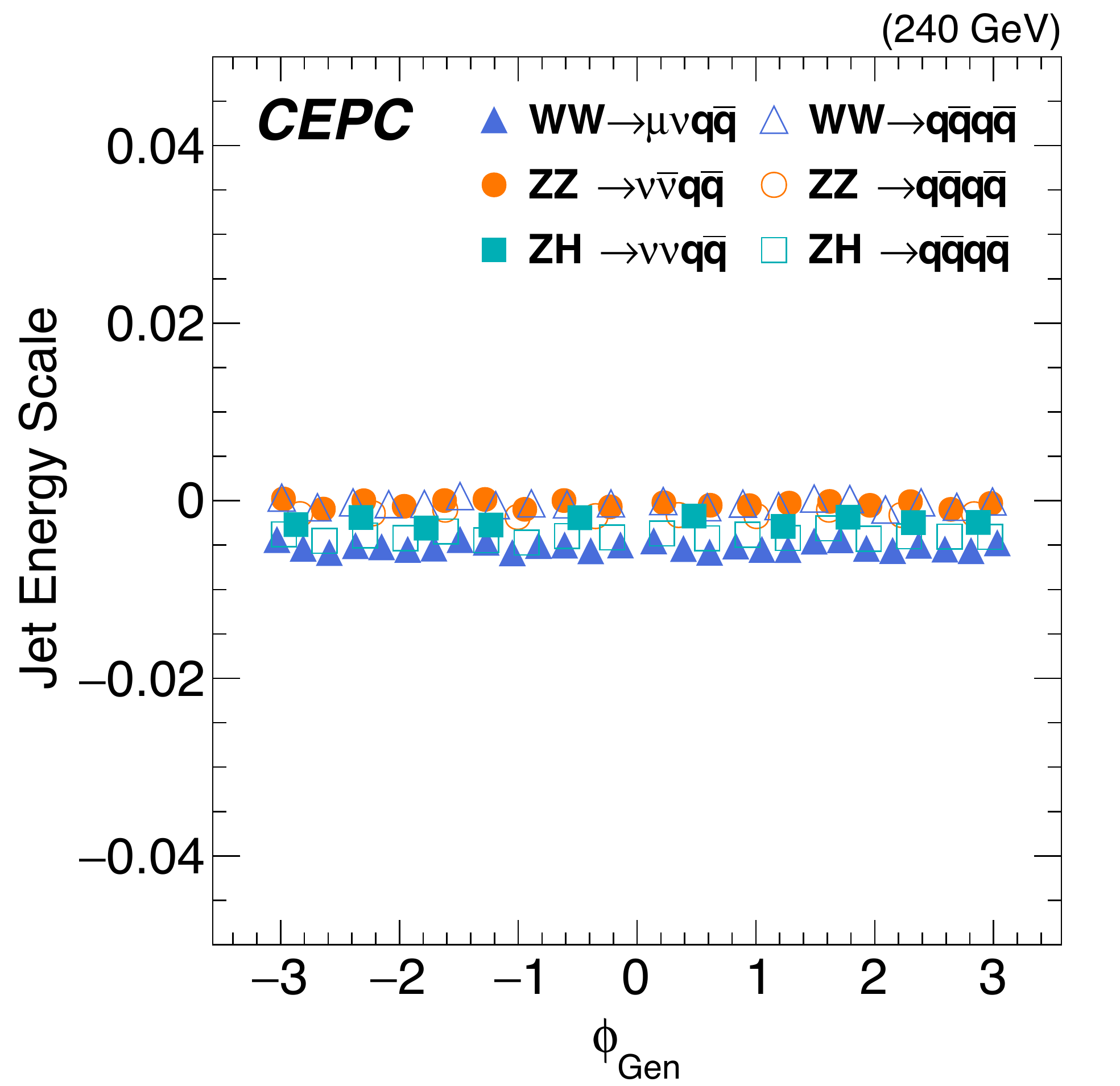}    
\end{minipage}%
}%
\vspace{-0.6cm}
\subfigure[]{
\begin{minipage}[t]{0.35\linewidth}
\includegraphics[width=1.0\columnwidth]{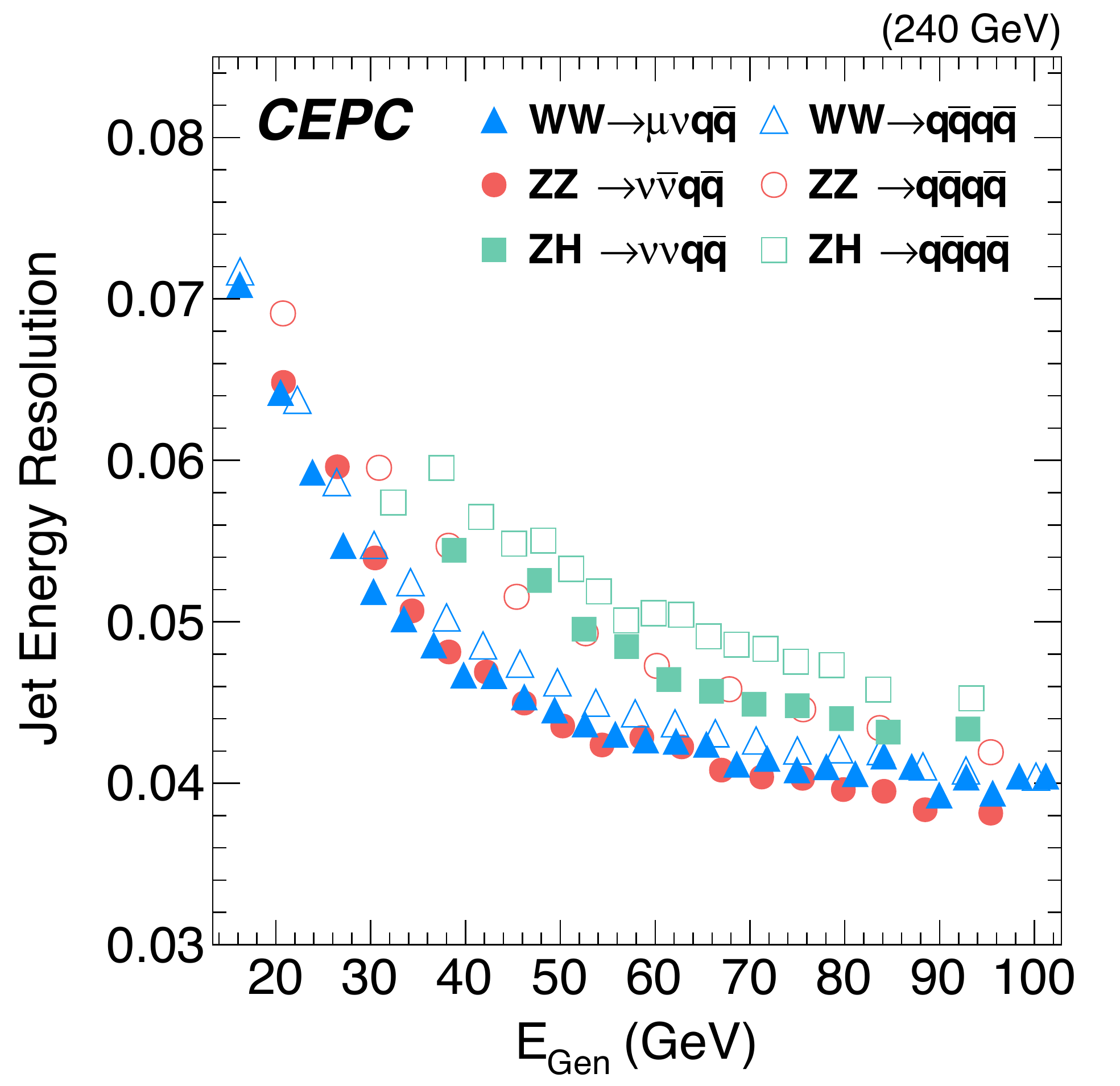}    
\end{minipage}%
}%
\subfigure[]{
\begin{minipage}[t]{0.35\linewidth}
\includegraphics[width=1.0\columnwidth]{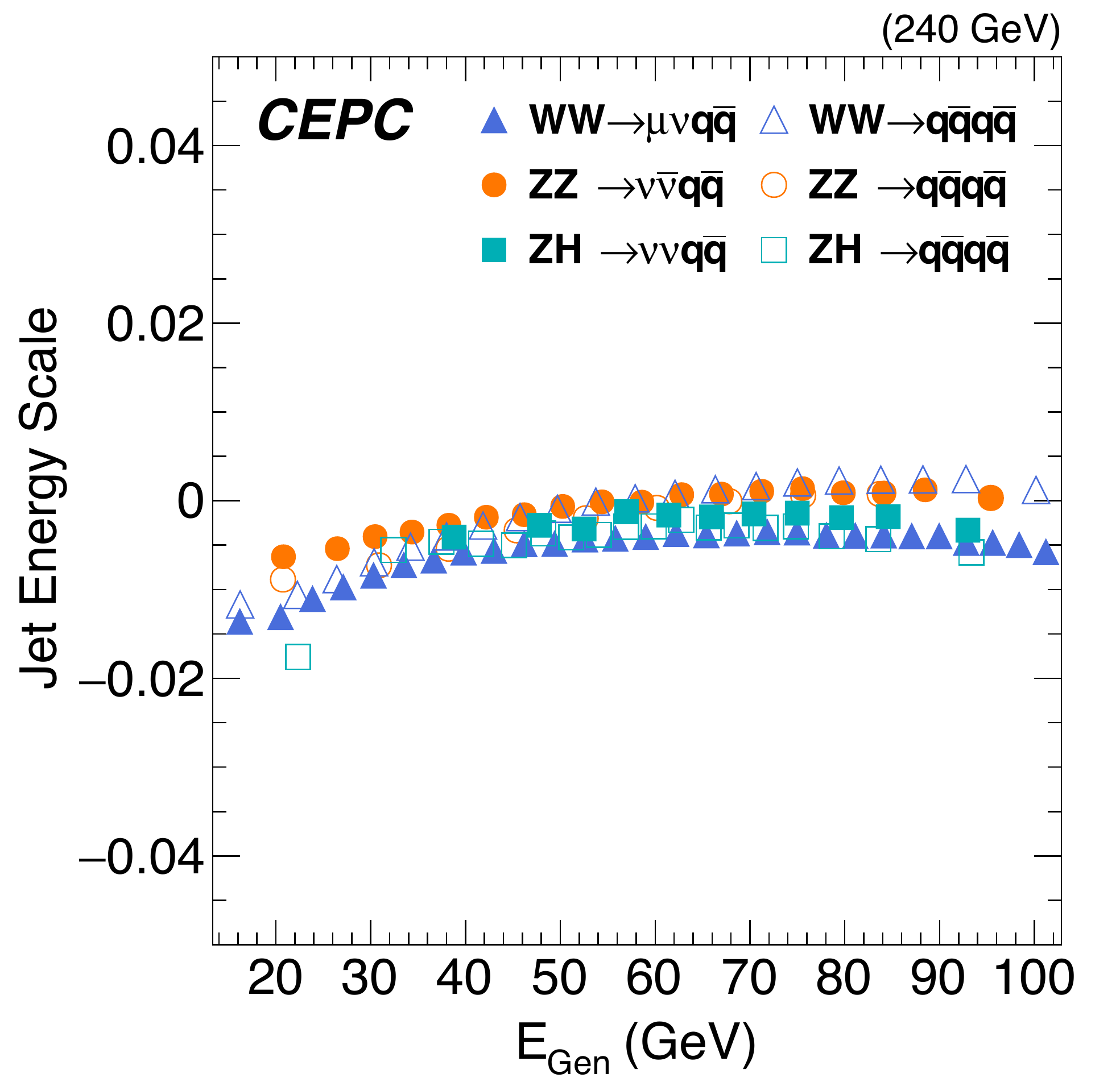}    
\end{minipage}%
}%
\caption{Jet energy resolution and scale as functions of (a-b) the $cos\theta_{Gen}$, (c-d) the azimuth angle, and (e-f) the GenJet energy for 2- (solid symbols) and 4-jet (open symbols) final states. The errors shown are only statistical. In (a) and (b), the jet energy resolutions grow and jet energy scales decrease because part of jet is out of the fiducial acceptance of the detector.}
\label{ee_kt_JER_cali_cos_RecoGen}
\end{figure}

Fig.~\ref{ee_kt_JER_cali_cos_RecoGen} shows the differential jet energy resolution as functions of the $cos\theta_{Gen}$, azimuth angle and GenJet energy of the above-mentioned physics benchmark processes. The JER is around 4-5$\%$ in the barrel region, while JES is controlled within $\pm 1\%$ after the jet energy is calibrated by the ratio of the Z-boson mass affected by double counting to the anticipated Z-boson mass from the ZZ semi-leptonic process. This is the first order method to calibrate the PFA double-counting effect.

Not surprisingly, the better JER is achieved for more energetic jets. This is because the energetic particles have less tendency to have multiple scattering while penetrating through the tracker system. Both the JER and JES are flat along the azimuth angle as the detector is, to the first order, symmetric along the azimuth direction, and increases away from the central of the barrel region because of the spatial resolution and the increasing amount of materials from the barrel to the endcaps. The rising JES around $|cos\theta| = 0.75$ is caused by the gap between the barrel and endcaps augmenting the Arbor PFA double counting which causes the RecoJet to be reconstructed with slightly greater energy than the correct one. On the other hand, the JER grows and JES decreases at $|cos\theta| = 1$ because part of jet is out of the fiducial acceptance of the detector.

The jet energy response at the CEPC baseline detector slightly depend on the number of final-state jets.
The JER of semi-leptonic (2-jet) events are systematically better than that of fully-hadronic (4-jet) events by roughly 10$\%$. A priori, these differences are caused by larger jet confusions (wrong grouping of jets) in 4-jet events. Moreover, the slight dependence on the process is due to the significant difference of the jet flavor composition of the Higgs bosons with the Z and W bosons, where the gluon jets from the Higgs boson possess more neutral hadrons than quark jets, causing the difference of JER/S among various processes.

\begin{figure}[!ht]
\centering
\includegraphics[width=0.45\columnwidth]{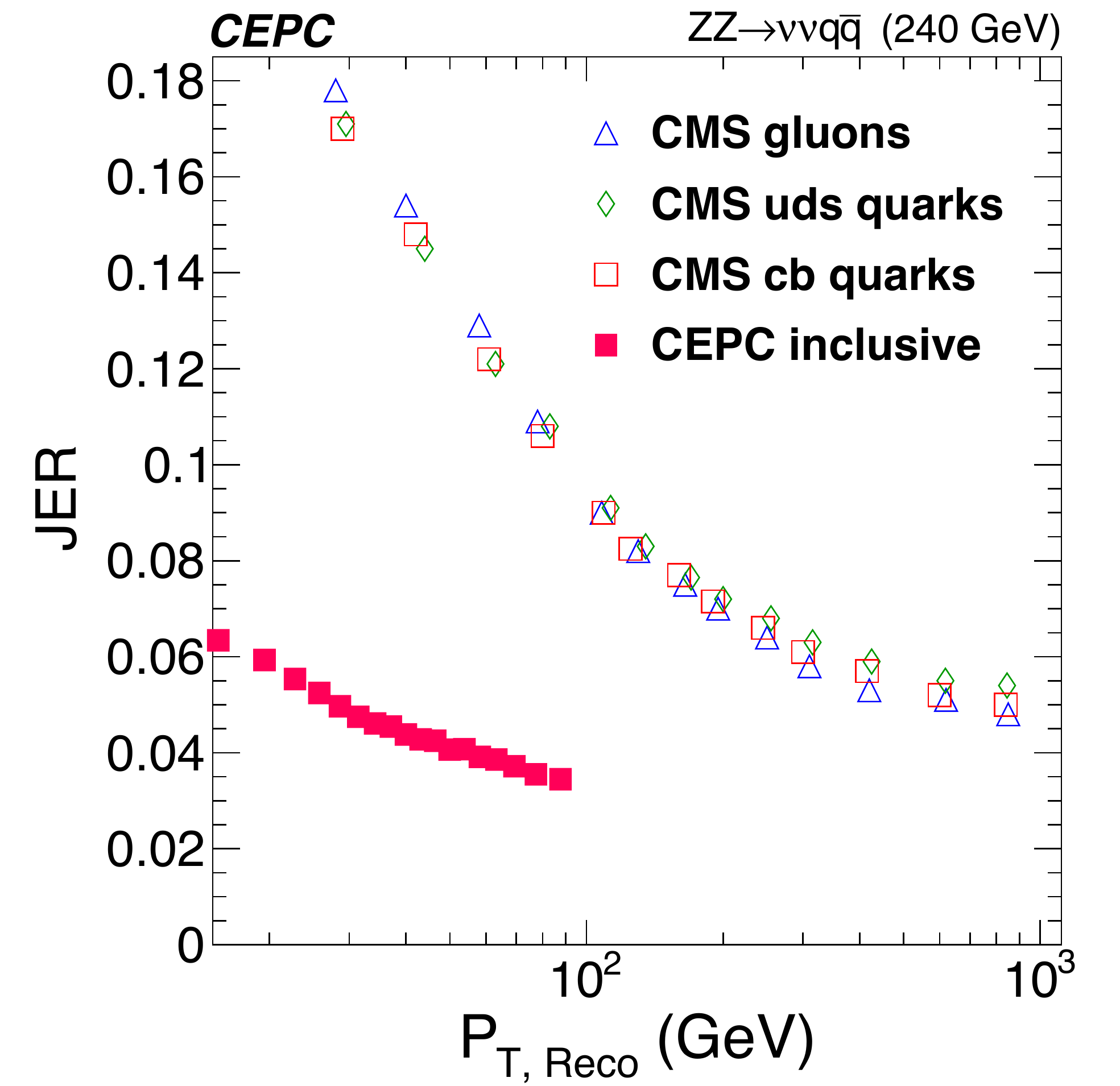}
\caption{CMS results~\cite{JER_ref} under zero pileup situation are compared to the inclusive quark jet results of $ZZ \rightarrow \nu \bar{\nu} q \bar{q}$ at the CEPC at 240 GeV. The errors shown are only statistical~\cite{the_paper_ref,R8}.} 
\label{JER_CEPC_CMS}
\end{figure}

Compared to the LHC experiment, the jet energy response of the CEPC is significantly improved. Compared to the JER of CMS~\cite{JER_ref} (and ATLAS~\cite{Aad_2013}, which has similar results), the CEPC has 2-4 times better JER in the same $P_{T}$ range~\cite{the_paper_ref,R8} (see Fig.~\ref{JER_CEPC_CMS}). In terms of the JES, the CEPC baseline setup reaches a JES < 1$\%$ with simple calibration, which is 3 times better than that of the LHC~\cite{JER_ref}, while more sophisticated calibration methods are expected to improve the JES considerably. 

\begin{figure}[!ht] 
\begin{minipage}{\columnwidth}
\centering
  \includegraphics[width=0.45\columnwidth]{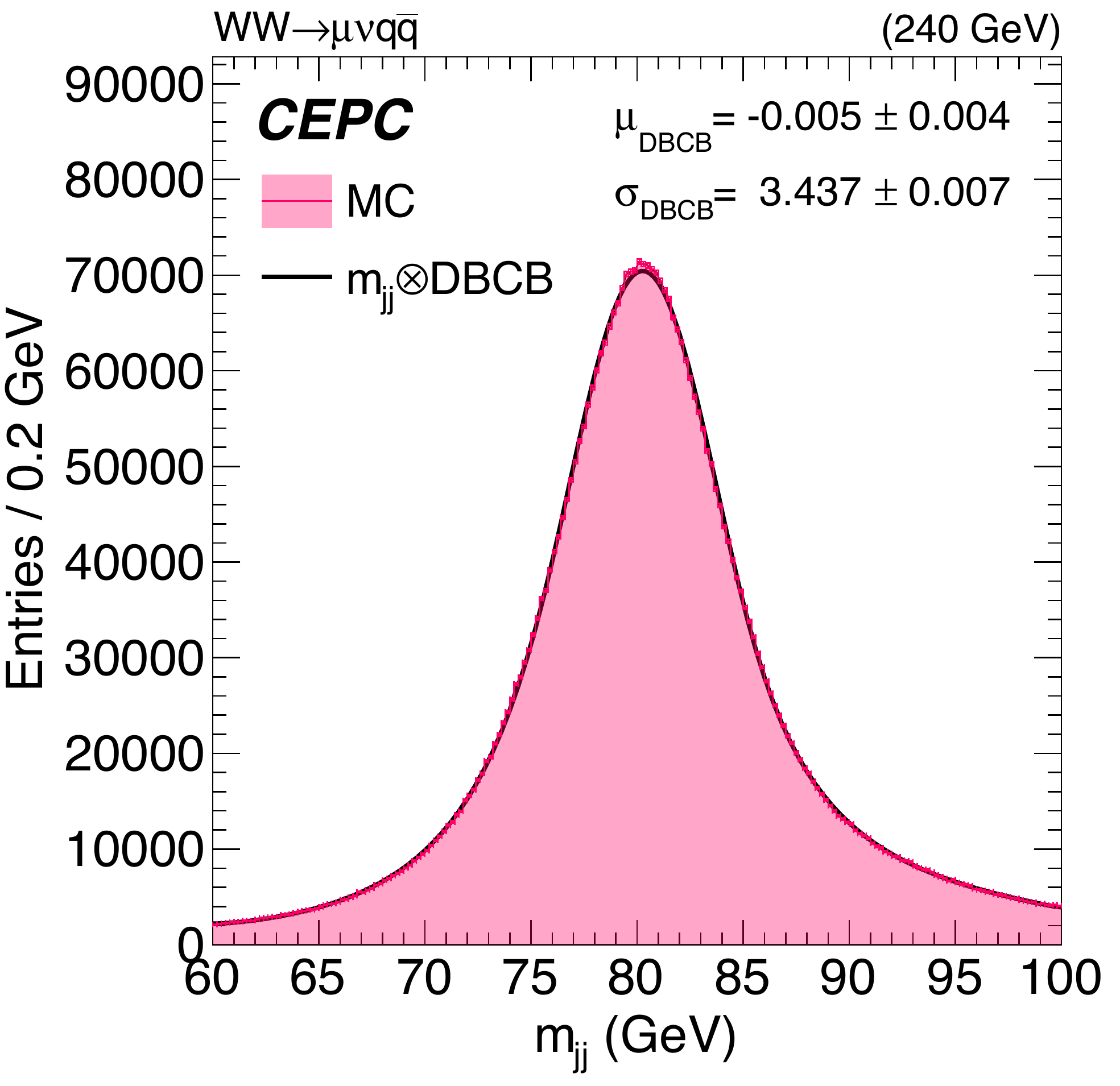}
\end{minipage}
\caption{The reconstructed invariant mass from the hadronic decay of the W boson in $\mu \nu qq$ events. Based on the differential RecoJet angle, energy, and flavor to apply jet differential energy corrections, the uncertainty of the W-boson mass difference between the reconstruction-level and true-level could be calibrated down to around 10 MeV, given by $\mu_{DBCB}$.}
\label{mW_mcp_templateFit}
\end{figure}

The differential jet response (as functions of $cos \theta$ and energy) is observed to be identical (within 1$\%$ difference) between the WW and ZZ processes. In other words, there exist a universal differential jet response which can be used to better calibrate and measure the resonance mass, for example, using the Z boson to calibrate the W-boson mass. After applying the differential jet energy calibration, the uncertainties of the boson masses are expected to be more precise. A template fit was used to quantify the calibrated boson mass uncertainty. The reconstructed boson masses were calibrated according to their respective di-parton invariant masses. Based on the differential RecoJet angle, energy, and flavor, the performance of jet differential energy corrections was derived from the mass difference between the reconstruction-level and the true-level Z boson. A preliminary statistical uncertainty on the massive boson masses could be calibrated down to around 10 MeV, which are elaborated in the Fig.~\ref{mW_mcp_templateFit}.

\subsection{Jet angular resolution and scale}
\label{3.2}

\begin{figure}[!ht]
\centering
\subfigure[]{
\begin{minipage}[t]{0.35\linewidth}
\includegraphics[width=1.0\columnwidth]{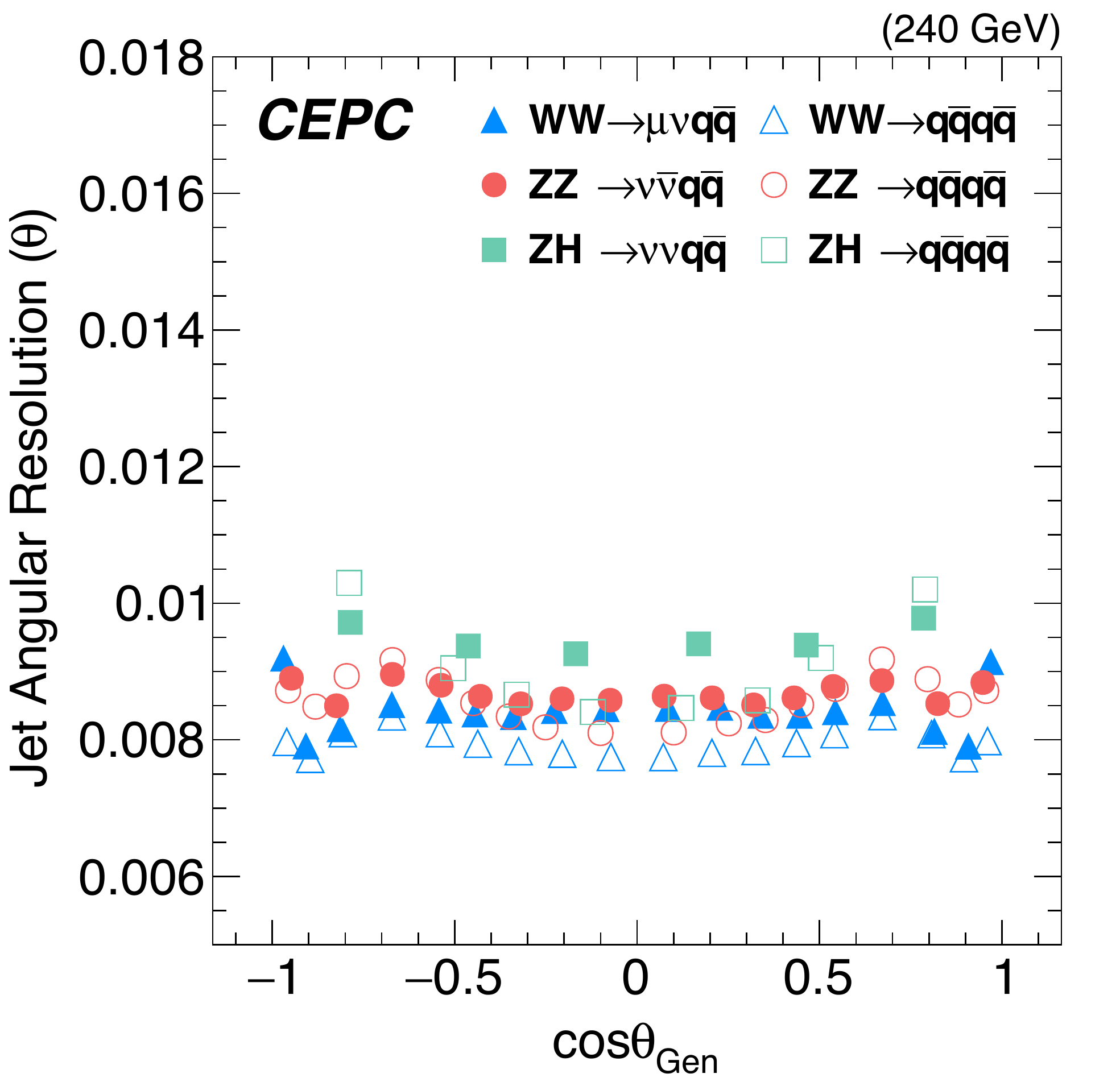}    
\end{minipage}%
}%
\subfigure[]{
\begin{minipage}[t]{0.35\linewidth}
\includegraphics[width=1.0\columnwidth]{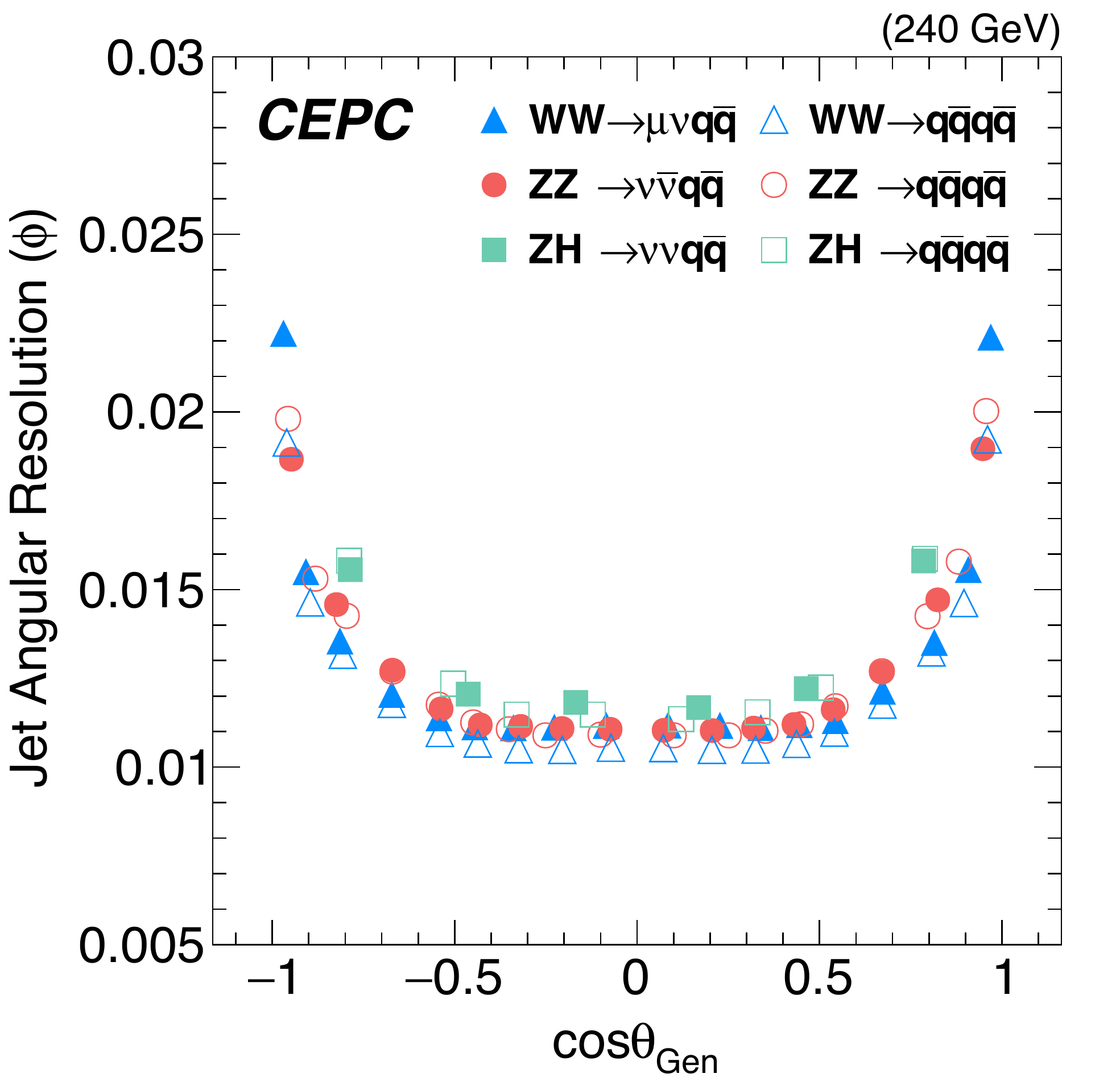}    
\end{minipage}%
}%
\vspace{-0.6cm}
\subfigure[]{
\begin{minipage}[t]{0.35\linewidth}
\includegraphics[width=1.0\columnwidth]{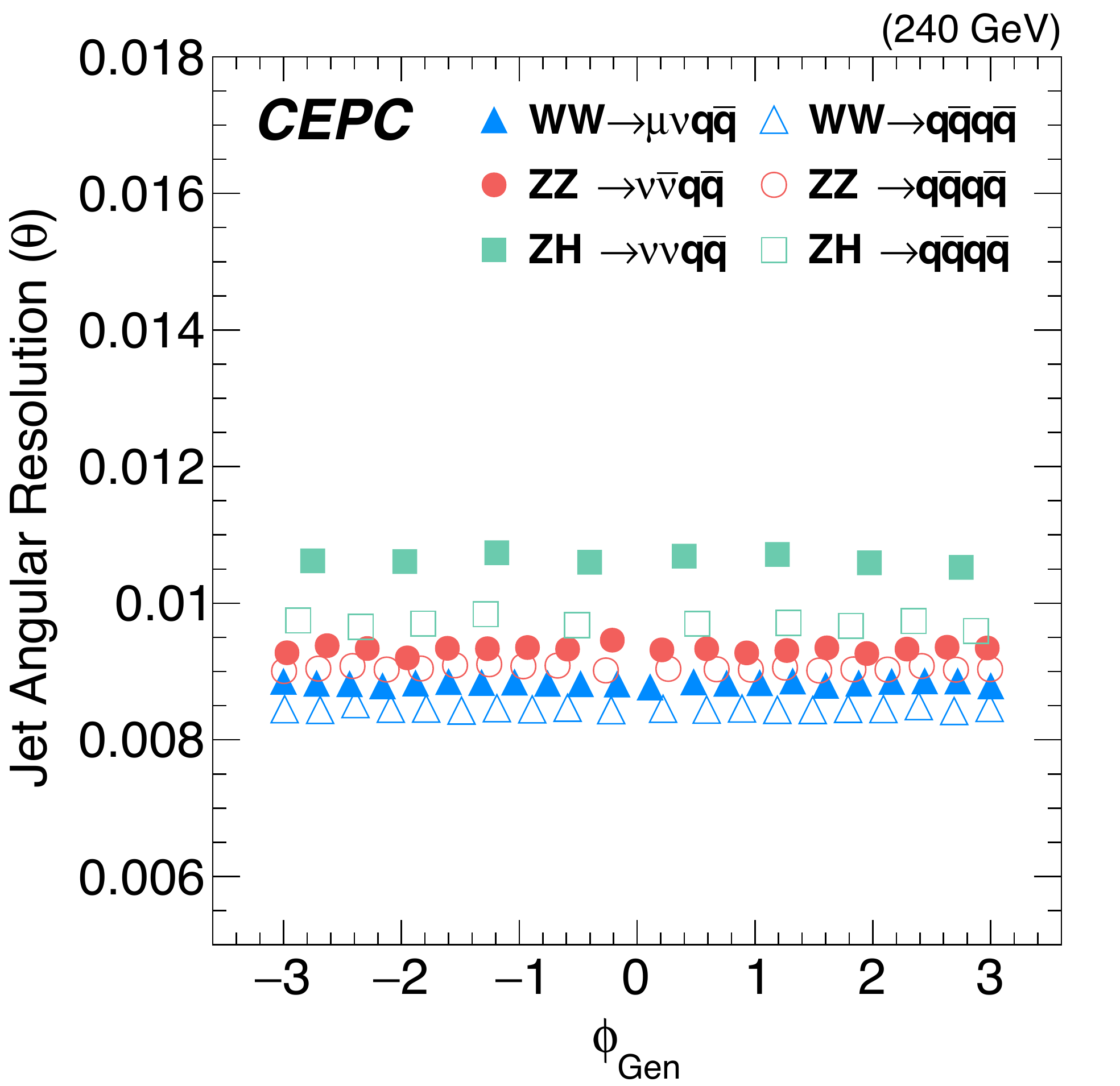}    
\end{minipage}%
}%
\subfigure[]{
\begin{minipage}[t]{0.35\linewidth}
\includegraphics[width=1.0\columnwidth]{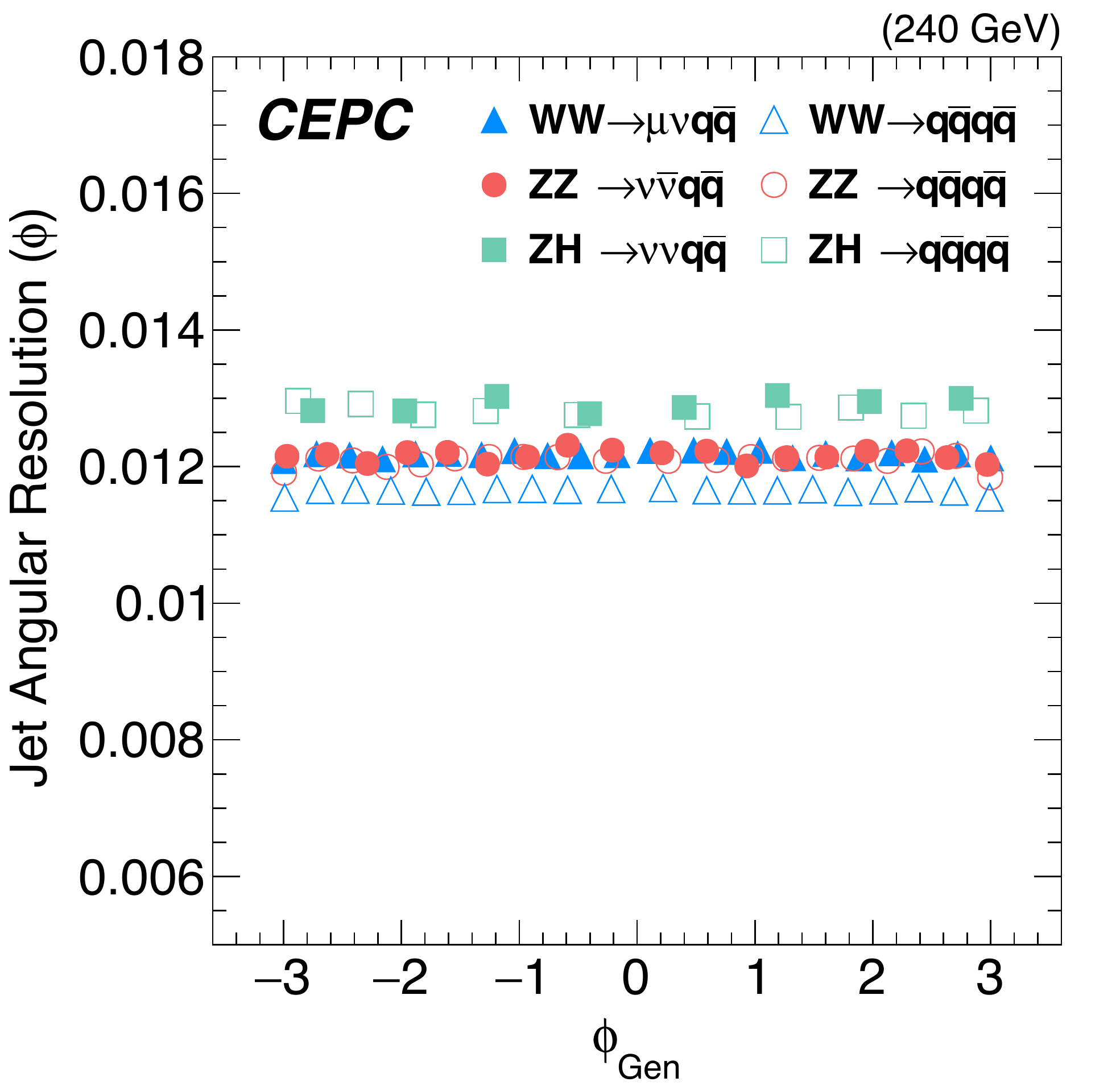}    
\end{minipage}%
}%
\vspace{-0.6cm}
\subfigure[]{
\begin{minipage}[t]{0.35\linewidth}
\includegraphics[width=1.0\columnwidth]{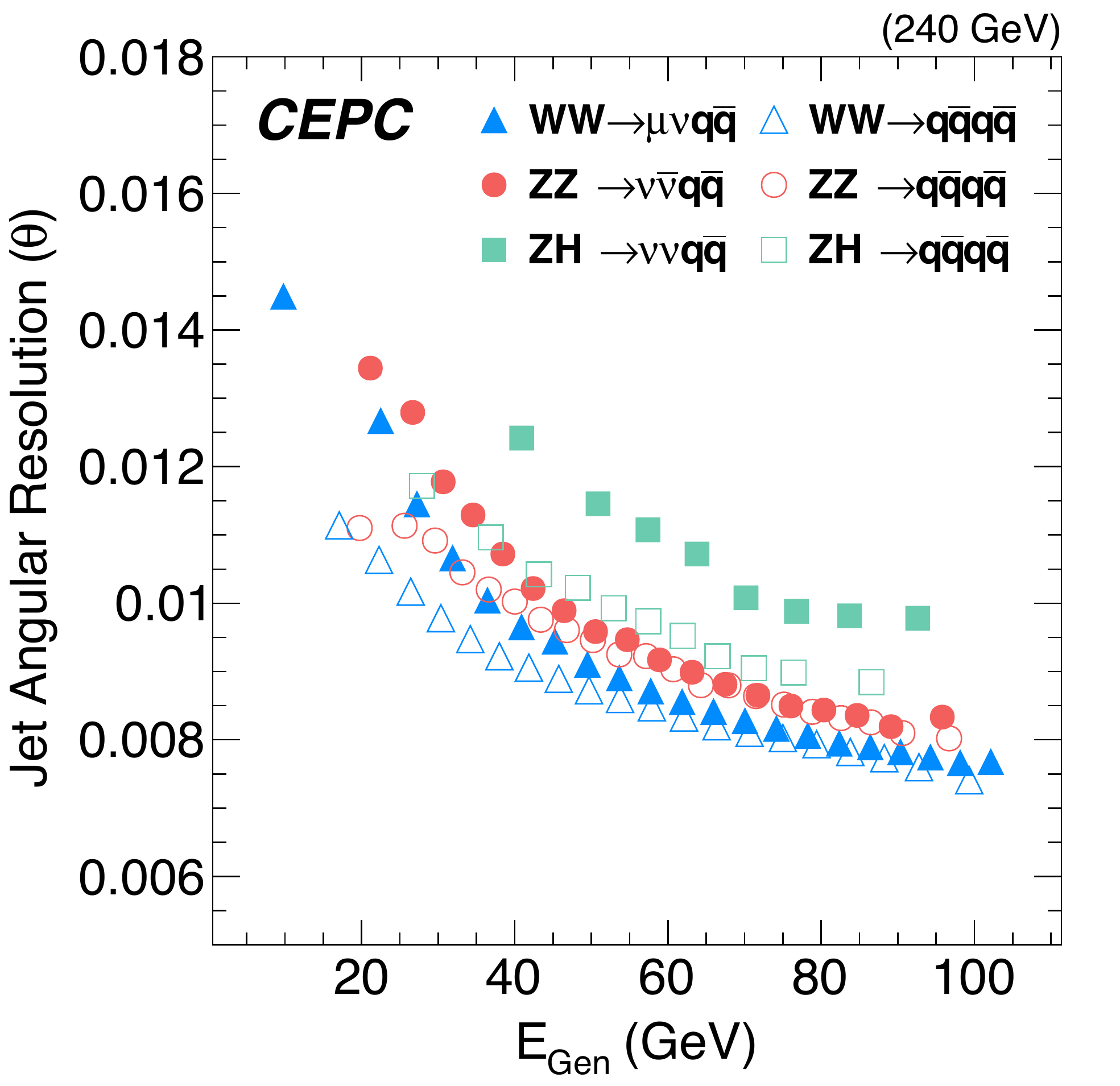}    
\end{minipage}%
}%
\subfigure[]{
\begin{minipage}[t]{0.35\linewidth}
\includegraphics[width=1.0\columnwidth]{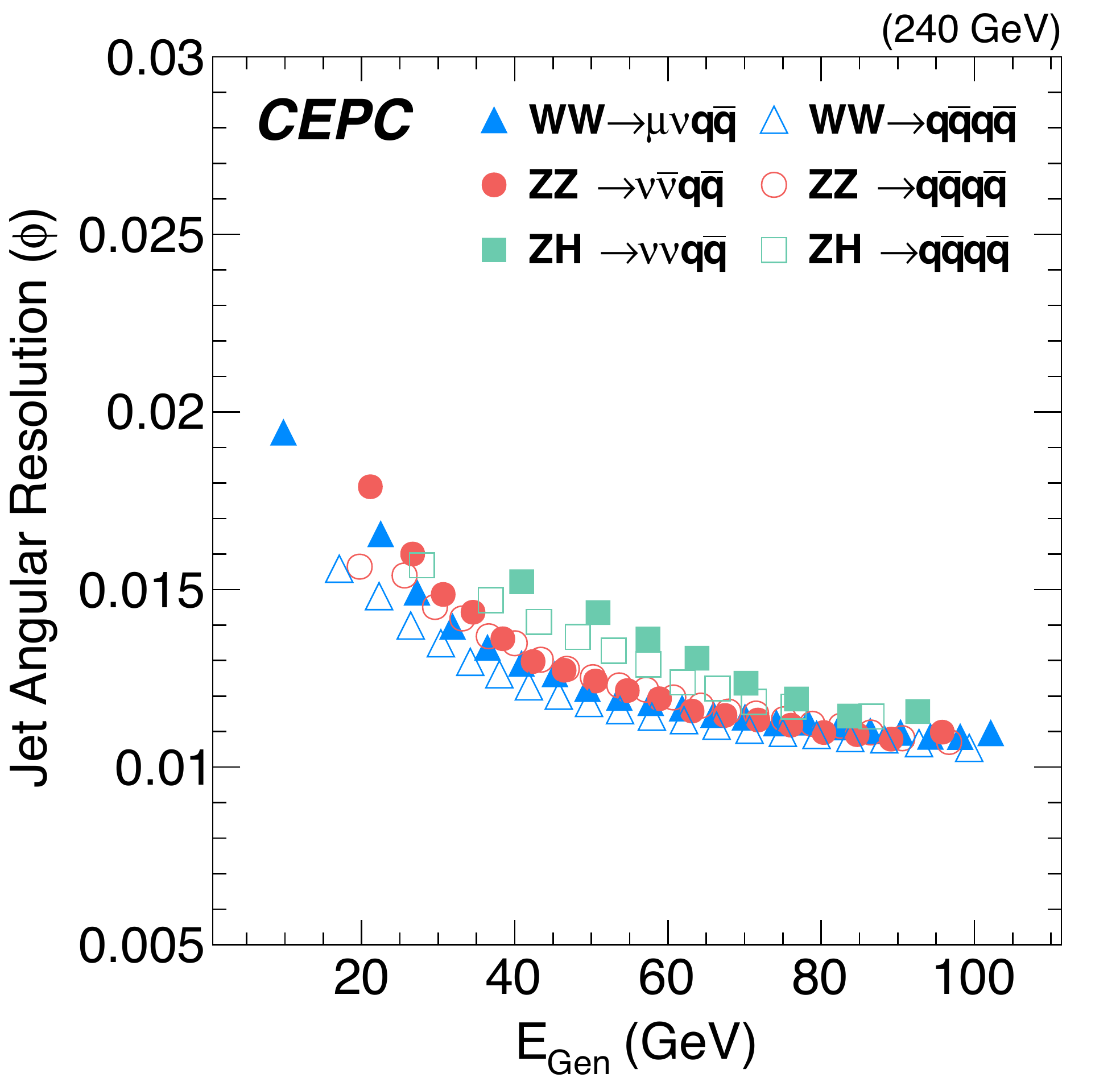}    
\end{minipage}%
}%
\caption{Jet angular resolution as functions of (a-b) the $cos\theta_{Gen}$, (c-d) the azimuth angle, and (e-f) the GenJet energy for 2- (solid symbols) and 4-jet (open symbols) final states. The errors shown are only statistical.}
\label{fig_ee_kt_JAR_cos_RecoGen}
\end{figure}

The JAR($\theta$) and JAR($\phi$) as functions of $cos\theta_{Gen}$, azimuth angle, and GenJet energy of the baseline reconstruction are illustrated in Fig.~\ref{fig_ee_kt_JAR_cos_RecoGen}. Generally, the JAR($\theta$) is from 0.8-1$\%$ and JAR($\phi$) is around 1.1$\%$. The performance differences of these processes are induced by their essential kinematic difference. In addition, the 4-jet JAR($\theta$) and JAR($\phi$) are 0.05$\%$ lower than 2-jet events because the splitting of calorimeter hits for energetic jets happen more frequently in 2-jet events.

\begin{figure}[!ht]
\centering
\subfigure[]{
\begin{minipage}[t]{0.35\linewidth}
\includegraphics[width=1.0\columnwidth]{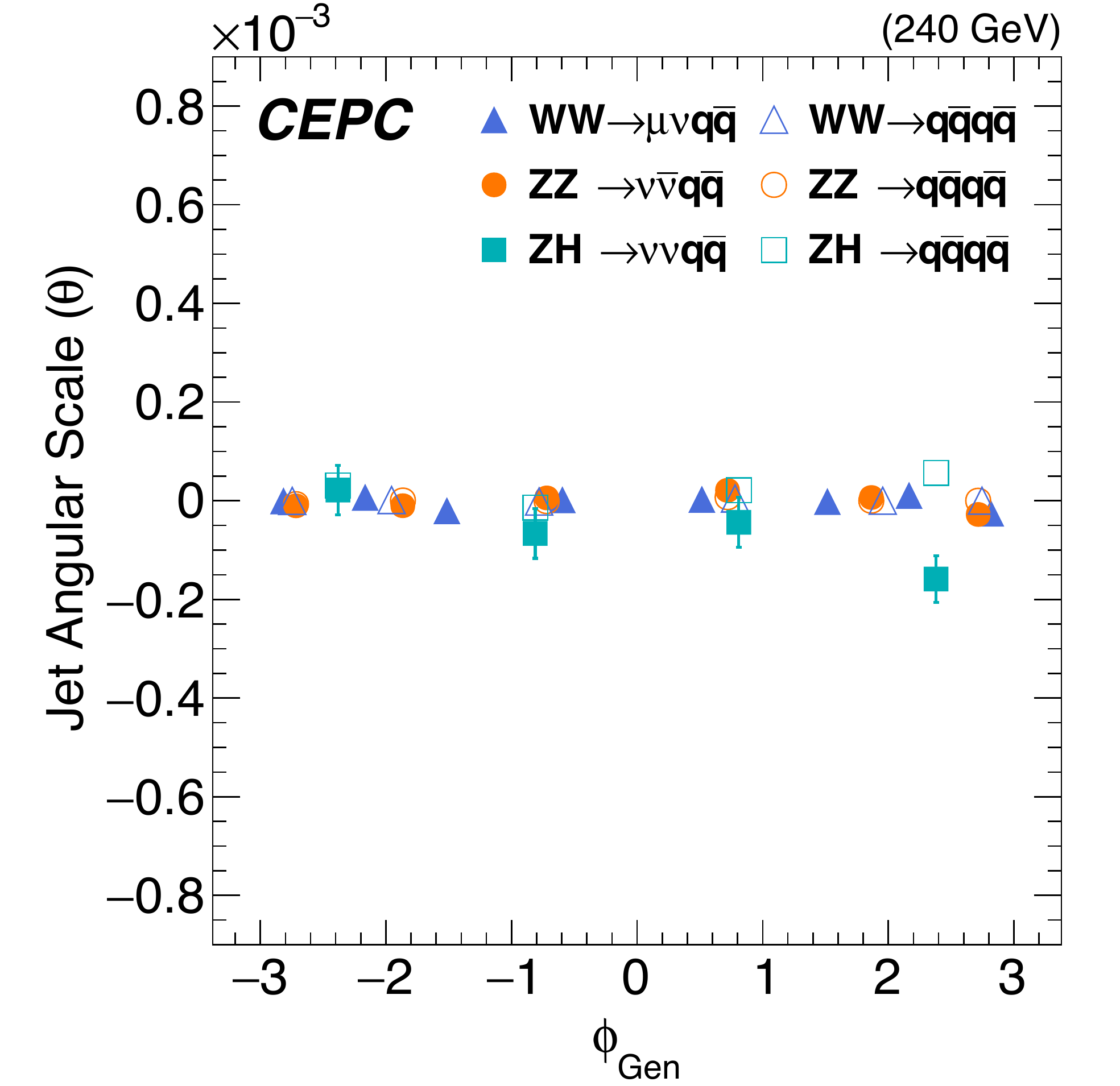}    
\end{minipage}%
}%
\subfigure[]{
\begin{minipage}[t]{0.35\linewidth}
\includegraphics[width=1.0\columnwidth]{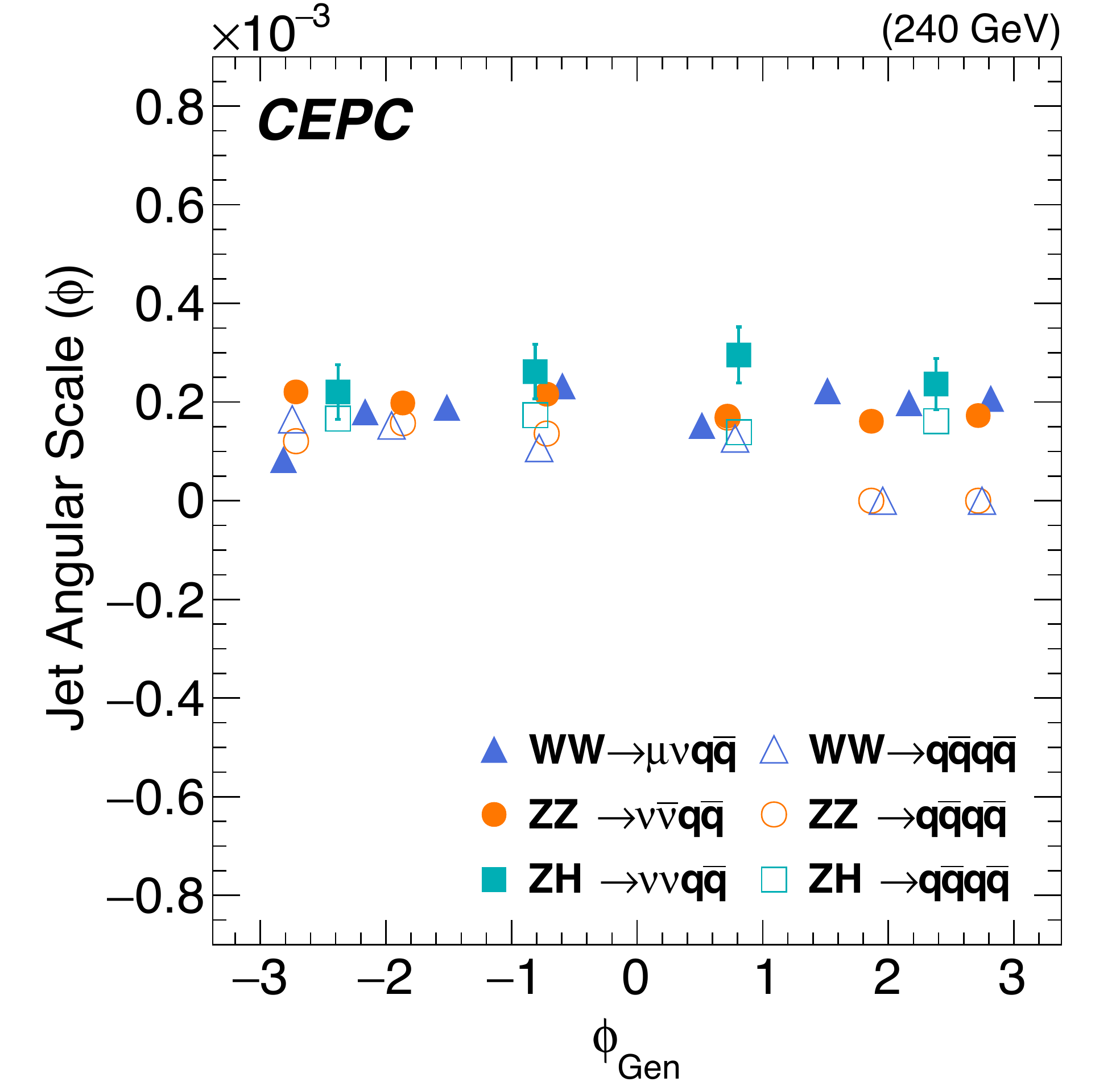}    
\end{minipage}%
}%
\vspace{-0.6cm}
\subfigure[]{
\begin{minipage}[t]{0.35\linewidth}
\includegraphics[width=1.0\columnwidth]{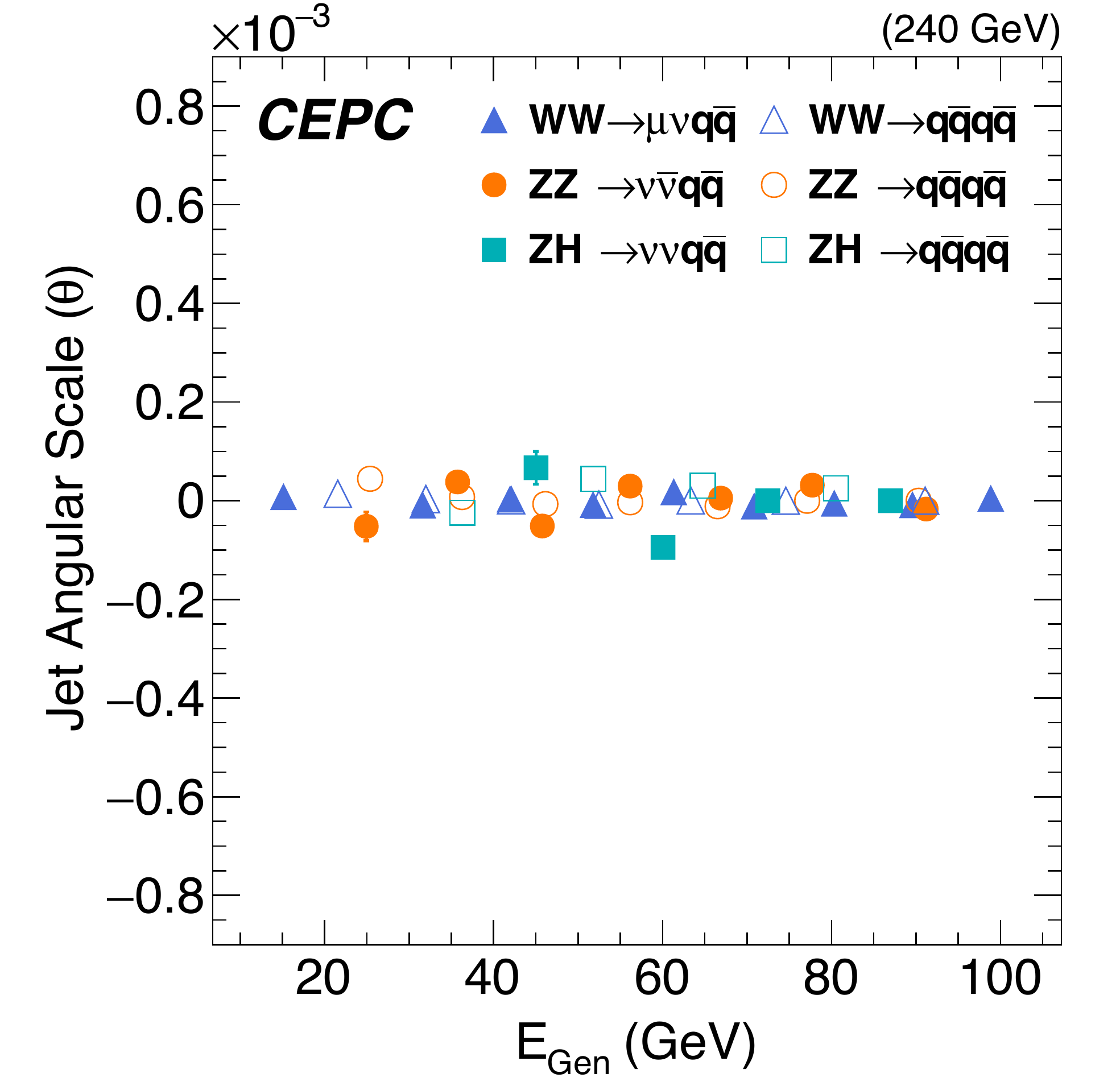}    
\end{minipage}%
}%
\subfigure[]{
\begin{minipage}[t]{0.35\linewidth}
\includegraphics[width=1.0\columnwidth]{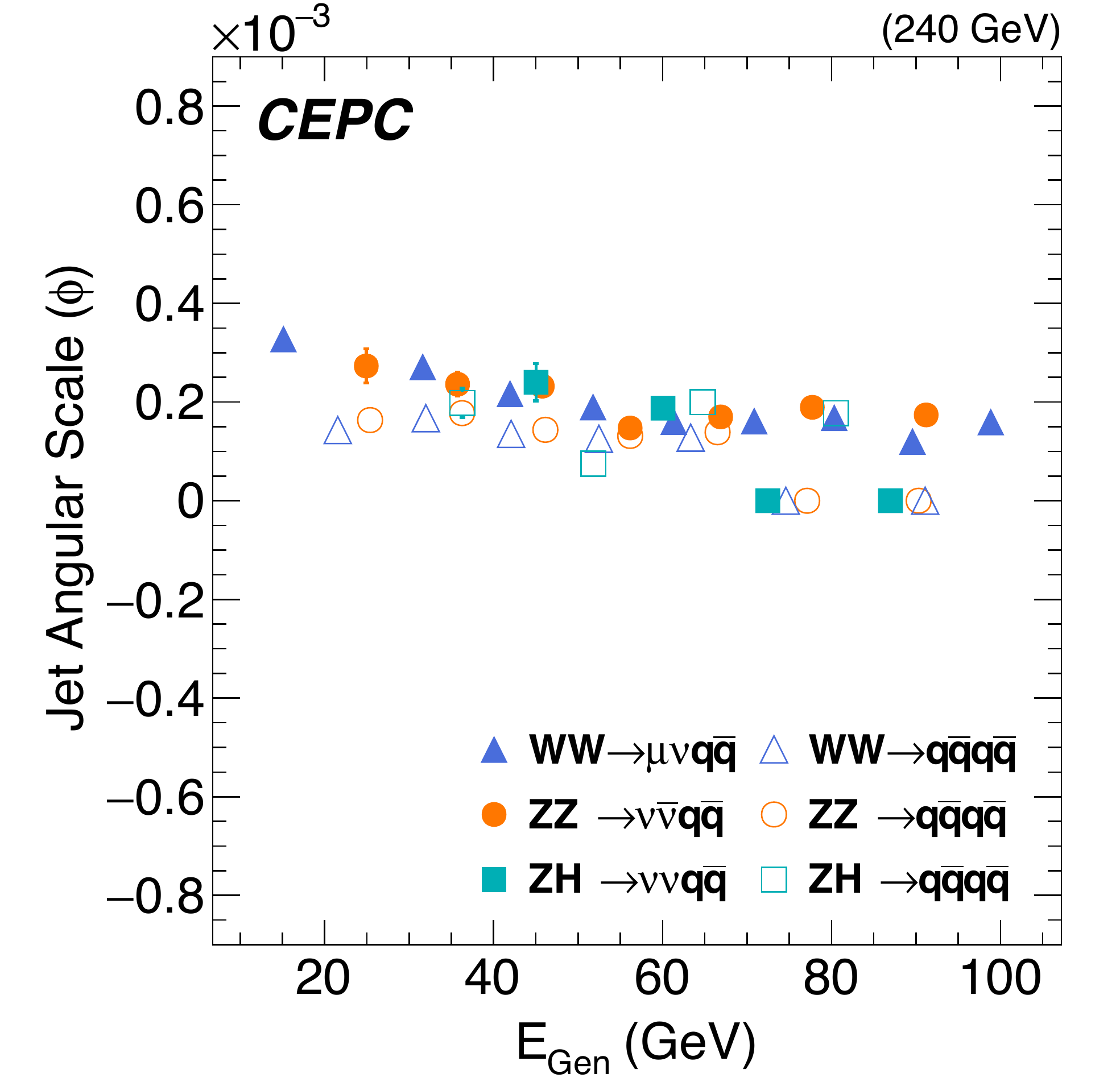}    
\end{minipage}%
}%
\vspace{-0.4cm}
\caption{Jet angular scale as functions (a-b) the azimuth angle, and (c-d) the GenJet energy for 2- (solid symbols) and 4-jet (open symbols) final states. The errors shown are only statistical. JAS($\theta$) is controlled to be near 0.02$\%$ while azimuth angular JAS($\phi$) is within 0.04$\%$, both with RMS around $10^{-5}$.}
\label{fig_ee_kt_JAS_cos_RecoGen}
\end{figure}

Similar to the case for JER, the JAR is also better for more energetic jets. JAR($\theta$) depends on the polar angle due to the amount of material and acceptance. JAR($\phi$) increases in the forward region since momentum resolution propagates to the azimuth angle definition ($\phi = tan^{-1}\frac{P_{y}}{ \sqrt{P_{x}^2 \Plus P_{y}^2}}$). Both JAR($\theta$) and JAR($\phi$) are independent of the azimuth angle as the detector is symmetric along the azimuth direction. Correspondingly, for the JAS results in Fig.~\ref{fig_ee_kt_JAS_cos_RecoGen}, the baseline JAS($\theta$) of the CEPC are well controlled to be near 0.02$\%$ and JAS($\phi$) within 0.04$\%$, both with root-mean-square (RMS) around $10^{-5}$.

The core fraction is utilized to examine how many events are included under the corresponding jet energy and angular resolution. Table~\ref{Cover_Fraction} summarizes the number of core fractions for each 2-/4-jet event. Generally, 4$\%$ JER contains around 60$\%$ of events and 1$\%$ JAR covers around 40-50$\%$ for 2-/4-jet processes. The 2-jet events have a higher core fraction than 4-jet events, where the fractions imply that both of the relative energy and angular distributions are close to a Gaussian distribution.

\begin{table*}
\caption{The table shows the core fraction of events within the jet energy and angular resolutions. ($\dagger$) One of the Z bosons is forced to decay to neutrinos in the final state.} 
\label{Cover_Fraction}
\begin{tabular*}{\textwidth}{@{\extracolsep{\fill}}clccc@{}}
\hline
Operation mode & \multicolumn{1}{l}{ Process } & \multicolumn{1}{c}{JER ($\%$)} & \multicolumn{1}{c}{JAR($\theta$) ($\%$)} & \multicolumn{1}{c}{JAR($\phi$) ($\%$)} \\
\hline
$Z$ 		& $e^{ \Plus }e^{ \Minus } \rightarrow Z \rightarrow q\bar{q}$		  																& 57.6 & 52.5 & 44.7\\
\hline
               & $e^{ \Plus }e^{ \Minus } \rightarrow WW \rightarrow \mu \nu q\bar{q}$                  								& 49.1 & 46.7 & 46.9\\
               & $e^{ \Plus }e^{ \Minus } \rightarrow WW \rightarrow q\bar{q}q\bar{q}$                  								& 58.2 & 50.9 & 49.9\\                           
$H$     	& $e^{ \Plus }e^{ \Minus } \rightarrow ZZ \rightarrow \nu \bar{\nu} q\bar{q}$            								& 56.4 & 48.3 & 40.6\\
               & $e^{ \Plus }e^{ \Minus } \rightarrow ZZ \rightarrow q\bar{q}q\bar{q}$                     								& 61.4 &	47.9 & 48.3\\
               & $e^{ \Plus }e^{ \Minus } \rightarrow ZH \rightarrow \nu \bar{\nu}\left( q\bar{q} \; or \; gg \right)\left(\dagger\right)$ 	& 62.5 &	44.1& 48.5\\
               & $e^{ \Plus }e^{ \Minus } \rightarrow ZH \rightarrow q\bar{q}\left( q\bar{q} \; or \; gg \right)$                    			& 63.3 & 43.5& 44.8\\       
\hline
\end{tabular*}
\end{table*}

\section{Performance with thrust-based algorithm}
\label{4}

In the current approach, jet clustering is an indispensable step to quantify the jet reconstruction performance. Given the fact that there are multiple jet clustering algorithms, it is interesting and necessary to study the dependence of the jet performance on the jet clustering algorithm. To evaluate the impact of the clustering method, the disparate studies of JER and JAR of the thrust-based algorithm are compared to the baseline algorithm in this section. The jet performances of the baseline algorithm are the same as those in Sect. \ref{3}. The WW, ZZ, and ZH semi-leptonic decay events during the Higgs factory operation mode and Z decays to 2-jet events during the Z factory operation mode are utilized. 
\subsection{Thrust-based jet clustering method}
\label{4.1}
The thrust-based jet clustering method is inspired from 2-jet events at the Z factory operation mode with a back-to-back topology which uses the classical event-shape variable, thrust~\cite{Becher_2008,Ali_2011,thrust_ref_1,thrust_ref_2,thrust_ref_3}, define in Eq.~\ref{eq:1} as:
\vspace{-0.8cm}
\begin{center}
\begin{equation} \label{eq:1}
T =  \mathop{max} \limits_{\overrightarrow{n}_{T}}  \left( \frac{ \sum_{i}^{N} |P_{i} \cdot n_{T}| } {\sum_{i}^{N} |P_{i}|} \right)
\end{equation}
\end{center}
where $P_{i}$ represents the momentum of each final-state particle in an event, $N$ is the number of the final-state particles, and $\overrightarrow{n}_{T}$ is the thrust axis, which is the unit vector that maximizes the sum of the projections of $P_{i}$. The four momenta of all the final-state particles are exploited to characterize the topology of the efflux of an event along the thrust axis $\overrightarrow{n}_{T}$. A perfect back-to-back $q\bar{q}$ system brings $T=1$ while an even distribution across the hemisphere leads to $T = 1 \Slash 2$. Using this property, 2-jet topologies can be clustered at the CEPC because those 2-jet final states only come from a single system after the $e^{ \Plus }e^{ \Minus }$ collision. Consequentially, the thrust-based algorithm is limited to 2-jet events, which is accessible since 2 jets can be well identified from 4- and 6-jet topologies.

The clustering program is as follows: Firstly, the collinear di-jet system is boosted to the back-to-back topology (for the boosted cases, such as $e^{ \Plus }e^{ \Minus } \rightarrow ZZ \rightarrow \nu\bar{\nu}q\bar{q}$), and all of the final-state particles are boosted to the rest frame where the system momentum is calculated without neutrinos at the GenJet level. Secondly, the trial axes are iterated until the thrust axis is found and then the event topology is divided into two hemispheres. Subsequently, the particles in each hemisphere are clustered into a jet. Finally, each particle is recovered back to the lab (original) frame. It is expected that the thrust-based algorithm has better performance than the baseline due to its ability to separate an event under the back-to-back topology.

\subsection{Jet energy resolution of the thrust-based algorithm}
\label{4.2}

\begin{figure}[!ht]
\centering
\subfigure[]{
\begin{minipage}[t]{0.35\linewidth}
\includegraphics[width=1.0\columnwidth]{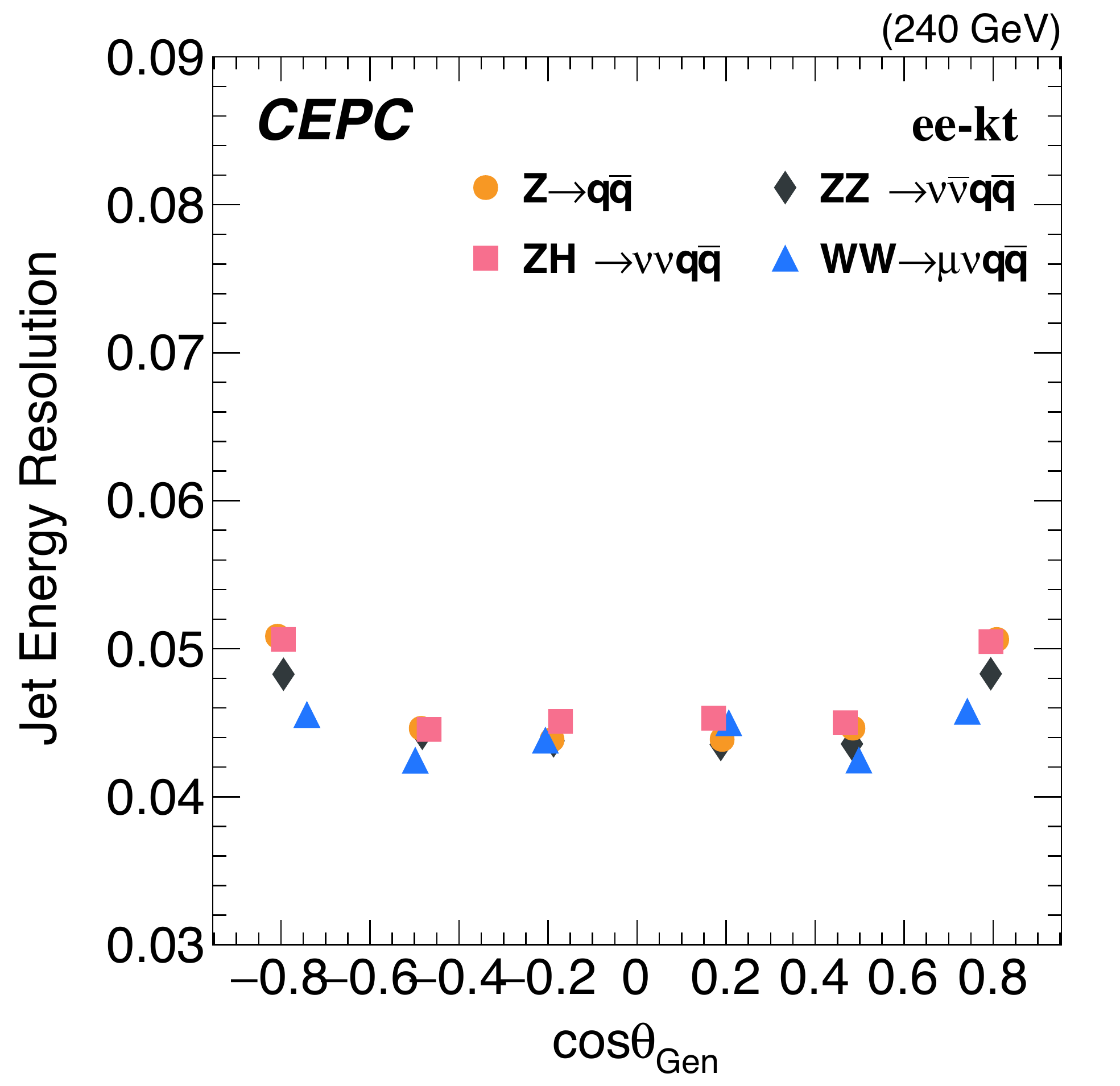}    
\end{minipage}%
\label{JER_ee_kt_Thrust_a}
}%
\subfigure[]{
\begin{minipage}[t]{0.35\linewidth}
\includegraphics[width=1.0\columnwidth]{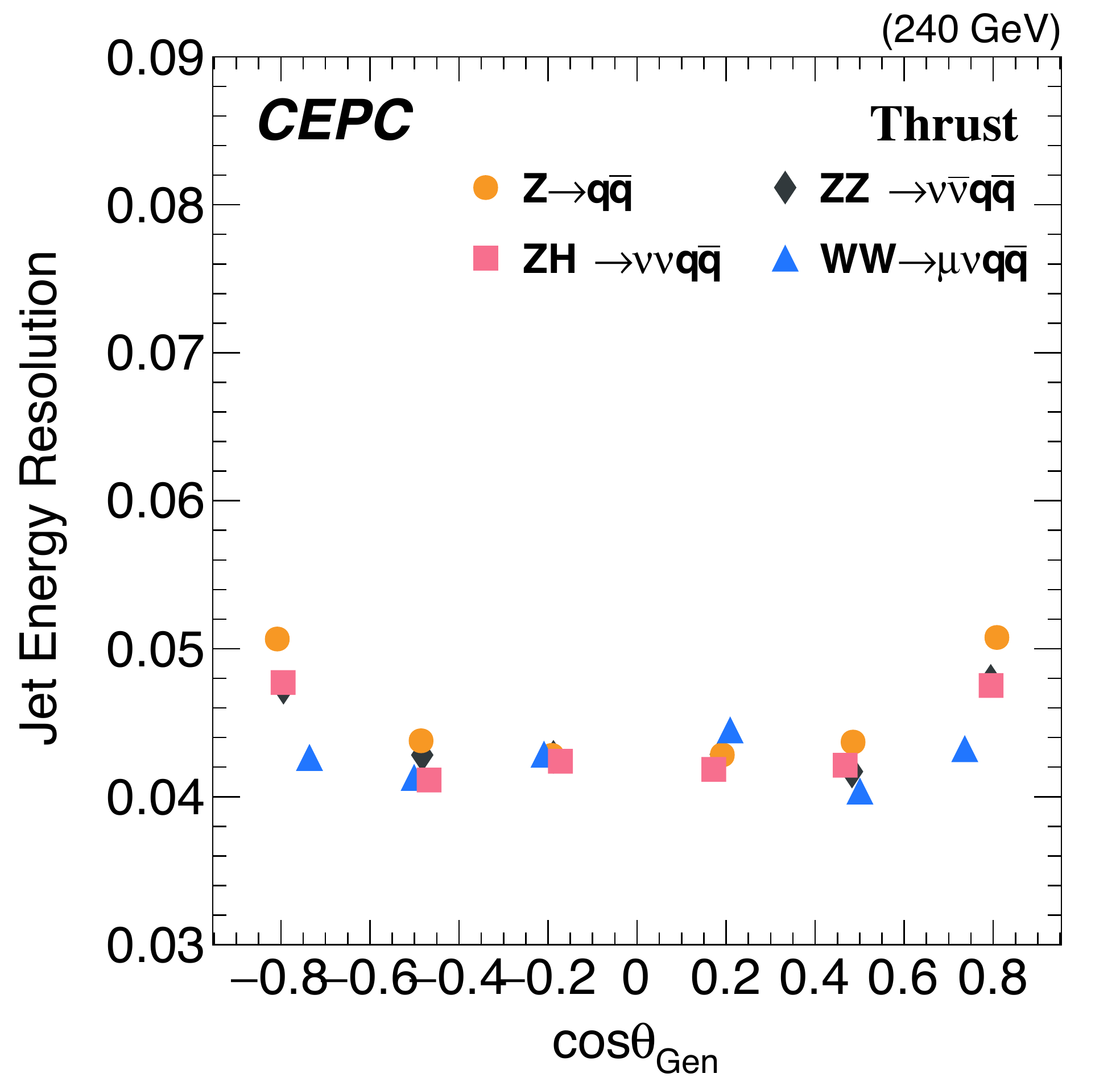}    
\end{minipage}%
\label{JER_ee_kt_Thrust_b}
}%
\\
\center
\vspace{-0.85cm}
\subfigure[]{
\begin{minipage}[t]{0.45\linewidth}
\includegraphics[width=1.0\columnwidth]{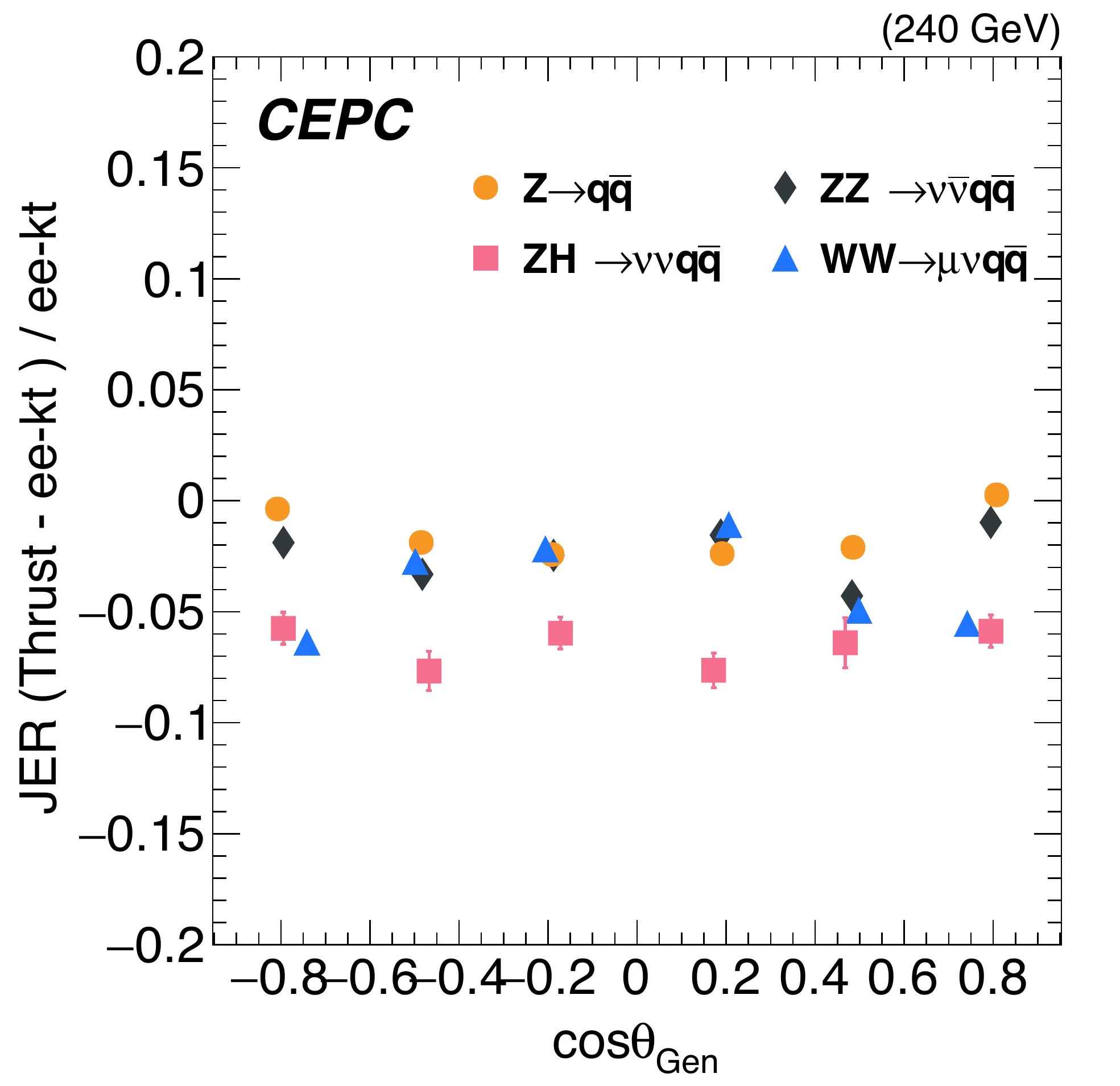}    
\end{minipage}%
\label{JER_ee_kt_Thrust_c}
}%

\caption{Performance of JER as a function of the GenJet angle, $cos \theta_{Gen}$, with (a) the baseline algorithm and (b) the thrust-based algorithm among several benchmark 2-jet final-state processes, and (c) the relative difference of JER between the two algorithms. The 2-jet results of WW, ZZ, and ZH using the baseline clustering algorithm are exactly the same as those in Fig. \ref{ee_kt_JER_cali_cos_RecoGen}. $Z \rightarrow q \bar{q}$ is operating with a different center-of-mass energy of 91.2 GeV. The errors shown are only statistical.}
\label{JER_ee_kt_Thrust}
\end{figure}

The JER as a function of the $cos\theta_{Gen}$ for the baseline and thrust-based algorithms is presented in Fig.~\ref{JER_ee_kt_Thrust_a} and Fig.~\ref{JER_ee_kt_Thrust_b}. The relative difference between the two algorithms is expounded in Fig.~\ref{JER_ee_kt_Thrust_c}. The improvement of JER is about 2-8$\%$ depending on the physics processes using the thrust-based algorithm compared to the baseline algorithm. 

\begin{figure}[!ht]
\centering
\subfigure[]{
\begin{minipage}[t]{0.45\linewidth}
\includegraphics[width=1.0\columnwidth]{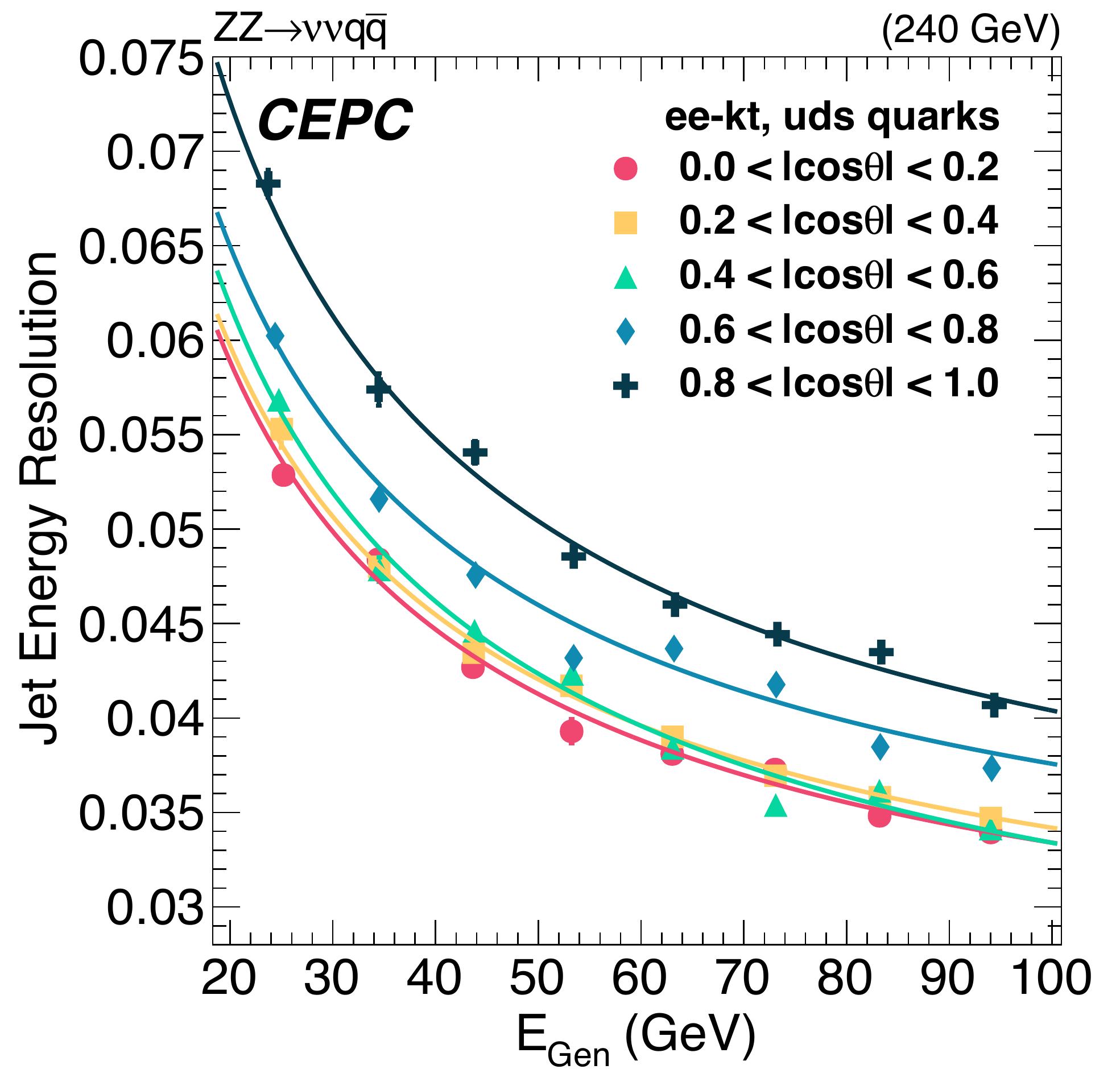}    
\end{minipage}%
}%
\subfigure[]{
\begin{minipage}[t]{0.45\linewidth}
\includegraphics[width=1.0\columnwidth]{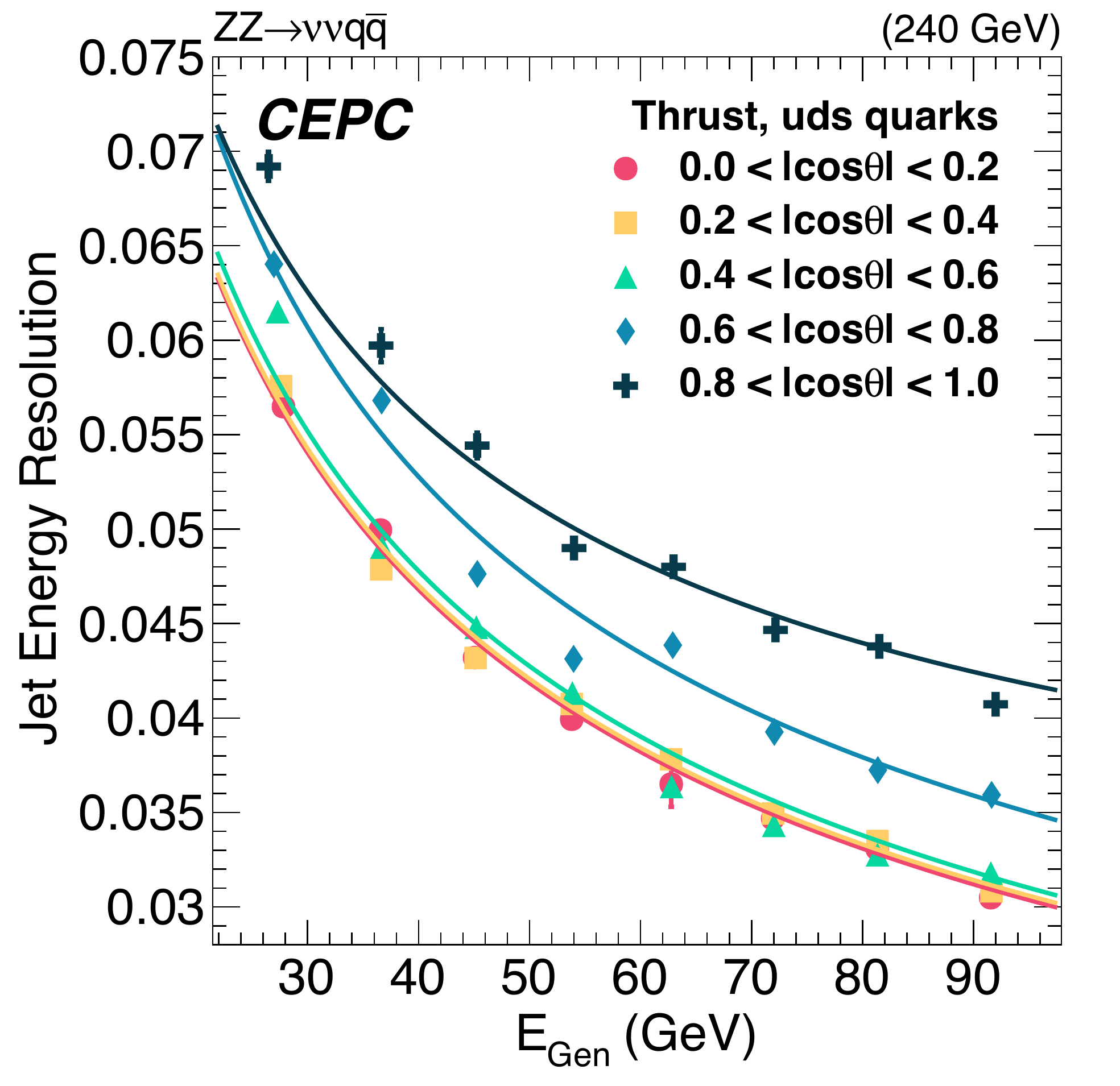}    
\end{minipage}
}%
\caption{The JER as a function of the GenJet energy of light-flavored (uds) jets from semi-leptonic decay are divided into five fiducial regions. (a) is clustered by the baseline algorithm while (b) is clustered by the thrust-based algorithm. The errors shown are only statistical.}
\label{JER_ee_kt_Thrust_fiducial}
\end{figure}

The comparisons between the baseline and thrust-based algorithms for the JER as a function of the GenJet energy are shown in Fig.~\ref{JER_ee_kt_Thrust_fiducial}. The JERs of light flavors (uds) from $ZZ$ semi-leptonic decay events are further divided into five fiducial regions. Results are fitted by the "$SC$" function to characterize the jet energy resolution, where $S$ is the stochastic fluctuations divided by $\sqrt{E}$, and $C$ is a constant term with no energy scaling. The 8$\%$ improvement of JER brought by the thrust-based algorithm is more significant when $E_{Gen}$ > 60 GeV within $|cos\theta|< 0.6$.

\subsection{Jet angular resolution of the thrust-based algorithm}
\label{4.3}

\begin{figure}[!ht]
\centering
\subfigure[]{
\begin{minipage}[t]{0.35\linewidth}
\includegraphics[width=1.0\columnwidth]{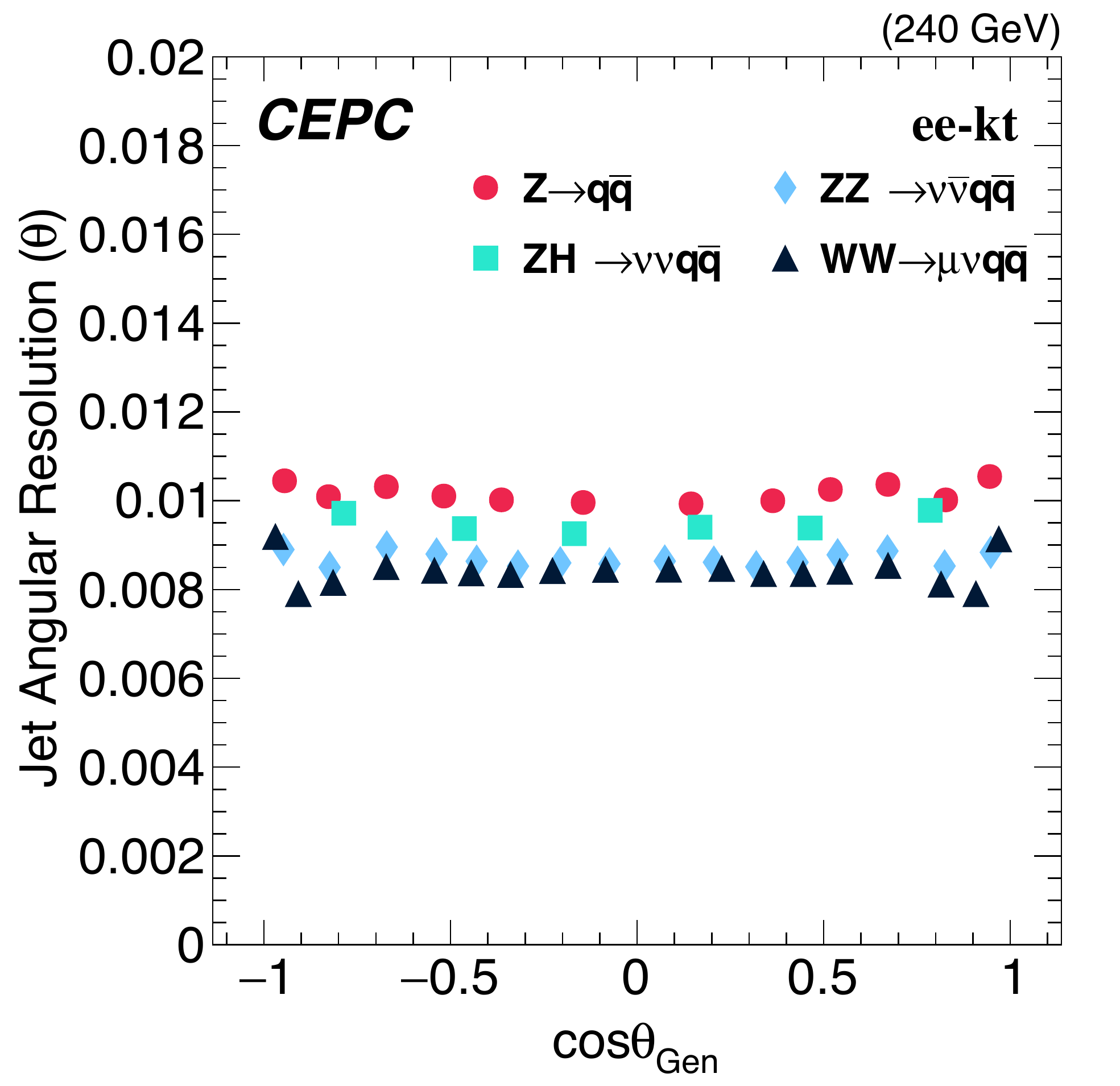}    
\end{minipage}%
}%
\subfigure[]{
\begin{minipage}[t]{0.35\linewidth}
\includegraphics[width=1.0\columnwidth]{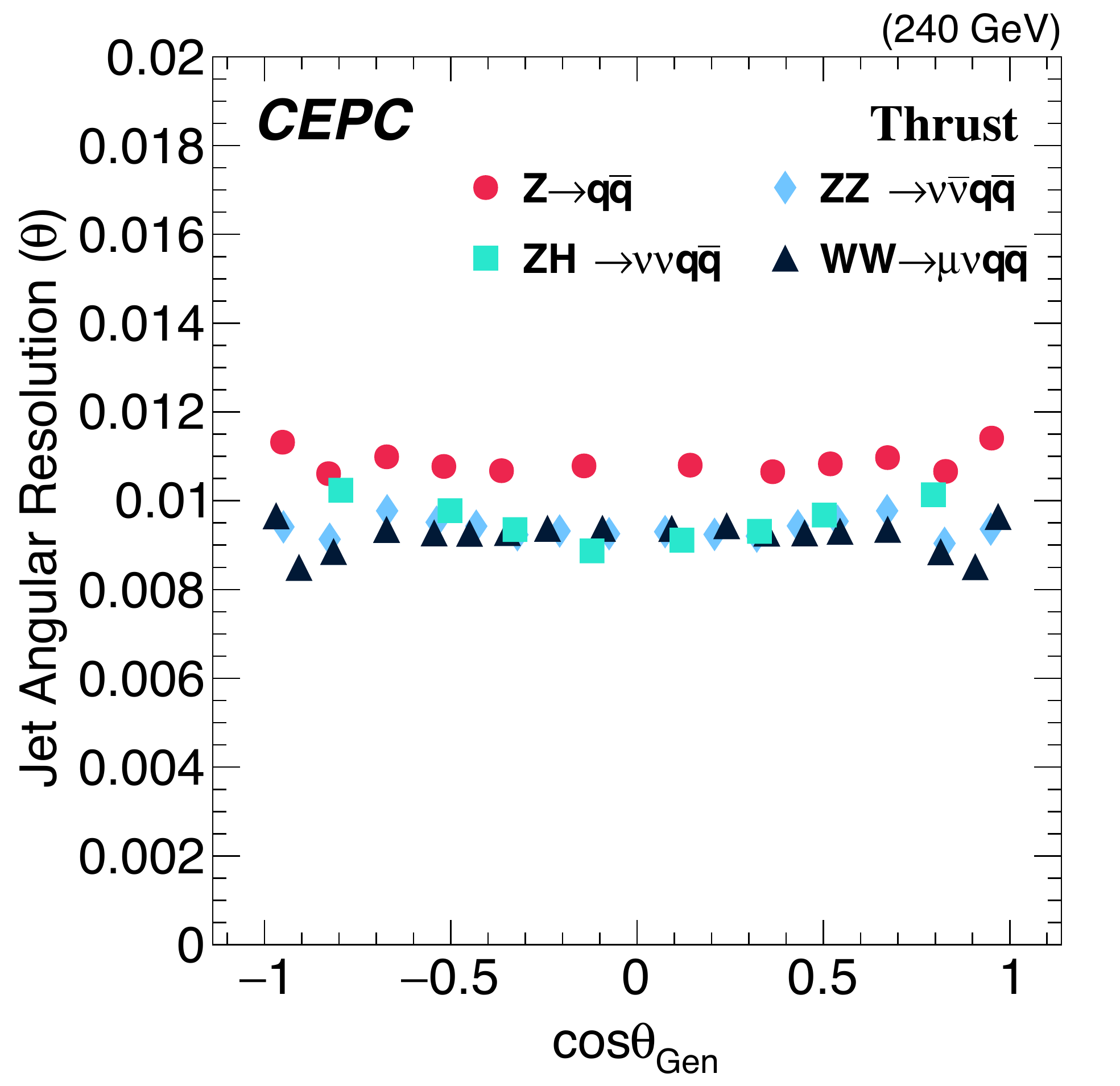}    
\end{minipage}%
}%
\vspace{-0.6cm}
\subfigure[]{
\begin{minipage}[t]{0.35\linewidth}
\includegraphics[width=1.0\columnwidth]{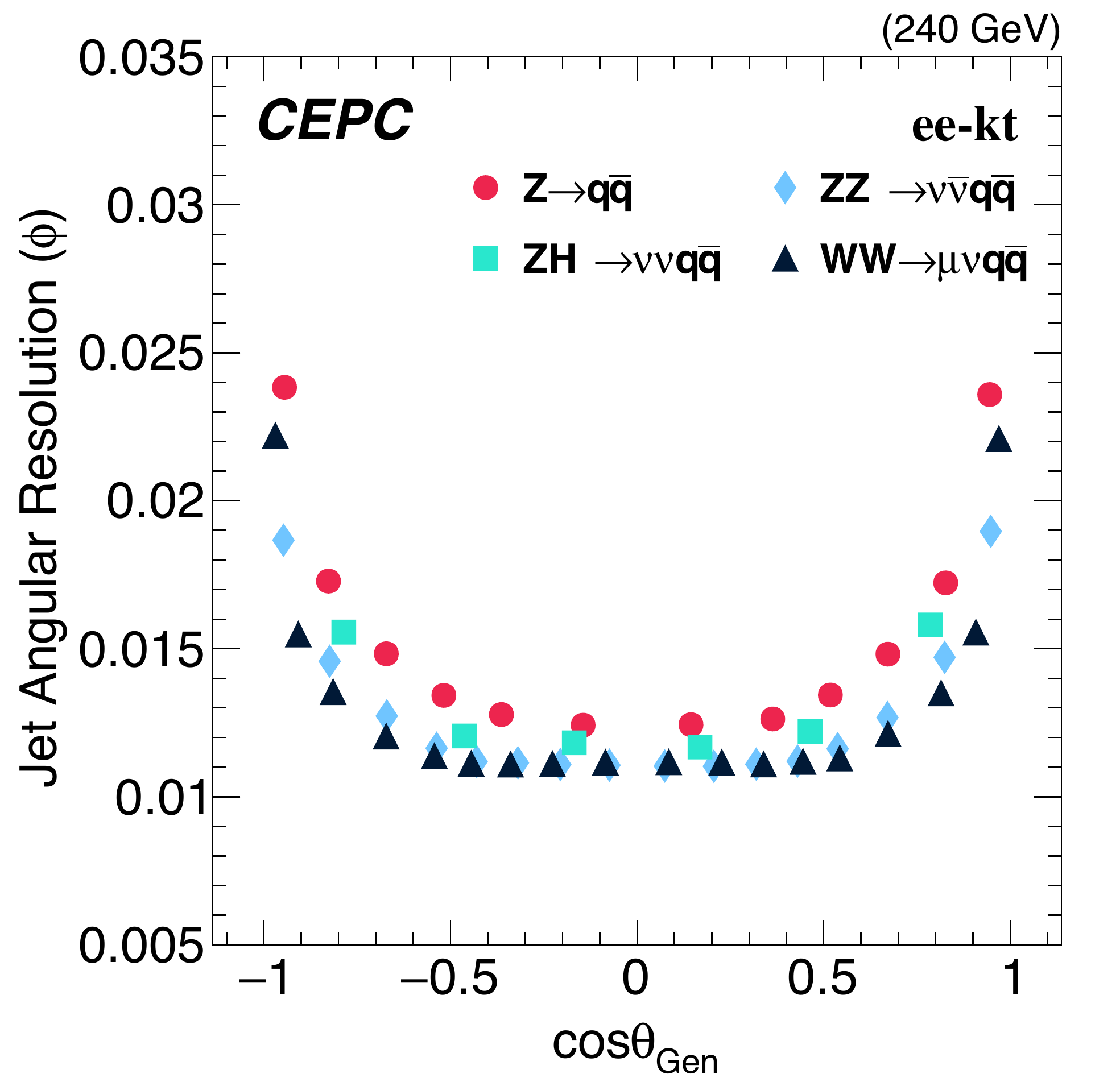}    
\end{minipage}%
}%
\subfigure[]{
\begin{minipage}[t]{0.35\linewidth}
\includegraphics[width=1.0\columnwidth]{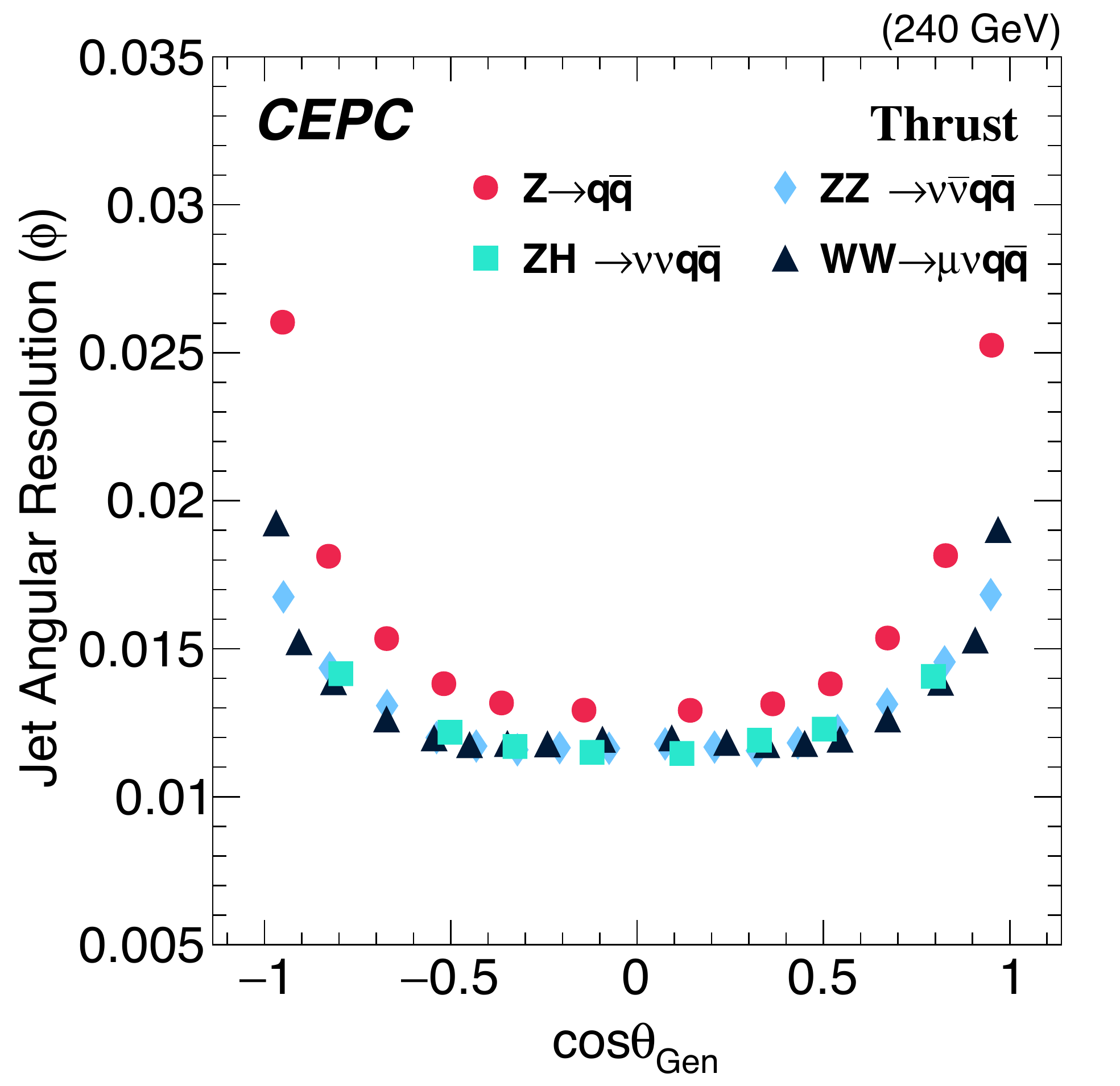}    
\end{minipage}%
}%

\vspace{-0.6cm}
\subfigure[]{
\begin{minipage}[t]{0.45\linewidth}
\includegraphics[width=1.0\columnwidth]{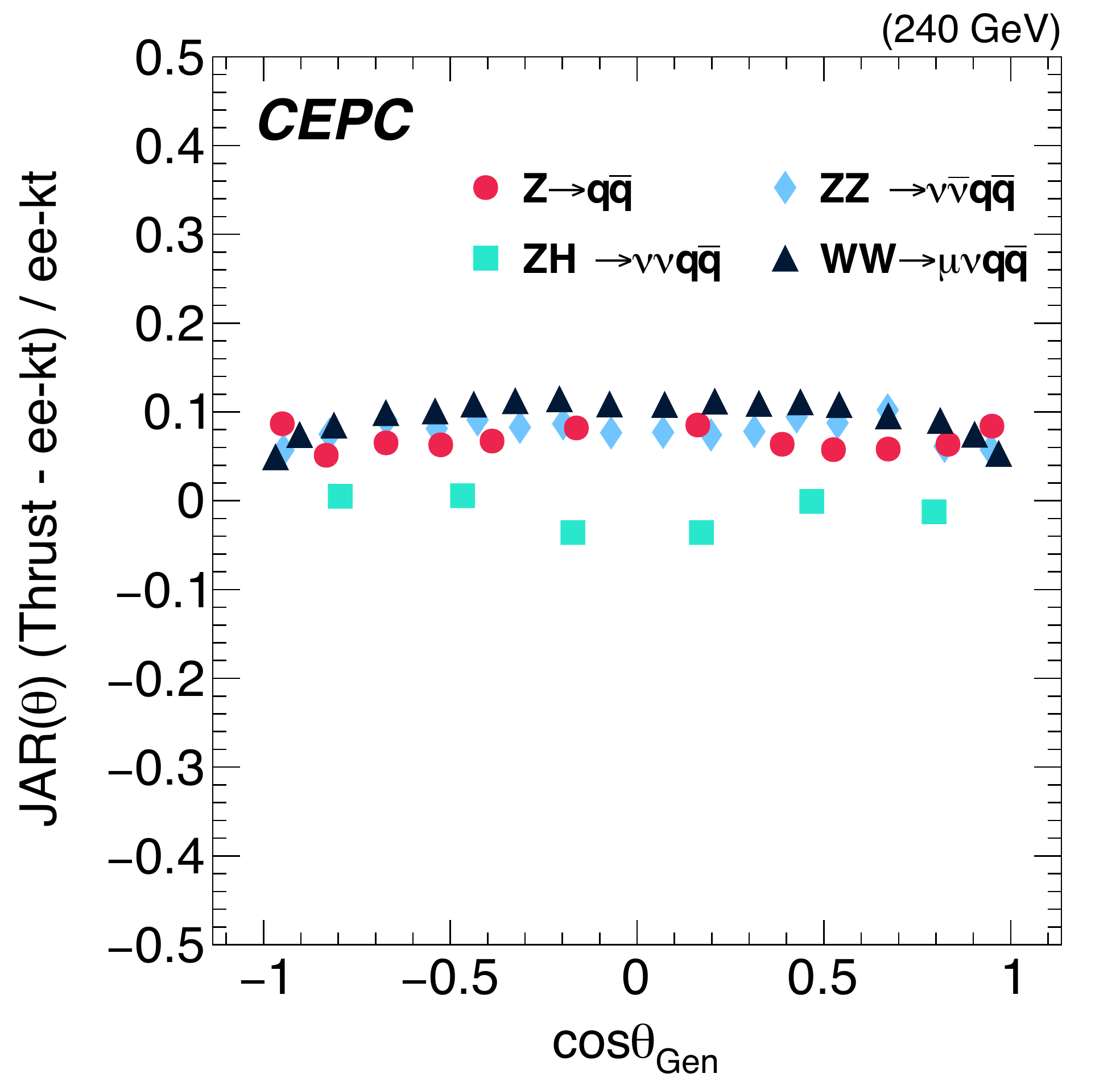}    
\end{minipage}%
\label{JAR_ee_kt_Thrust_diff1}
}%
\subfigure[]{
\begin{minipage}[t]{0.45\linewidth}
\includegraphics[width=1.0\columnwidth]{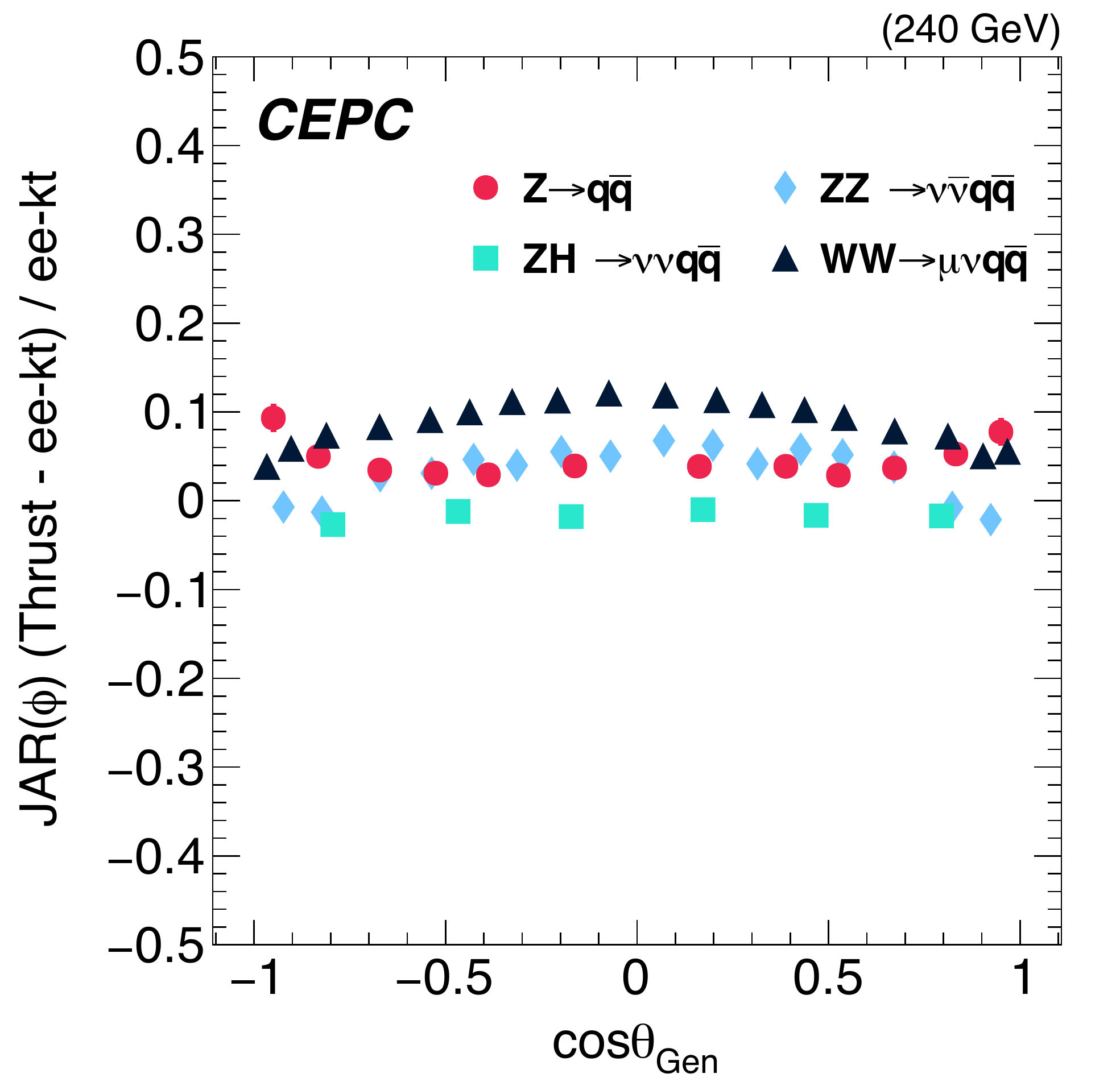}    
\end{minipage}%
\label{JAR_ee_kt_Thrust_diff2}
}%

\caption{The comparison of the polar and azimuth angular resolutions as a function of the GenJet $cos \theta$ in several benchmark 2-jet events clustered by the (a, c) baseline and (b, d) thrust-based algorithms and the relative difference of the (e) polar and (f) azimuth angular resolutions between the baseline algorithm and thrust-based algorithm. The results of WW, ZZ, and ZH using the baseline clustering algorithm are exactly the same as those in Fig. \ref{fig_ee_kt_JAR_cos_RecoGen}. $Z \rightarrow q \bar{q}$ is operating with a different center-of-mass energy of 91.2 GeV. The errors shown are only statistical.}
\label{JAR_ee_kt_Thrust}
\end{figure}

The comparisons of the baseline and thrust-based algorithms in terms of JAR($\theta$) and JAR($\phi$) as a function of the GenJet $cos\theta$ are presented in Fig.~\ref{JAR_ee_kt_Thrust}. Both JAR($\theta$) and JAR($\phi$) are degraded by about 5-10$\%$ using the thrust-based algorithm compared to the baseline algorithm, as shown in Fig.~\ref{JAR_ee_kt_Thrust_diff1} and Fig.~\ref{JAR_ee_kt_Thrust_diff2}. Since the thrust-based algorithm considers the event-shape variable of an event, the thrust JAR/JAS is more sensitive to the effects of the Arbor PFA double-counting behavior, driven dominantly by high-multiplicity regions, than the baseline algorithm.

\clearpage
\section{Conclusion}
\label{5}

A profound understanding of the hadronic system and the jet reconstruction performance is essential for the CEPC physics potential and detector requirement-optimization studies.
Many critical CEPC physics measurements rely on processes with fully-hadronic and semi-leptonic final states. 
As an electron-positron collider, the CEPC is free of QCD background, providing large amounts of clean samples of these events. 
Compared to the LHC, the physics measurements with fully-hadronic and semi-leptonic final states are one of the major comparative advantages of electron-positron colliders, especially as a Higgs factory.
The jets can be identified with high efficiency even for very low-$P_{T}$ jets (see Fig~\ref{JER_CEPC_CMS}). A 4-6$\%$ JER is anticipated for 20-100 GeV jets. The reconstruction performance of the hadronic system can be characterized by the relative mass resolution of massive bosons decaying to hadronic final states, the boson mass resolution (BMR), which is roughly only two times worse than the $H \rightarrow \gamma \gamma$ channel. Based on the current baseline detector, we already achieved a Higgs boson mass resolution of 3.8$\%$ in di-jet final states~\cite{R8}.

We analyze the jet responses for major SM fully-hadronic and semi-leponic processes using full simulation samples processed with the official CEPC software. These samples include 17 million 2-jet processes (Z, WW, ZZ, and ZH) and 7.7 million 4-jet processes (WW, ZZ, and ZH) with 91.2/240 GeV center-of-mass energy. The samples are simulated with the CEPC baseline detector geometry and reconstructed with the Arbor Particle Flow algorithm~\cite{R12}.
The MC/reconstructed particles are clustered into Gen/Reco jets with the same jet clustering algorithm, $e^{\Plus}e^{\Minus}k_{t}$, and matched to each other using a minimal angle--oriented matching algorithm. 
The Arbor with the particle flow algorithm--oriented detector was used in the CEPC feasibility study and the CDR~\cite{R8}. The relative energy/angular difference of the jet pairs characterize the jet response, which is then quantified into JER/JES and JAR/JAS. 

For 2-jet processes, JERs are 4$\%$, and JARs are 1$\%$ in the barrel region; JERs are 3-5.5$\%$ and JARs are 0.8-1.1$\%$ within a jet energy of 20 to 100 GeV. A better jet energy resolution can be achieved for more energetic jets. Moreover, around 60$\%$ and 40 $\%$ of core fractions are covered by the energy and angular resolutions, correspondingly. 

The CEPC baseline detector has a JES of 1$\%$, with clear dependence on the jet polar angle and jet energy, and therefore can be corrected, while the JAS reaches an RMS of $10^{-5}$. A direct application of these differential dependences is to calibrate the W-boson mass down to 10 MeV in the semi-leptonic final states. Compared to the JER of CMS~\cite{JER_ref} (and ATLAS~\cite{Aad_2013}) detector at the LHC, the CEPC has 2-4 times better JER in the same $P_{T}$ range~\cite{the_paper_ref, R8}. 

The impact of jet clustering algorithms is analyzed via the comparison of two different jet clustering algorithms for the 2-jet processes. We introduced a thrust-based algorithm and compared its performance to the baseline $e^{ \Plus }e^{ \Minus }k_{t}$ algorithm. The thrust-based algorithm can improve the jet energy resolution by about 8$\%$ which is more significant for light-flavored jets with $E_{Gen}$ > 60 GeV in the $|cos\theta|< 0.6$ region with respect to the $e^{ \Plus }e^{ \Minus }k_{t}$ algorithm. However, the thrust-based algorithm is limited to 2-jet processes, and it degrades the angular resolution by about 5-10$\%$ with respect to the baseline. This paper quantifies the single jet response, and sets a new benchmark on the hadronic system reconstruction at the CEPC.

\acknowledgments
We very much appreciate Gang Li and Cheng-dong Fu for the official CDR samples production, which are intensively used in this study, and Bo Liu, Dan Yu, and Yongfeng Zhu for the technical support.

We acknowledge in particular the support of the Beijing Municipal Science and Technology Commission Project (Grant No. Z191100007219010), International Partnership Program of Chinese Academy of Sciences (Grant No. 113111KYSB20190030), Skackleton Program, Muon Imaging for Particle Geophysics and the Next Generation of High Energy Physics Experiment (Grant No. MOST 108-2638-M-008-001-MY2), and General Research Program of Taiwan (Grant No. MOST 106-2112-M-001-023).

\appendix
\section{Matching approaches}\label{app}
\begin{figure}[!ht]
\centering
\subfigure[]{
\begin{minipage}[t]{0.35\linewidth}
\includegraphics[width=1.0\columnwidth]{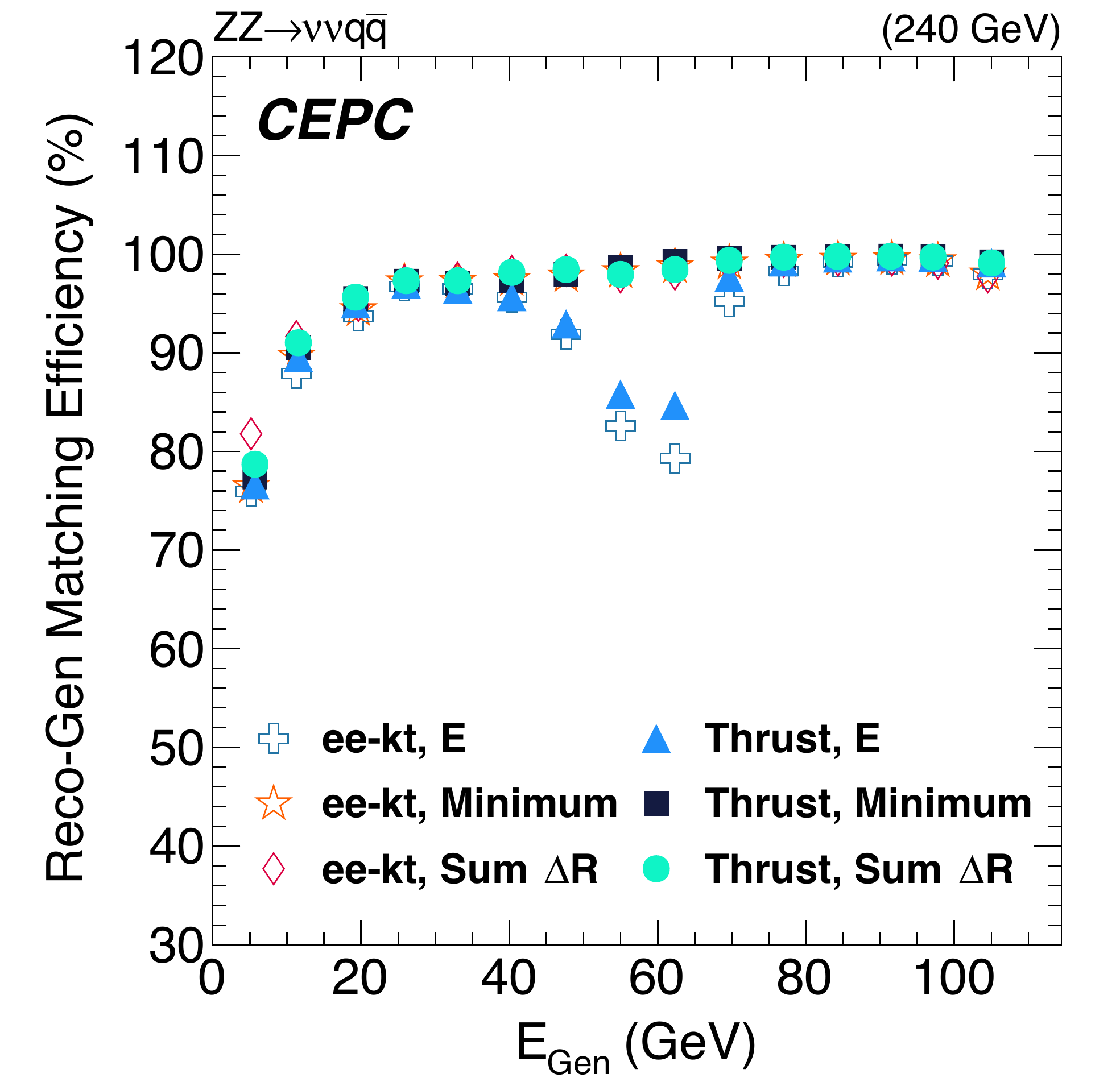}    
\end{minipage}%
}%
\subfigure[]{
\begin{minipage}[t]{0.35\linewidth}
\includegraphics[width=1.0\columnwidth]{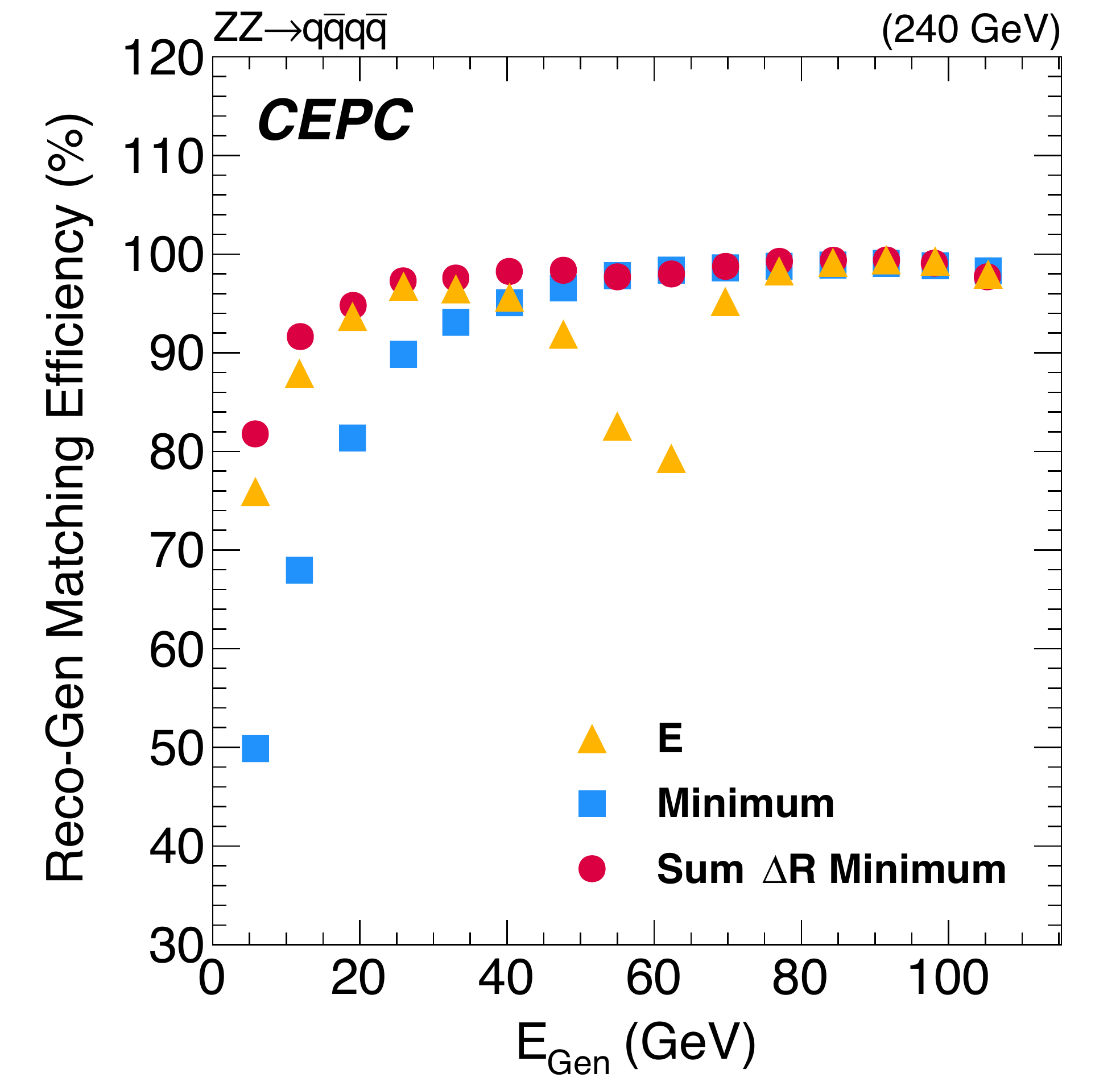}    
\end{minipage}%
}%
\vspace{-0.65cm}
\caption{The matching efficiency of the three methods as a function of the GenJet energy of the ZZ process to (a) semi-leptonic and (b) fully-hadronic decay events. The errors shown are only statistical. Because the energies of jets are similar at 60 GeV, the jet cannot be matched well using the energy, marked as "E" in the plots. Sum $\Delta R$ Minimum method has the highest matching efficiency in most cases.}
\label{Match_eff}
\end{figure}

Three matching approaches between the reconstructed jet and the true-level jet are studied. In the first approach, the RecoJets are paired to the GenJets according to their energy, with the leading RecoJet paired to the leading GenJet. This is labeled "E" in Fig.~\ref{Match_eff}. For the second approach, the most energetic RecoJet is paired to the GenJet that yields the minimum $\Delta R$, which is defined as $\sqrt{ (\Delta \theta)^2 \Plus (\Delta \phi)^2}$. Subsequently, the sub-leading RecoJet will match to the rest of the remaining GenJets with minimum $\Delta R$, and so on. If there are more than one GenJets with $\Delta R < 0.4$ to the RecoJet, the minimum $ |\Delta E|$, which is defined as the absolute value of relative energy difference between the RecoJet and GenJet, is employed. This approach is named as "Minimum" in Fig.~\ref{Match_eff}. Next, the third approach is to calculate the discriminating criteria for every possible matching combination since there must be a correct matching amid them. There are 2/24 possible combinations between GenJet-RecoJet under 2-/4-jet situations. The combination having the minimum sum $\Delta R$ value is used to quantify the jet performance. This approach is named as "Sum $\Delta R$ Minimum" or "Sum $\Delta R$" in Fig.~\ref{Match_eff}.

\begin{figure}[h]
\centering
\subfigure[]{
\begin{minipage}[t]{0.35\linewidth}
\includegraphics[width=1.0\columnwidth]{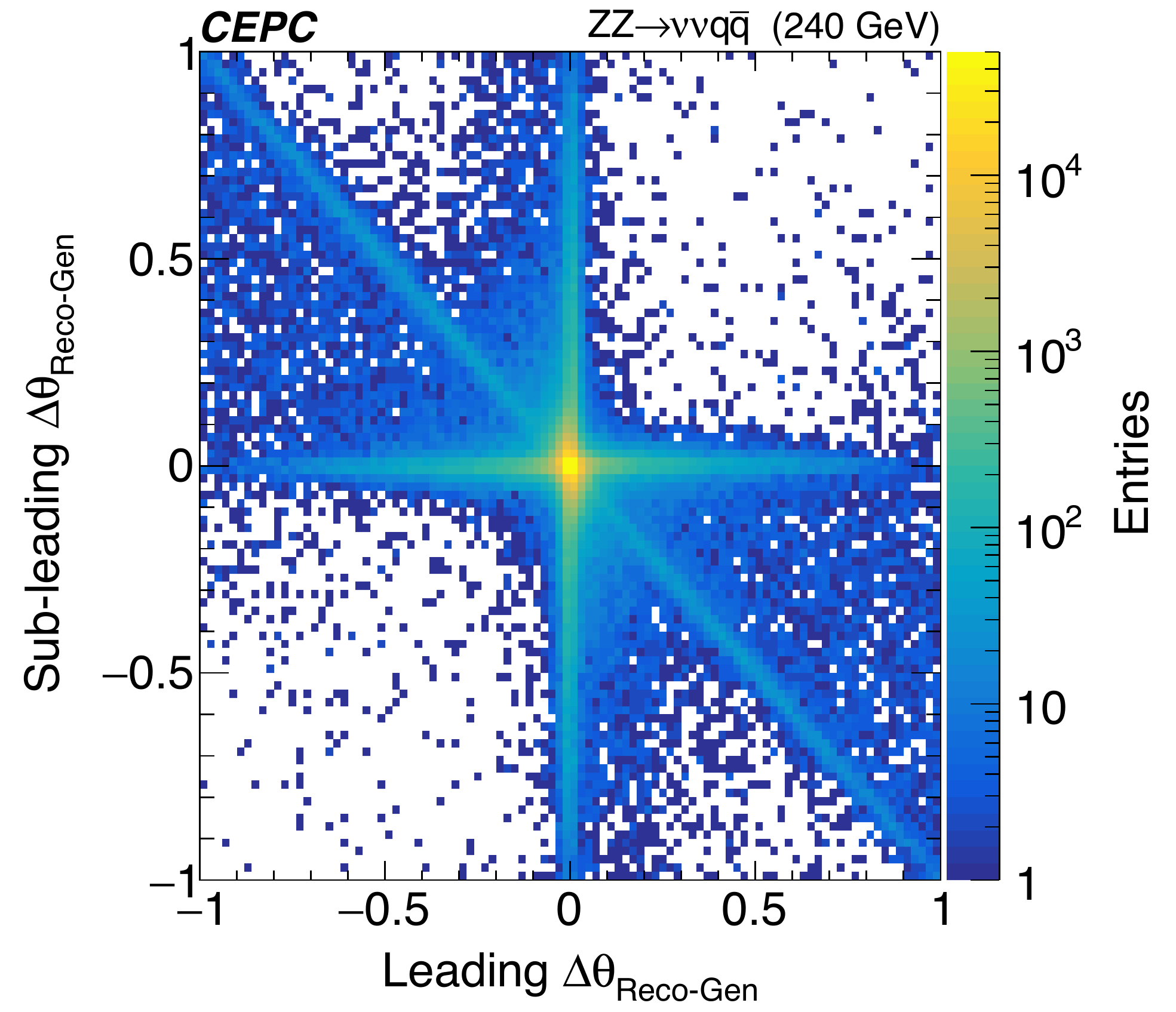}    
\end{minipage}%
\label{Match_demo_a}
}%
\subfigure[]{
\begin{minipage}[t]{0.35\linewidth}
\includegraphics[width=1.0\columnwidth]{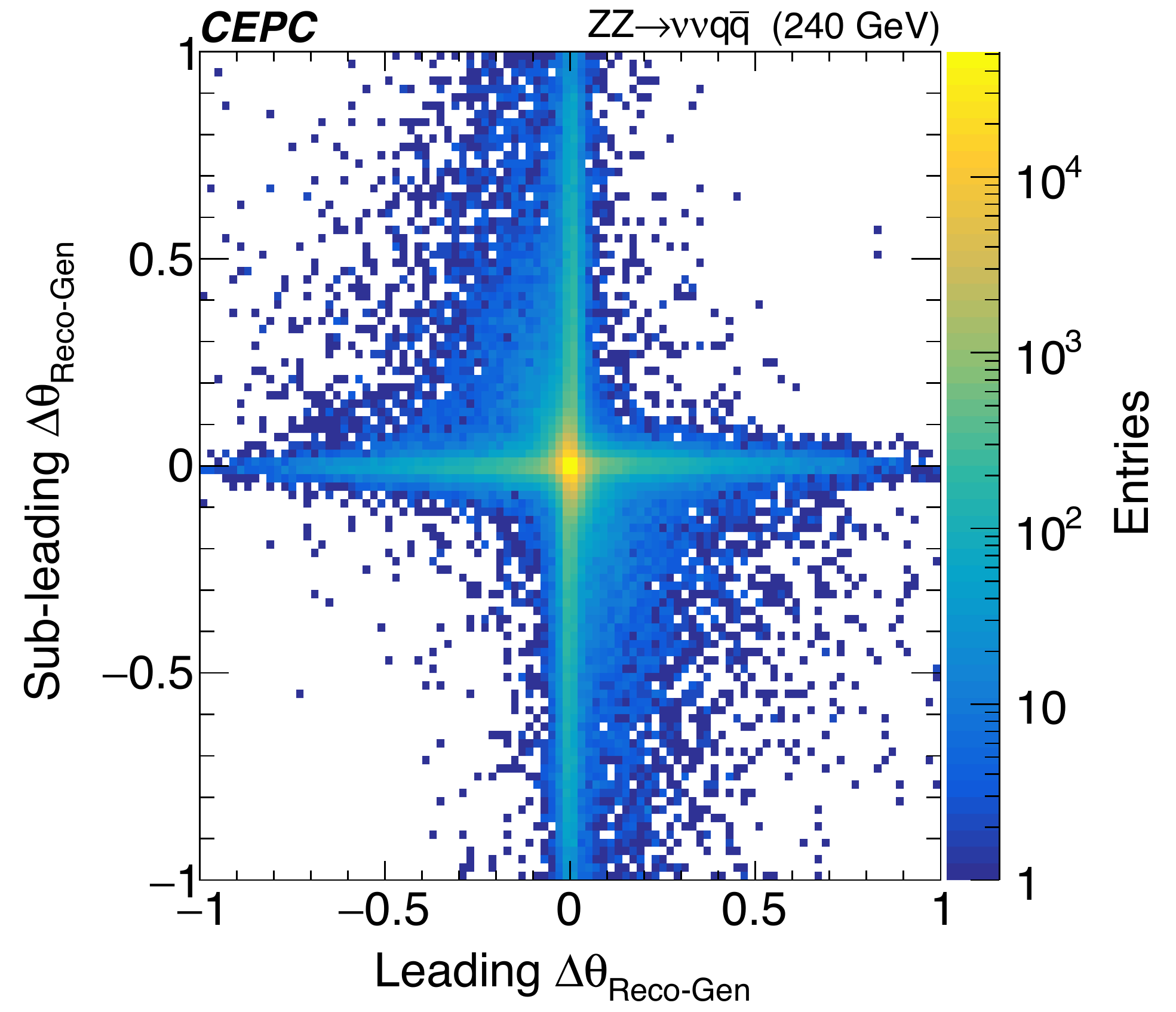}    
\end{minipage}%
\label{Match_demo_b}
}
\center
\vspace{-1.0cm}
\subfigure[]{
\begin{minipage}[t]{0.35\linewidth}
\includegraphics[width=1.0\columnwidth]{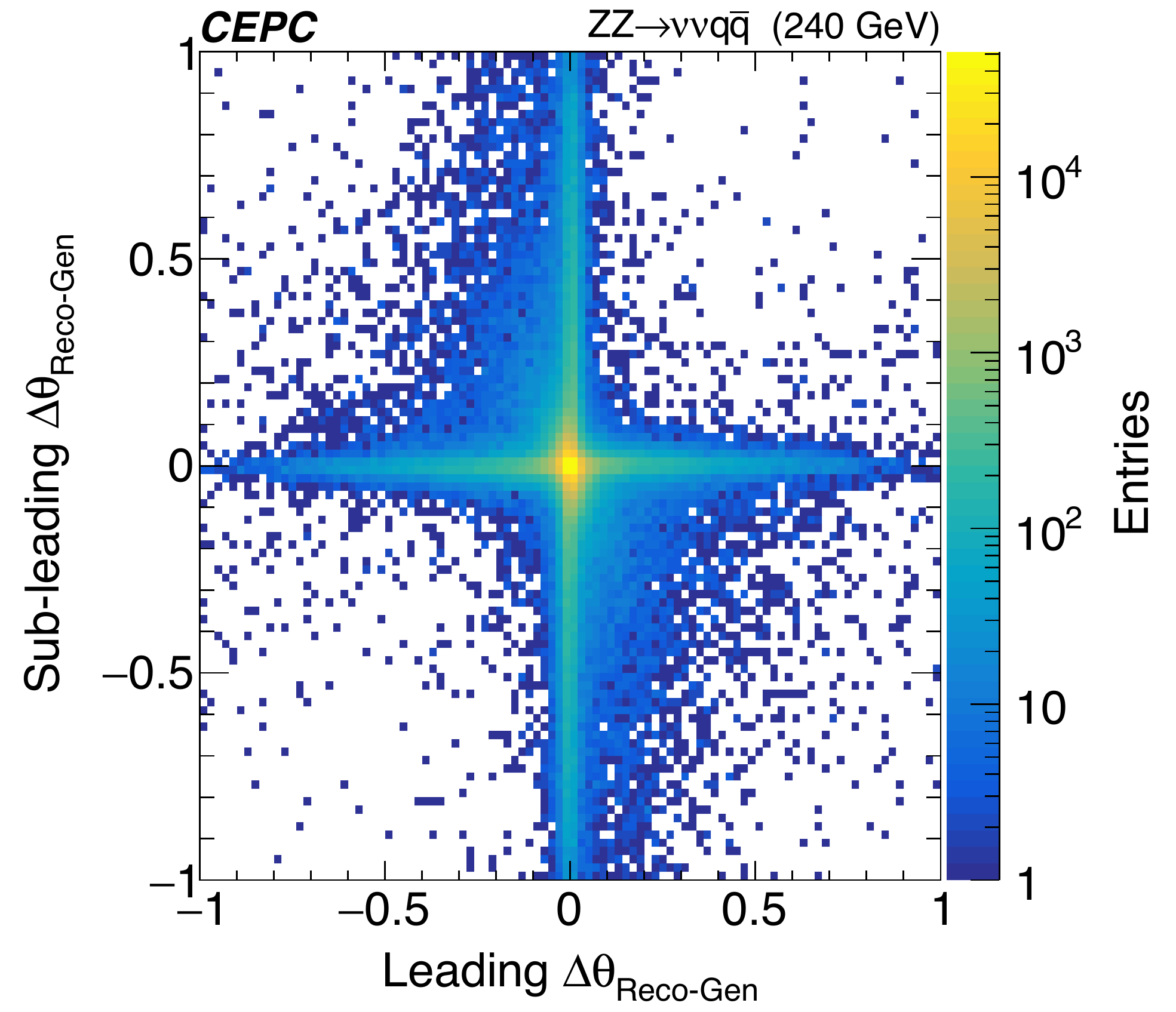}    
\end{minipage}%
\label{Match_demo_c}
}%
\vspace{-0.65cm}
\caption{The matching demonstration in terms of $ \Delta \theta $ the leading and sub-leading RecoJet and GenJet pairs for the ZZ process, where (a) is the Matching according to the jet energy, (b) is the Minimum, and (c) is the Sum $\Delta R$ Minimum matching approaches. The wrong matching is embodied by the diagonal.}
\label{Match_demo}
\end{figure}

To the first order, inappropriate matching methods can induce a wrong jet pairing between the RecoJet and GenJet which would embody in their $\Delta \theta$. After passing the clean selection~\cite{the_paper_ref}, several physical effects are removed. Fig.~\ref{Match_demo} shows the $\Delta \theta$ of the leading and sub-leading RecoJet and GenJet pairs for the three approaches. The diagonal shown in Fig.~\ref{Match_demo_a} is caused by the wrong jet matching which is not observed in the Sum $\Delta R$ Minimum and Minimum methods, Fig.~\ref{Match_demo_b} and Fig.~\ref{Match_demo_c}. 

To further discriminate the Sum $\Delta R$ Minimum and Minimum methods, the matching efficiency, the ratio between the number of RecoJet matched to GenJet within $\Delta R < 0.4$ and the total number of the GenJet in the particular energy bin, is characterized in Fig.~\ref{Match_eff} for all three methods. Because the spatial resolution of the jet is better than the energy resolution, the Sum $\Delta R$ Minimum method eventually has the highest matching efficiency for 2- and 4-jet events as shown in Fig.~\ref{Match_eff}. As a result, the Sum $\Delta R$ Minimum method is utilized in the jet performance studies. 

\clearpage

\bibliographystyle{JHEP} 
\bibliography{myfile}

\end{document}